\documentclass[10pt,a4paper]{article}

\usepackage{graphicx}
\usepackage{amssymb}
\usepackage{amsmath}

\usepackage[utf8]{inputenc}
\usepackage[T1]{fontenc}

\newcommand{\COMMENT}[1]{}

\usepackage[colorlinks]{hyperref}

\def\DREIECK#1{{\def\bull{}%
\count1=0
\loop
\edef\bull{$\bullet$\bull}
\ifnum\count1<#1
\advance\count1 by 1
\centerline{\bull}
\vskip-7.7pt
\repeat
\vskip 7.7pt\relax}}
\begin{document}

\setlength{\overfullrule}{10pt}

\title{Forming Point Patterns by a Probabilistic Cellular Automata Rule}

\author{
Rolf Hoffmann
\footnote{Technische Universit\"at Darmstadt, Darmstadt, Germany} 
}
\maketitle
%
\begin{abstract}
The objective is to find a Cellular Automata rule that can form a 2D point pattern with a maximum number of points (1-cells). Points are not allowed to touch each other, they have to be separated by 0-cells, and every 0-cell can find at least one point in its Moore-neighborhood. Probabilistic rules are designed that can solve this task with asynchronous updating and cyclic boundary condition. The task is considered as a tiling problem, where point tiles are used to cover the space with overlaps. A point tile consists of a center pixel (the kernel with value 1)  and 8 surrounding pixels forming the hull with value 0. The term pixel is used to distinguish the cells of a tile from the cells of a cellular automaton. For each of the 9 tile pixels a so-called template is defined by a shift of the point tile. In the rule application, the 9 templates are tested at the actual cell position. If all template pixels (except the central reference pixel) of a template match with the corresponding neighbors of the actual cell under consideration, the cell's state is adjusted to the reference pixel’s value. Otherwise the cell is set to the random value 0 or 1 with a certain probability. The hull pixels are allowed to overlap. In order to evolve a maximum of points, the overlap between tiles has to be maximized. To do that, the number of template hits is counted. Depending on the hit-number, additional noise is injected with certain probabilities. Thereby optimal patterns with the maximum number of points can be evolved. The behavior and performance of the designed rules is evaluated for different parameter settings.

\end{abstract}


\noindent\textbf{Remark:} This paper is an extension of the presentation ``Forming Point Patterns by a Probabilistic Cellular Automata Rule''
given at Summer Solstice Conference on Complex Systems in Dresden, Germany, July 15 -- 17, 2019
\normalsize
%
\newpage
\tableofcontents
\newpage

\section{Introduction}

\begin{figure}[b] 
\centering
\includegraphics[width=8cm]{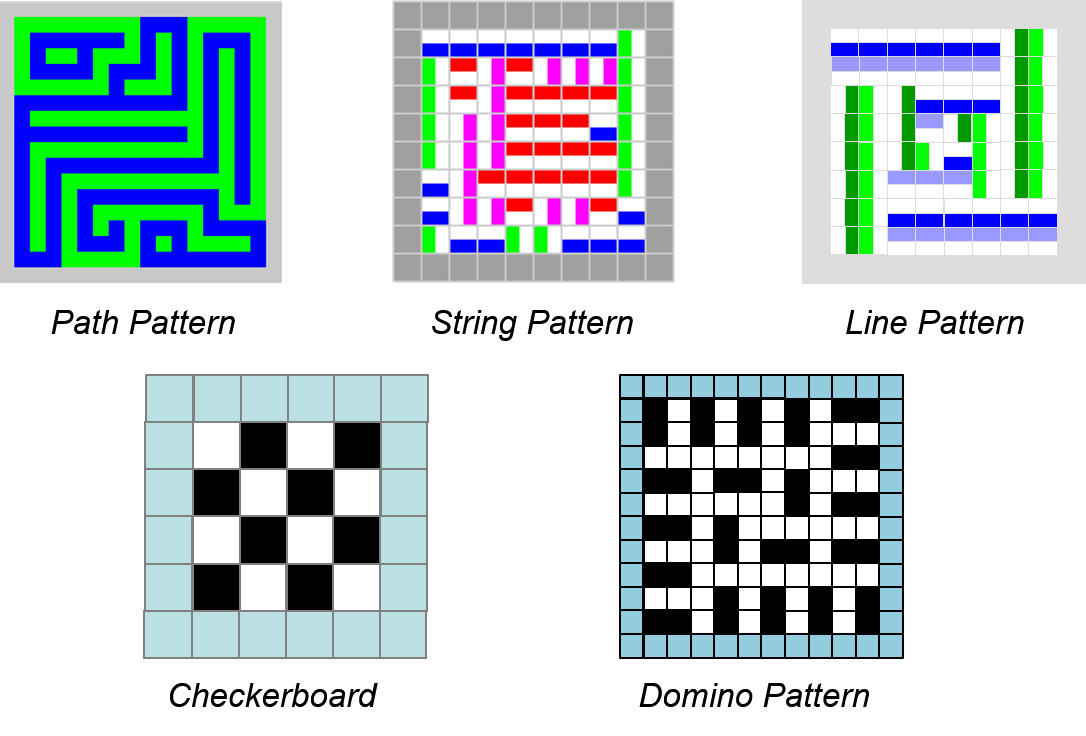}		
\caption{
Patterns evolved by CA agents.
}
\label{AgentPattern}
\end{figure}

This paper is an extension of the presentation ``Forming Point Patterns by a Probabilistic Cellular Automata Rule''
given at Summer Solstice Conference on Complex Systems in Dresden, Germany, July 15 -- 17, 2019 (see Appendix).
The introduced novel approach is methodical and will be explained here in more detail. 

\COMMENT{
\bibitem{2014-Hoffmann-ACRI-Pattern}
Hoffmann, R. 
(2014).
How Agents Can Form a Specific Pattern.
ACRI Conf. 2014, LNCS 8751, pp. 660-669

\bibitem{2016-Hoffmann-Polonia-PathPattern}
Hoffmann, R. 
(2016).
Cellular Automata Agents form Path Patterns Effectively.
Acta Physica Polonica B Proceedings Supplement, Vol. 9 (2016) No.1

\bibitem{2016-Hoffmann-D-ACRI-LinePattern}
Hoffmann, R. and  D{\'e}s{\'e}rable, D.
(2016).
Line Patterns Formed by Cellular Automata Agents.
ACRI Conf. 2016, LNCS 9863, pp. 424-434

\bibitem{2017-Hoffmann-D-PACT-MaxDomino-Agents}
Hoffmann, R. and  D{\'e}s{\'e}rable, D.
(2017).
Generating Maximal Domino Patterns by Cellular Automata Agents. 
Lecture Notes in Computer Science, PaCT Conference
}

In previous work 
\cite{
2014-Hoffmann-ACRI-Pattern,
2016-Hoffmann-Polonia-PathPattern,
2016-Hoffmann-D-ACRI-LinePattern,
2017-Hoffmann-D-PACT-MaxDomino-Agents}
different patterns were generated by moving Cellular Automata agents.
Such pattern are depicted in Fig. \ref{AgentPattern}.
The behavior of an agent was defined by an embedded finite state machine (FSM).
The agent's FSM was trained offline by a genetic algorithm (GA) to form a specific patterns. 
The patterns were locally defined by small pattern templates.
The number of matching templates was used to define the fitness function for the GA.
A population of agents with different FSMs was optimized by testing their performance through simulation:
Agents were favored that produced global patterns best by counting the local pattern matches . 

The effort to train such agents is quite high, especially to find agents that work on any field size. 
In order to avoid such a computational effort, a novel approach to construct directly the required CA rule 
was proposed that is described here. 
It has the potential to be applied to more complex pattern formations. 
Whereas the local matching templates are hidden in the FSMs of the moving agents in the former work, 
now the local matching templates are directly used in the definition of a classical uniform CA rule. 

This new approach was successfully applied 
to place a maximal number of dominoes in a 2D field 
\cite{2019-pact-domino},
to find a sensor point pattern (to cover an area by active sensors)
\cite{2020-HoffmannSeredynski-SensorPoint},
to place a maximal number of dominoes in the diamond
\cite{Hoffmann-2021},
and to cover a space with a minimal number of dominoes 
\cite{Hoffmann-2021-MinimalDominoPact}.

The objective is to find a point pattern with a maximal number of points by a CA rule. 
The cell's state is $\in\{0,1\}$ where `1' represents a point and `0' some material / space between points. 
Points are not allowed to touch each other, they have to be separated by 0-cells, and every
0-cell finds at least one point in its Moore-neighborhood.

To solve this problem we consider it as a tiling problem with overlapping \textit{point tiles}. 
A point tile is a $3\times3$ pixel array where  the center is `1' and the other pixels are `0'.
The task is to cover the space with overlapping point tiles without gaps.

Our problem is one of the diverse covering problems 
\cite{Snyder2011}
and it is related to the NP-complete \textit{vertex cover problem} introduced by Hakimi  \cite{Hakimi1965} in 1965. 
A vertex cover is a set of nodes in a graph such that every edge of the graph has at least one end point in the set. A minimum cover is a vertex cover which has the smallest number of nodes for a given graph.
Hakimi proposed a solution  method based on Boolean functions, 
later integer linear programming \cite{Gomesa2006}, branch-and-bound, genetic algorithm, and local search \cite{Richter2007} were used, among others. 
Other related problems are
the \textit{Location Set Covering  Problem} 
\cite{Church1976}
and the 
\textit{Central Facilities Location Problem} 
\cite{Mehrez2016}.
These problems aim to find the locations for \emph{P} facilities  that can be reached within a weighted distance from demand points, minimizing the number of \emph{P}, or minimizing the average distance, or maximizing the coverage. 
For covering problems there are a lot of applications, in economy, urban planning, engineering, etc.

\begin{figure}[tb] 
\centering
\includegraphics[width=0.32\textwidth]{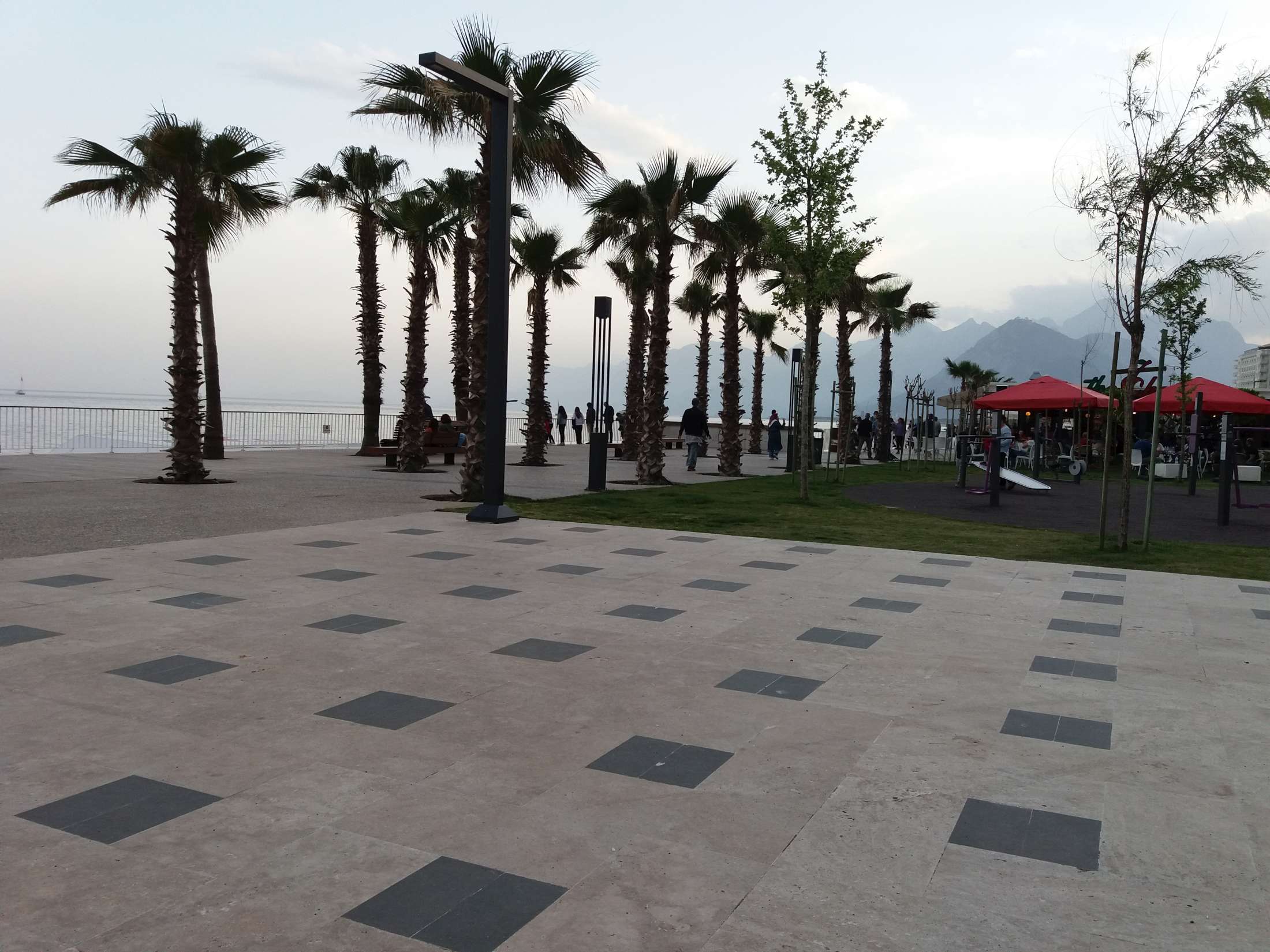}	
\includegraphics[width=0.32\textwidth]{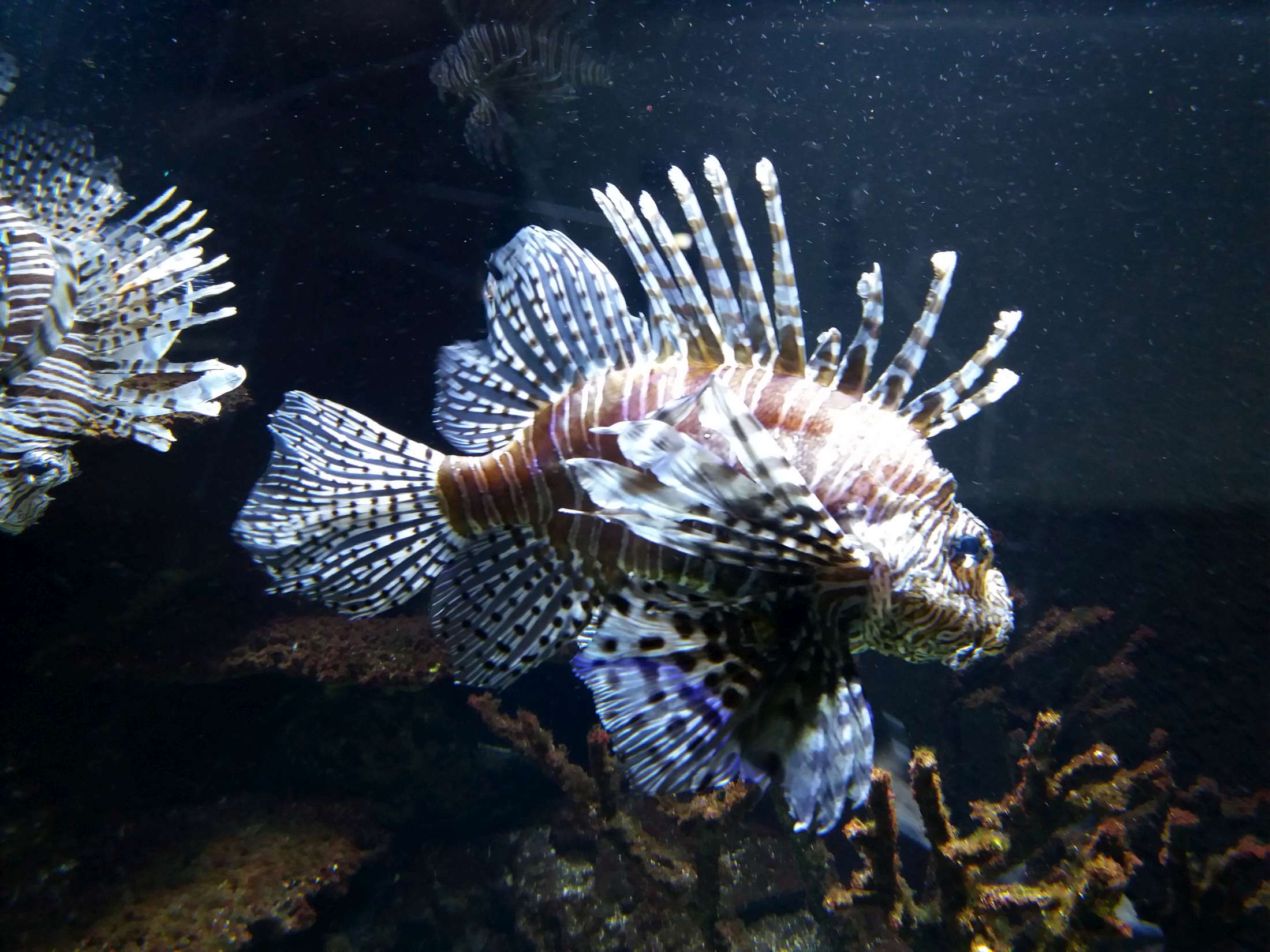}	
\includegraphics[width=0.32\textwidth]{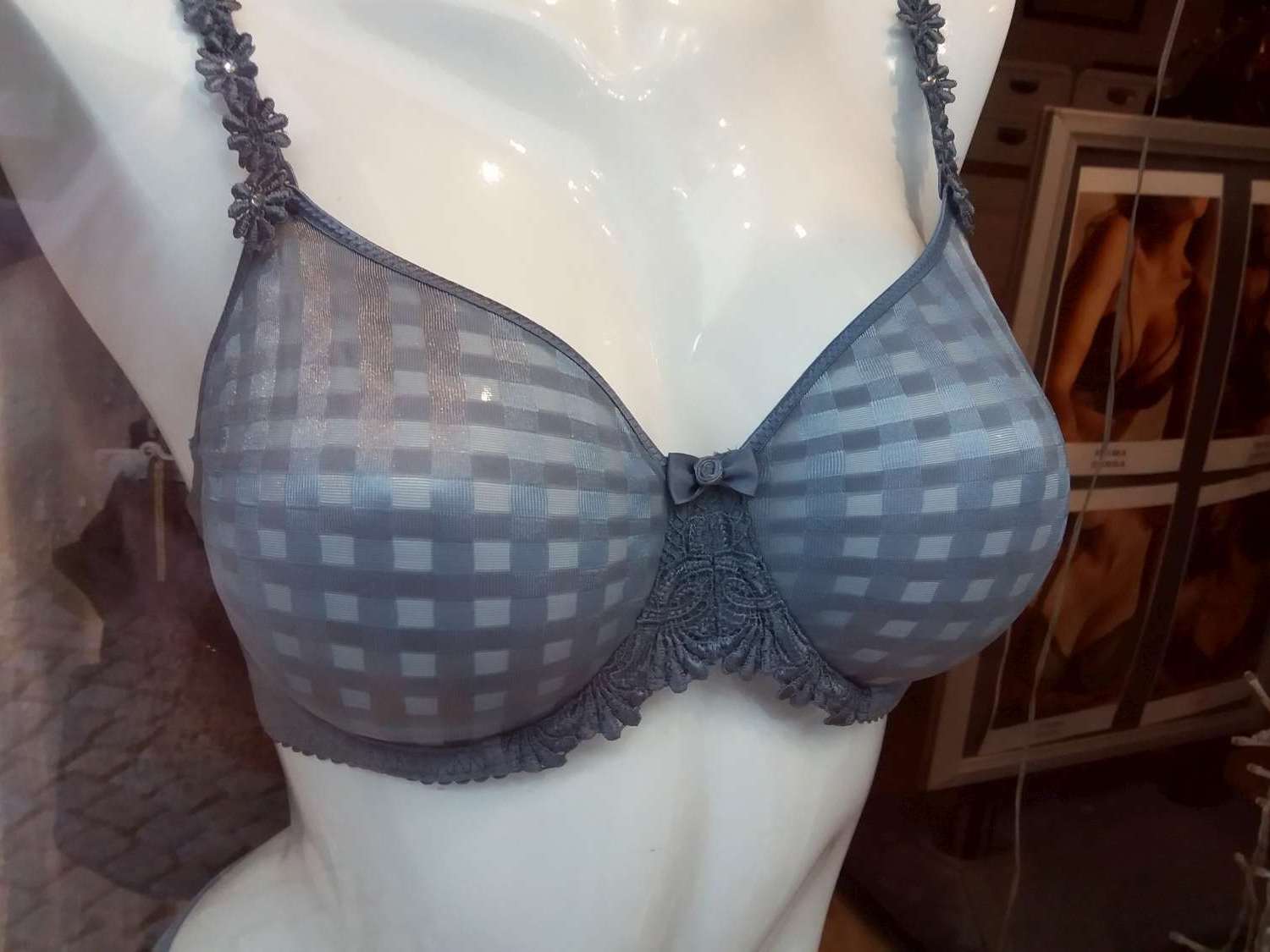}		
\caption{
Everyday's point patterns.
}
\label{Antalya2200}
\end{figure}

Sample applications for maximal point patterns are:
constructing a sieve with a maximum number of holes keeping a high stiffness,
constructing an effective brush,
dense packing of square parcels, or
attracting particles minimizing their total energy.
We can observe point patterns everyday, like the ones depicted in Fig. \ref{Antalya2200}

\begin{figure}[tb] 
\centering
\includegraphics[width=7cm]{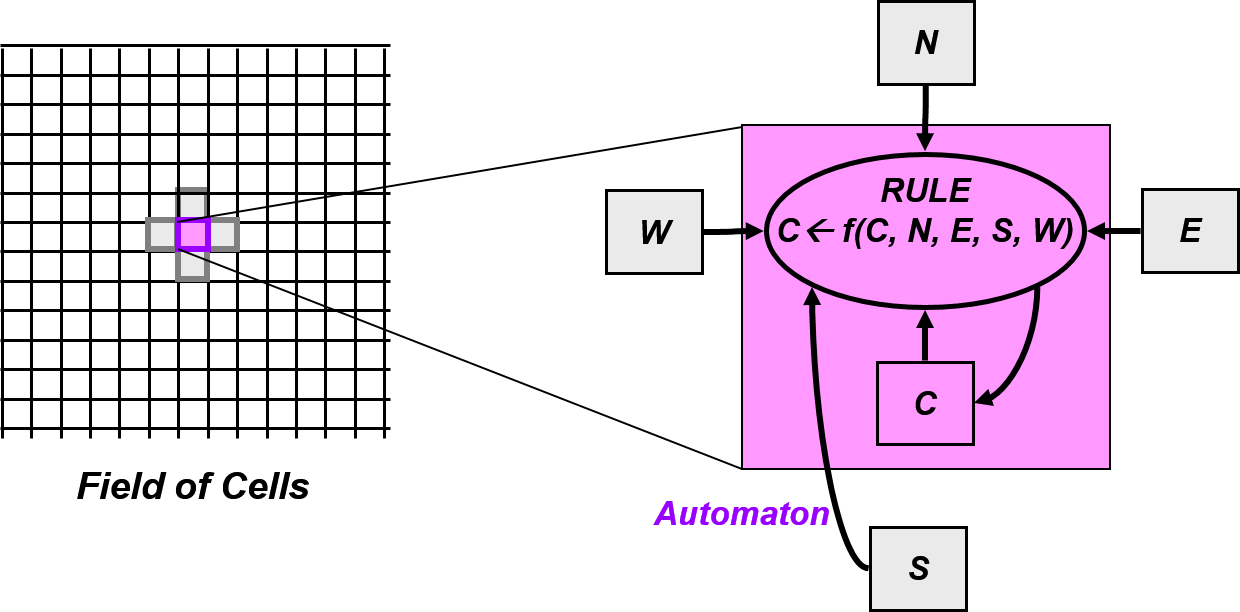}		
\caption{
Each cell is an Automaton, connected to its neighbors. 
C: Center Cell to be modified. N, E, S, W: neighboring cells
}
\label{4-Automaton}
\end{figure}
\begin{figure}[tb] 
\centering
\includegraphics[width=7cm]{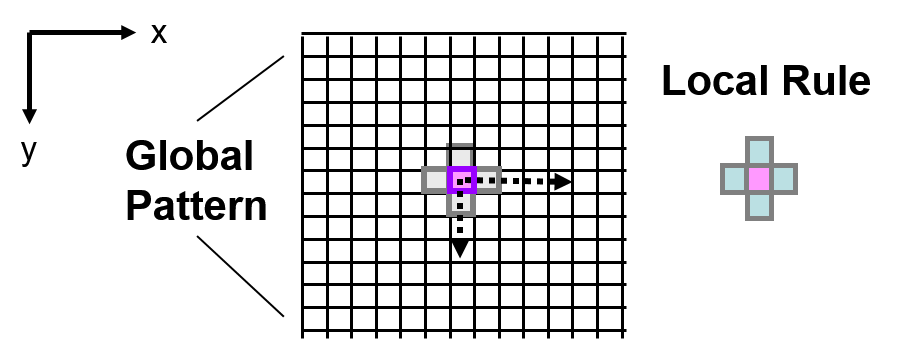}		
\caption{
A local rule is applied at every site $(x, y)$ of the array (field of cells). The rule is simple but it can induce a  complex global pattern.
}
\label{3-GlobalPattern-LocalRule}
\end{figure}

Cellular Automata (CA) is a well known modeling and decentralized computing paradigm
\cite{WolframNewKind}.
A CA is a field / array of cells placed in a grid with local connections 
(Fig. \ref{4-Automaton}).
The next cell's state is computed by a local rule $f$ taking the neighboring states into account:
$C\leftarrow f(C,N, E,S,W)$.
Every cell applies the same simple local rule which may result in a complex global pattern / configuration change
(Fig. \ref{3-GlobalPattern-LocalRule}). 

\section{The Problem and Solutions for \textit{n $\times$ n} Fields}
\subsection{The Problem}

Given is a field of  $n \times n$ cells with cyclic boundary conditions.
The cell's state is $s \in \{0,1\}$ where 0 will be represented by green or white and 1 by blue or black.
Initially we assume a random configuration. 

The objective is to find a CA rule that  successfully forms a \textit{Point Pattern}.
A point pattern consists of \textit{point} cells ($s=1$) and  \textit{zero} cells ($s=0$).   
There are two constraints that have to be fulfilled for any cell at site $(x,y)$ of the field.

\begin{enumerate}
\item
The 8 cells surrounding a point at $(x,y)$ in the Moore-neighborhood have to be zero. 
So the first constraint is:

$
s(x,y)=1 \wedge
s(x,y\pm1)=0 \wedge
s(x\pm1,y)=0 \wedge
s(x\pm1,y\pm1)=0~.
$

\item
At least one of the 8 cells surrounding a zero cell at $(x,y)$ in the Moore-neighborhood has to be a point. 
The second constraint is:

$s(x,y)=0 \wedge \exists  s(x',y')=1$ 

where $(x',y') \in \{   (x,y\pm1),(x\pm1,y),(x\pm1,y\pm1)  \}$~.

\end{enumerate}

The two conditions define allowed positions between zeroes and ones.
Simply speaking, points should be near to each other but are not allowed to touch. 
We call patterns that fulfill these constraints \textit{valid}.
We aim at valid patterns with a \textit{maximal} number of possible points $p_{max}$,
therefore we call our problem also the \textit{max point pattern problem}:

\vspace{10pt}
$p \rightarrow p_{max}$ where $p$ is the number of points in the pattern. 
\vspace{10pt}

\noindent
In addition we favorite a fast CA rule that rapidly produces a valid point pattern. 
Note that another problem would be to aim at patterns with a \textit{minimal} number of points,
the \textit{min point pattern problem}.  
This problem needs more discussion and will not be treated here further. 

\subsection{Optimal Solutions for \textit{n} Even} 

\begin{figure}[htb] 
\centering
\includegraphics[width=7cm]{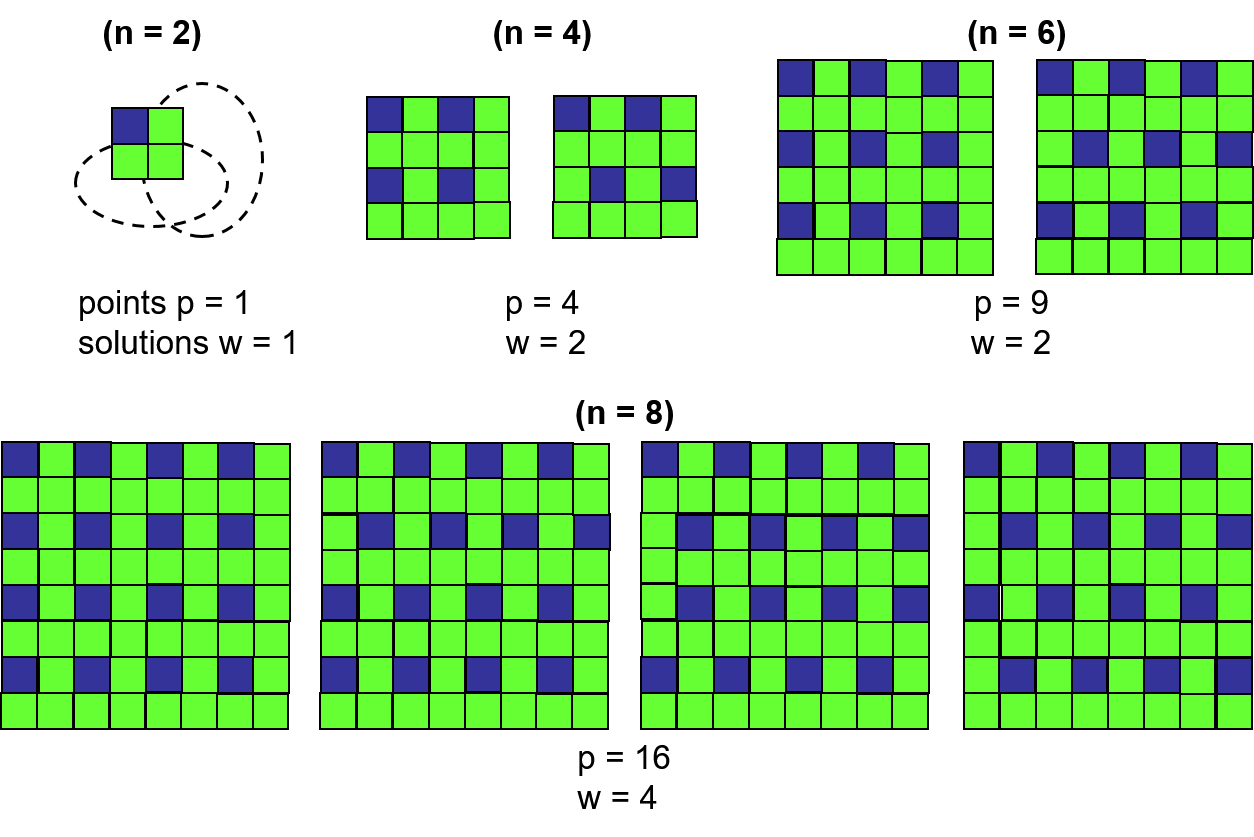}		
\caption{
Optimal patterns for even $n=2,4,6,8$. 
}
\label{SolutionsEven}
\end{figure}

First we look at the possible optimal solutions for $n \times n$ fields, where $n$ is even 
(Fig. \ref{SolutionsEven}).
We consider solutions as equivalent if they are equal under cyclic shift, rotation and mirroring. 
Let $w$ denote the number of different pattern solutions.
For $n=2$ there is only one solution with one point.
For $n=4$ we find $w=2$ solutions with $p=4$ points.
For $n=6$ we have 2 solutions with 9 points.
For $n=8$ we get 4 solutions with 16 points. 

In general the \textit{maximal} number of points is $p_{max}=n^2/4$
because the square cell field can be tiled into small non-overlapping blocks / tiles of size $2 \times 2$, each with one black cell / point. 

\section{Rule Design Issues}

\subsection{Updating Scheme}
We may use
\textit{synchronous} or  \textit{asynchronous} updating
and
a \textit{deterministic} or a \textit{probabilistic} rule.
This makes four options:

\begin{enumerate}
\item
 synchronous updating \& deterministic rule
\item
synchronous updating \& probabilistic rule
\item
asynchronous updating \& deterministic rule
\item
asynchronous updating \& probabilistic rule
\end{enumerate}

Our goal is to find a rule that produces valid point patterns and preferably converges to an optimal solution (\textit{max pattern}),
or finds several or even all optimal solutions during its evolution in time.  

\textit{Option 1:} 
Until now it was not possible to design such a rule.
The problem is that the evolving pattern may get stuck in sub-optimal local fixed or oscillating structures such as we know from the \emph{Game of Life}.
It remains an open questions if it is possible to find such a rule.

\textit{\textit{Options 2 -- 4:}}
These options are related because the computation of a new configuration is stochastic. It seems that they can be transformed into each other to a certain extent. 

We decided to use option 4 (\textit{asynchronous updating \& probabilistic rule}).
With asynchronous updating we don't need buffered storage elements and a central clock for synchronization which is closer to the modeling of natural processes.

Basically asynchronous updating can be considered as a sequential process where only one cell of the array is updated at a certain (micro) time-step. 
Nevertheless it may be possible to update cells of a subset in parallel if there is no influence between outputs and inputs within the subset. An example is to update odd and even cells alternately. 
We want to use the following  sequential updating scheme:

\begin{itemize}
\item
A cell is selected by a certain \textit{Selection Strategy}.

\item
The new cell's state-value is computed by the rule and immediately assigned to the cell's state memory
during each micro time-step 
$\tau \rightarrow \tau+1$,
i.e. $s^{\tau+1} \leftarrow Rule(s, neighbors' states)^\tau$.
\item
Each time-step $t \rightarrow t +1$ consists of $N = n^2$ micro time-steps.
This means that $n^2$ updates are performed during one time-step. 
\item
Configurations are considered / observed (logged, sampled) at time-steps. 
\end{itemize}

There are several possible selection strategies: 

\begin{enumerate}
\item 
\textbf{Pure Random}, 
also called \textit{random select, pure asynchronous, fully asynchronous, independent random ordering}.

For each new micro time-step, a cell is selected uniformly at random and then immediately updated.
Thereby a cell may not be updated during a time-step, or several times.

\item
\textbf{Random Sequence}, also called \textit{random new sweep}.

The update sequence is given by a random permutation of the cell indexes (defining the order), changed for every new time-step.
Thereby each cell is updated exactly once during one time-step.

\item
\textbf{Deterministic Sequence.}
The permutation of cell indexes is fixed. 
Examples: strict index order, first select odd cells then even cells, order in which the distance between cells is large or small on average.
\end{enumerate}

Experiments with the later defined CA rules showed that any of these selection strategies is working well with a relatively small influence on the performance.
Therefore an update strategy can be chosen that is easy to implement or which can work better in parallel. A deeper investigation is necessary to confirm this conclusion. 

\begin{figure}[htb] 
\centering
\includegraphics[width=9cm]{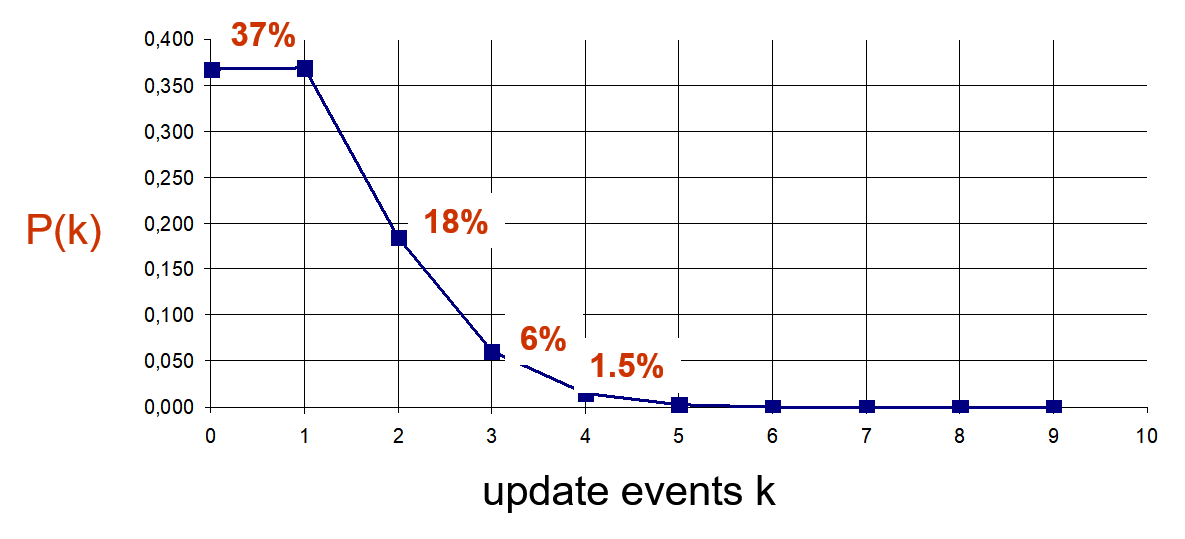}		
\caption{
The probability how often a cell is updated during a time-step with $N$
micro time-steps. $N = 16 \times 16$.
}
\label{UpdateEvents}
\end{figure}

\begin{figure}[htb] 
\centering
\includegraphics[width=9cm]{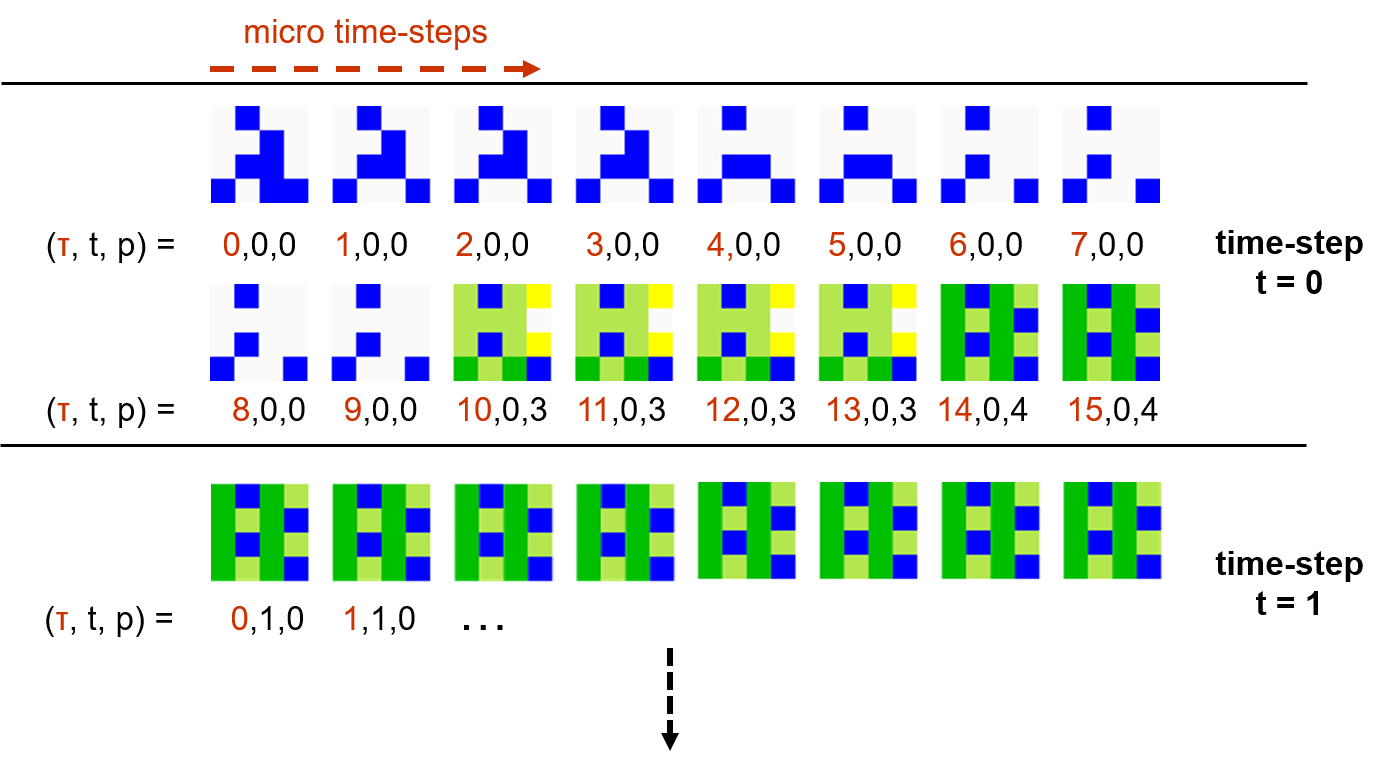}		
\caption{
Example of an evolution of a $4 \times 4$ 4-point pattern. The configurations at each micro time-step are shown. 
A time-step $t \rightarrow t +1$  consists of $n \times n$ micro time-steps $\tau \rightarrow \tau +1$ .
The update strategy was pure random.
Points are colored in blue, state = 1.
Colors for state = 0 and for cover levels: 
0 (white), 1 (yellow), 2 (light green), 3 (green), 4 (dark green).
}
\label{MicroSteps}
\end{figure}

Let us consider the strategy \textit{Pure Random}.
The probability $P(k)$ of $k$ update events for a specific cell during altogether $N$ events ($n^2$ micro-steps during one time-step) is given by the 
binomial distribution

\vspace{10pt}
$P(k)=(1-q)^{N-k}q^k \binom{N}{k}$ where $q=1/N$.
\vspace{10pt}
 
Fig. \ref{UpdateEvents} depicts the probability how often a cell is updated during a time-step, for $N = 16 \times 16$.
For larger fields with $N>16$ cells the graph changes only marginally. 
Notice that the probability of updating a cell never or once is $P(0)= P(1)=37\%$, and to update twice is $P(2)=18\%$.

Fig. \ref{MicroSteps} shows a sample evolution of a $4 \times 4$ 4-point pattern, micro step by micro step using 
pure random updating. 

\subsection{Tiling Problem}
\begin{figure}[htb] 
\centering
\includegraphics[width=9cm]{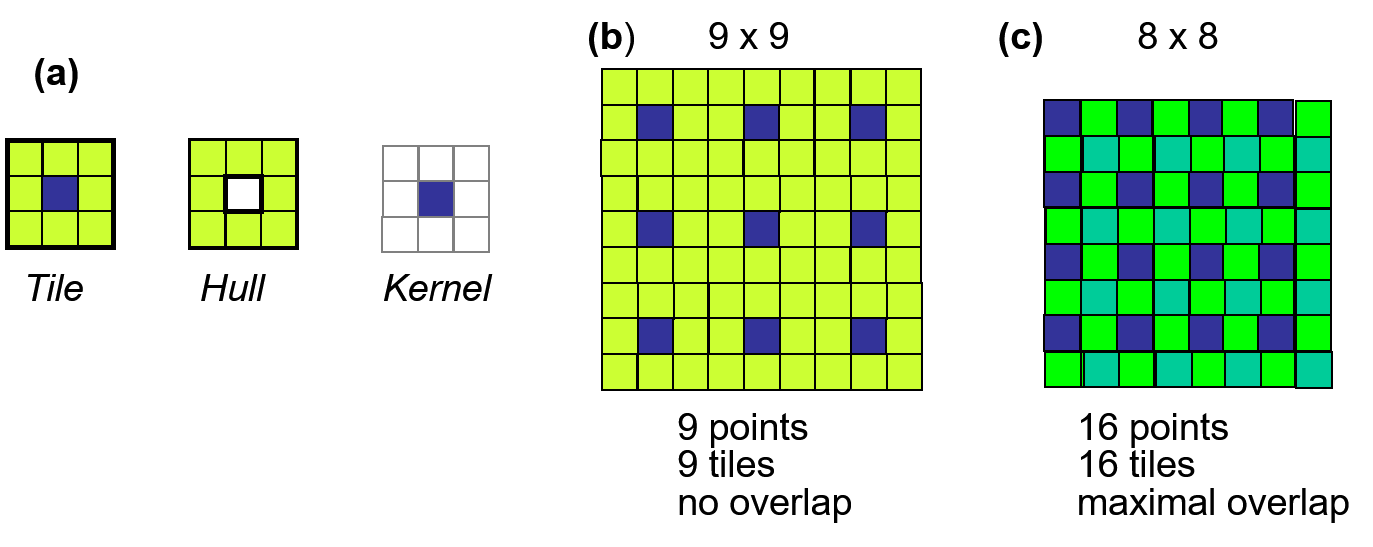}		
\caption{
(a) The used \textit{point tile} with its hull and kernel. 
(b) A cyclic $9 \times 9$ field can be tiled by 9 tiles without overlapping, a min pattern.
(c) A cyclic $8 \times 8$ field can be tiled by 16 tiles with maximal overlapping, a max pattern.
Overlapping  tile pixels are colored in green (cover level = 2) or dark green (cover level = 4).
}
\label{Tile}
\end{figure}

\begin{figure}[htb] 
\centering
\includegraphics[width=9cm]{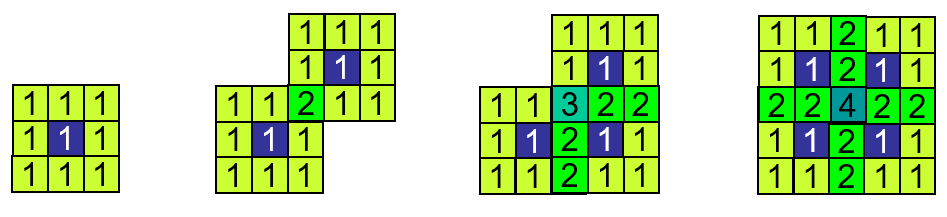}		
\caption{
Tiles may overlap. The numbers give the cover level $v$, the number of overlapping tiles / pixels.
}
\label{CoverLevel}
\end{figure}

The  basic idea is to consider the problem as a tiling problem.
Given is a certain tile, in our case we call it \textit{point tile} (Fig. \ref{Tile}a).
The point tile consists of 9 elements, we call them ``pixels'' and not ``cells'' 
because we reserve the word ``cell'' for cellular automata cells. 
The tile is partitioned into a hull and a kernel. 
The hull consists of 9 pixels with the value 0, and the kernel is the center pixel with value 1.

Now we want to cover a given cell field by (potentially) overlapping tiles and thereby find a solution of our problem.
Fig. \ref{Tile}b shows how a $9 \times 9$ field (cyclic boundary condition) can be covered with the minimum of 9 tiles without overlapping.
Fig. \ref{Tile}c shows how a cyclic $8 \times 8$ field can be covered with the maximum of 16 tiles with high overlapping.

In order to find a valid solution we have to allow that the tiles may overlap (Fig. \ref{CoverLevel}).
Gaps (uncovered cells) between tiles are not allowed.
The cover level $v$ gives the number of tiles / tile pixels that overlap at a certain site of the cell field. 
The kernel pixel is not allowed to overlap with any other pixel from another tile. 
So the cover level for the kernel is fixed to 1. 
Only hull pixels may overlap with other hull pixels, up to 4.
By covering each cell by overlapping tile pixels we obtain a \textit{valid} pattern fulfilling the above stated constraints.

We call a pattern \textit{max/min pattern} if the \textit{number of used tiles} (equals the \textit{number of points} $p$) is maximal resp. minimal.
In this paper we focus on the problem  how to produce max patterns, the \textit{max point pattern problem}.

Manually a valid pattern can be found by moving  tiles around in a given grid field and asserting the constraints.
Also a computer program could find valid patterns by moving tiles around, checking the constraints and 
maximizing or minimizing the number of tiles.  
But here we want to solve the problem in another way by using so-called \textit{templates} which are shifted tiles
or truncations of them.

\subsection{Templates}

\begin{figure}[htb] 
\centering
\includegraphics[width=9cm]{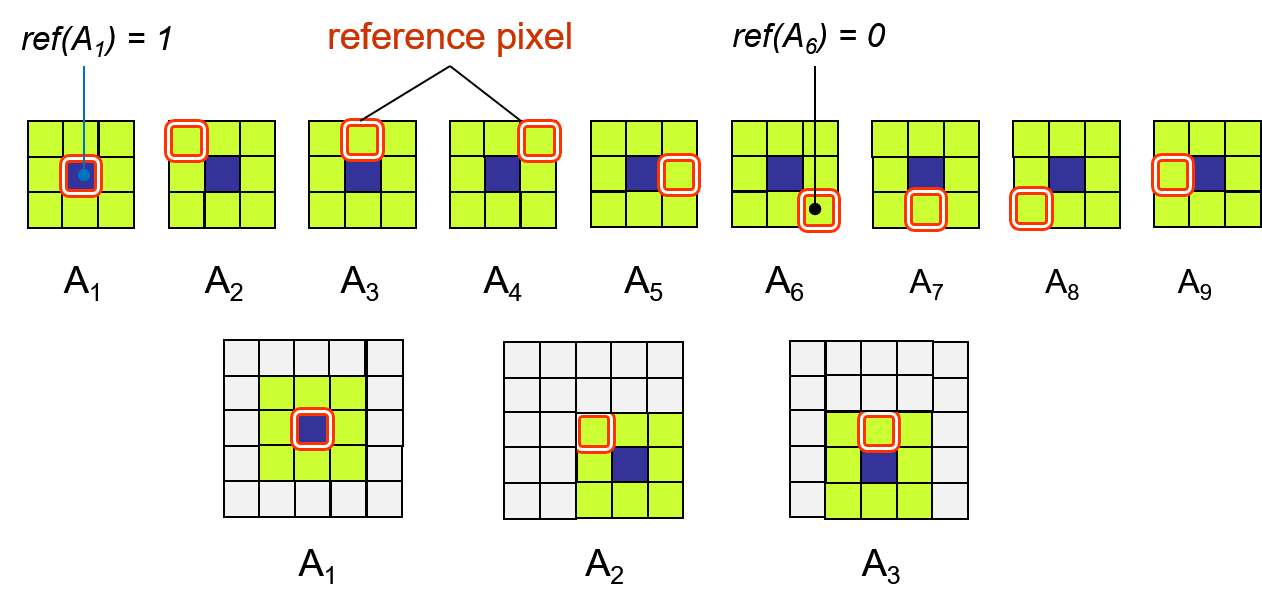}		
\caption{
The 9 templates derived from the point tile. For each pixel (encircled in red) a template is defined. The red marked pixel is called \textit{reference pixel}. The reference pixel serves as the center of a template. The value of the reference pixel is \textit{ref}($A_k$).
}
\label{Templates}
\end{figure}

Now we want to design CA rules. 
We define  ``templates'', small matching patterns that will be used in the later described CA rules. 
They a systematically derived from the point tile 
(Fig. \ref{Templates}).
Let the point tile be stored in a $3\times3$ pixel array $[G(i,j) ]$ 
with $i,j \in \{-1,0,+1\}$ and $G(i,j)\in \{0,1\}$. 

We reserve  template arrays $[A_k(i,j)]$ $(k=1 ~...~ 9)$,
with $i,j \in \{-2,0,+2\}$ and $A_k(i,j)\in \{\#,0,1\}$. 
The pixel value '\#' is used for unfilled ``don't care pixels''.
The size of the array has to be large enough to hold all ``real'' tile pixels  $G(i,j) \in \{0,1\}$.

For each pixel $G(i,j)$
a template  $[A_k(i,j) ]$ is derived by shifting $G$ 
in a way that each of the nine tile pixels $G(i,j)$ ends up as the center pixel of the \textit{k-th} template: 
$(i,j)_{\textit{of}~G}\rightarrow (0,0)_{\textit{of}~A_k}$.
The \textit{k-th} pixel $(i,j)$  of $G$ is called \textit{reference pixel} 
because it defines the center $(0,0)$ of one of the \textit{k} templates $A_k$.
Its value $G(i,j)$ is called \textit{reference value} $\textit{ref}(A_k)=A_k(0,0)$. 

For a more precise definition we use the shift operator 
$\textit{shift}_{(\Delta x, \Delta y)}(E)$ 
that
shifts the matrix $E$ in \emph{x}-direction by $\Delta x$
and in \emph{y}-direction (here defined downwards) by  $\Delta y$.
The 9 templates are generated by iterating over all tile pixels $(i,j)$:

\vspace{8pt}
\begin{tabular}{lll}
$A_1 \leftarrow \textit{shift}_{(0,0)}(E)$ 
&
$A_2 \leftarrow \textit{shift}_{(1, 1)}(E)$ 
&
$A_3 \leftarrow \textit{shift}_{(0, 1)}(E)$\\

$A_4 \leftarrow \textit{shift}_{(-1, 1)}(E)$ 
&
$A_5 \leftarrow \textit{shift}_{(-1, 0)}(E) $
&
$A_6 \leftarrow \textit{shift}_{(-1, -1)}(E)$\\ 

$A_7 \leftarrow \textit{shift}_{(0, -1)}(E)$ 
&
$A_8 \leftarrow \textit{shift}_{(1, -1)}(E) $
&
$A_9 \leftarrow \textit{shift}_{(1, 0)}(E)$ 
\end{tabular}
\vspace{8pt}

The reference values are `1' for $A_1$, and `0' for $A_2, A_3,\ldots, A_9$.

We have to be aware that the array size for storing templates $A_k$ may be larger than the array size of the point tile because of the shift operation.
An array of size $(3+\textit{Xshift}) \times (3+\textit{Yshift}) $ is sufficient large, 
where \textit{Xshift} / \textit{Yshift}  is the maximal needed shift count in  $x$- / $y$-direction,
i.e. $|i|_{max}$ resp. $|j|_{max}$.
Thus we need a $5 \times 5$ array to store any of the point templates. 
The don't care symbol '\#' is used for pixels that are not relevant in order to complete the whole template array.
For example, the templates $A_1, A_2, A_3$ can be encoded and stored as follows

\vspace{9pt}
\renewcommand{\baselinestretch}{.6}\normalsize
\begin{minipage}[h]{.9\textwidth}
\begin{verbatim}
      #####              #####              #####
      #000#              #####              #####
A1 =  #010#        A2 =  ##000        A3 =  #000#
      #000#              ##010              #010#
      #####              ##000              #000#   .
\end{verbatim}
\end{minipage}
\renewcommand{\baselinestretch}{1}\normalsize
\vspace{9pt}

By rotation, the templates $A_4, A_6, A_8$ can be generated from $A_2$.
Similarly $A_5, A_7, A_9$ can be generated from $A_3$.

For our later designed CA rule we use also so-called ``\textit{neighborhood templates}''. 
We yield the neighborhood template $A_i^*$ from the template  $A_i$ by setting the center to '\#'.
Note that this pixel is the reference pixel. 

\vspace{9pt}
\renewcommand{\baselinestretch}{.6}\normalsize
\begin{minipage}[h]{.9\textwidth}
\begin{verbatim}
      #####              #####              #####
      #000#              #####              #####
A1* = #0#0#        A2* = ###00        A3* = #0#0#
      #000#              ##010              #010#
      #####              ##000              #000#  
\end{verbatim}
\end{minipage}
\renewcommand{\baselinestretch}{1}\normalsize
\vspace{9pt}

\begin{figure}[htb] 
\centering
\includegraphics[width=8cm]{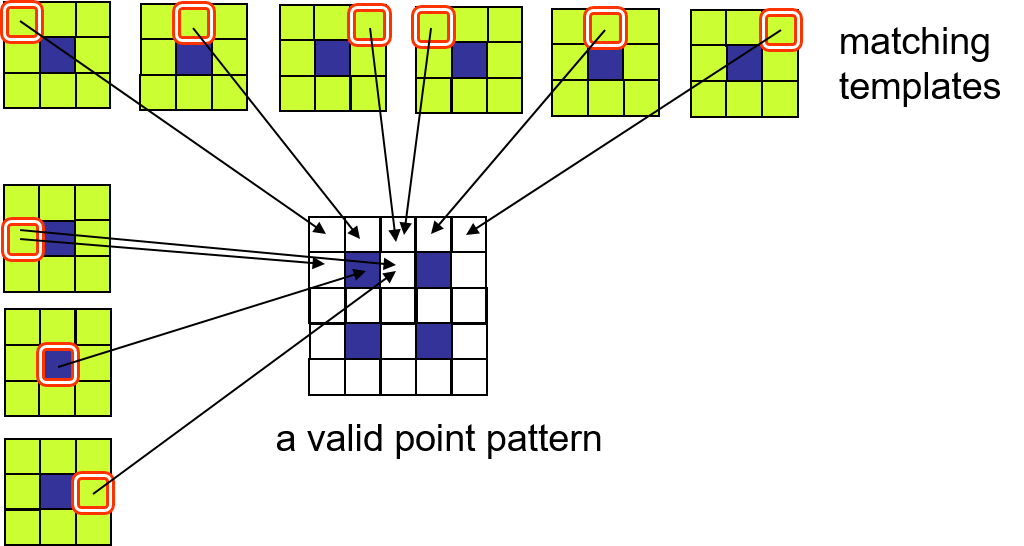}		
\caption{
For each cell of a valid point pattern exists at least one matching template (a hit).
}
\label{MatchingTemplates}
\end{figure}

Note that for each cell of a valid point pattern there exists at least one matching template, a hit,
see example in Fig. \ref{MatchingTemplates}.

\subsection{Reduced Templates}
It turned out by simulations of the CA rules (following Sect. \ref{RuleDesign}) 
that the template size of  $5 \times 5$ can be reduced to $3 \times 3$.

\vspace{9pt}
\renewcommand{\baselinestretch}{.6}\normalsize
\begin{minipage}[h]{.9\textwidth}
\begin{verbatim}
      000              ###              ###
A1 =  010        A2 =  #00        A3 =  000
      000              #01              010			
\end{verbatim}
\end{minipage}
\renewcommand{\baselinestretch}{1}\normalsize
\vspace{9pt}

It is even possible to further simplify $A_2$ and $A_3$ (as well as their symmetrical ones).
The reason is that it is sufficient to test for neighboring zeroes where there is a 1 (constraint 1) 
and to test for at least one 1-neighbor where there is a 0 (constraint 2).

\renewcommand{\baselinestretch}{.6}\normalsize
\begin{table}[h!]
\begin{verbatim}
      ###       ###            ###      ###   
A2 =  #00  ->   #0#      A3 =  000  ->  #0# 
      #01       ##1            010      #1#  
\end{verbatim}			
\end{table}			
\renewcommand{\baselinestretch}{1}\normalsize
\vspace{-9pt}

\section{The Designed Rules}
\label{RuleDesign}

We present now four designed rules that can be applied one after the other in the same time slot.

\begin{itemize}
\item
\textbf{Rule A.}
The current cell state is adjusted to the reference value if a neighborhood template matches.

\item
\textbf{Rule B.}
Noise is injected if there is no hit (match). This rule ensures that no cells remain uncovered.

\item
\textbf{Rule C1.}
In order to maximize the number of points, noise is injected if there is one hit. This rule drives
the evolution to higher cover levels

\item
\textbf{Rule C2.}
This rule drives the evolution to stable max patterns if $n$ is even. 

\end{itemize}

\subsection{Basic Rule A: Adjust}

\begin{figure}[htb] 
\centering
\includegraphics[width=9cm]{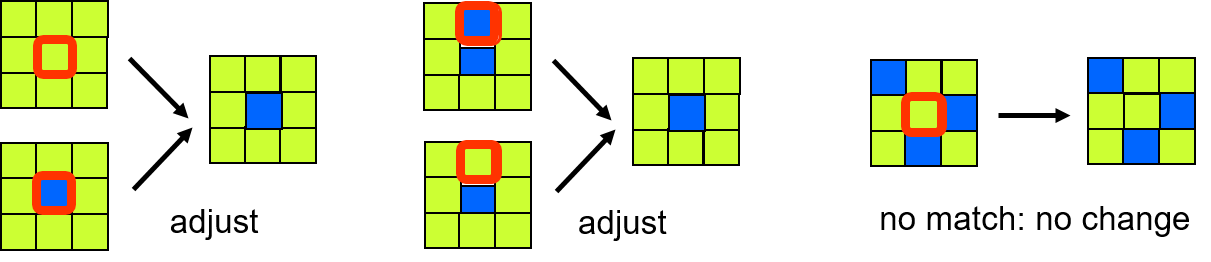}		
\caption{
Example for Rule A. A CA cell is adjusted if a neighborhood template match.
Otherwise the cell's state remains unchanged.
}
\label{Adjust}
\end{figure}

The templates can be used to test a pattern for validity.
In fact, the templates are defined in such a way that 
for each cell $(x,y)$ of a valid pattern there exists at least one matching template  
(Fig. \ref{MatchingTemplates}).
The basic idea is to adjust templates, make them complete.
When the tested cells in the neighborhood of cell $(x,y)$ are equal to the template except the center (the reference pixel of the template),
then cell  $(x,y)$ is adjusted  to the correct value (the reference value). 
So we may use the neighborhood templates $A_k^*$ for testing against the corresponding CA neighbors and adjust anyway.
This is performed by the following Rule A (case (a)).
Otherwise the new state is equal to the old state (case (b)). 

\[
s'(x,y) =
\left
\{
\begin{array}{lll} 

\textit{ref}(A_i)  &\textbf{if} ~\exists A_i^*~\textit{that matches with CA neighbors at $(x,y)$} &(a) \\ 
                      
s(x,y)                &\textbf{otherwise} \textit{~no change}                                                         &(b) \\
\end{array}
\right. .
\]

This rule evolves very fast stable point patterns. 
During the process of the CA rule application the pattern may not be valid or only partially.
In partially valid pattern some tiles are detected through  template testing, but there 
exist some noisy or uncovered cells. 
The number of template matches are called ``hits''.

\begin{figure}[htb] 
\centering
\includegraphics[width=9cm]{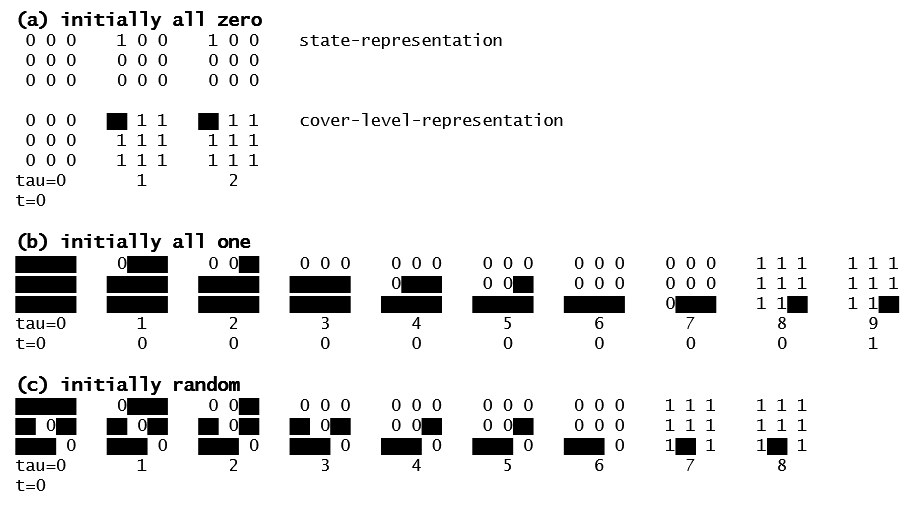}		
\caption{
The evolution of a $3 \times 3$ field under different initial conditions and strict index update order.
}
\label{SimulationRuleA}
\end{figure}

Fig. \ref{SimulationRuleA}
shows simple simulations of a  $3 \times 3$ field with strict index update order,
with different initial states: (a) all zero, (b) all one, and (c) random.

\begin{itemize}
\item[(a)]
 We observe that a point appears already at micro time-step $\tau=1$.
This happens because already for the first cell the neighborhood template $A_1^*$ is fulfilled, `0's are detected in the whole
Moore-neighborhood (taking cyclic boundary condition into account).
The pattern remains stable, because all other cells match with a template (a `1' is detected in the neighborhood).

\item[(b)]
 8 micro time-steps are needed to evolve a stable point pattern (with one point).
At $\tau=0$ the first cell is checked for matching templates. 
Since all 8 reduced neighborhood templates $A_{i\geq 1}^*$ match (there exist `1'-neighbors),
the active cells  are set to `0', step by step in strict index order until only one `1' (the point) is left over.

\item[(c)]
 7 micro time-steps are needed to evolve a stable point pattern, starting with a random configuration.
The number of time-steps depends on the actual initial configuration. 
Averaging over 100  random initial configurations we get 
$\tau_{average}= 6.5 (2  ~min - 8 ~max)$.
\end{itemize}

\begin{table}[htb] 
	\caption{
	Rule A applied on a field of size $3 \times 3$.	
	Micro time-steps needed for different update orders, starting with an all one configuration. 	
	All 100 evolved patterns are stable and contain 1 point.}	
\begin{center}
  \begin{tabular}{ |c | c | c | c | c |}
    \hline		
     $3\times 3$ \textit{init all one}     & $\tau_{average}$   & $\tau_{min} - \tau_{max}$  & \textit{time-steps}=   \\
                                           &                    &                            & $\tau_{average}/9$      \\
    \hline		
    strict index order     & 8       & 8  - 8      & 0.89  \\
		
		\hline
    random sequence  & 17.03   & 10 - 32    & 1.89  \\
		
		\hline
    pure random  & 19.90       & 12 - 27    & 2.21  \\
    \hline

  \end{tabular}
\end{center}
\label{TableRuleA3x3}
\end{table} 

\begin{table}[htb] 
	\caption{
	Rule A applied on a field of size $4 \times 4$.	
	Micro time-steps needed for different update orders, starting with different initial configurations. 
	All 100 evolved patterns are stable and contain 4 points.} 
\begin{center}
  \begin{tabular}{ |c | c | c | c |}
     \hline
     $4\times 4$ \textit{init all zero}   & $\tau_{average}$   & $\tau_{min}  - \tau_{max}$  & \textit{time-steps}=   \\
                                           &                    &                             & $\tau_{average}/16$     \\
    \hline
    strict index order     & 11     & 11  - 11  & 0.69 \\ 
		\hline
    random sequence  & 25.64   & 7 - 87 &    1.60  \\ 
		\hline
    pure random     & 21.67   & 6 - 52   & 2.17 \\  
    \hline
%
    
    \hline		
        \hline		
     $4\times 4$ \textit{init all one}          & $\tau_{average}$   & $\tau_{min}  - \tau_{max}$  & \textit{time-steps}=\\ 	
                                                 &                    &                             & $\tau_{average}/16$     \\
    \hline		
    strict index order     & 26      & 26  - 26  & 1.63  \\ 
		\hline
    random sequence  & 50.23   & 16 - 102 & 3.14    \\ 
		\hline
    pure random     & 49.68   & 28 - 81   & 3.11 \\  
    \hline

    
    \hline		
        \hline		
     $4\times 4$   \textit{init random }       & $\tau_{average}$   & $\tau_{min} - \tau_{max}$  & \textit{time-steps}=\\ 
                                                 &                    &                             & $\tau_{average}/16$     \\
    \hline		
    strict index order      & 24.06         & 16 - 28      & 1.50 \\ 
		\hline	
    random sequence     & 45.29      & 13  - 136      & 2.83 \\ 		
		\hline
    pure random         & 43.29       & 7 - 126    & 2.71  \\
    \hline
  \end{tabular}
\end{center}
\label{TableRuleA4x4}
\end{table} 

Another test was performed on a  $3 \times 3$ field in order to assess the influence of the update order.
Starting with an initial configuration of all ones, we get the results shown in Table \ref{TableRuleA3x3}.
 For averaging, 100 runs were performed. 
A stable point pattern evolved fastest with strict order updating. 
Though this is only a small example it shows that strict sequential updating can perform very well what we can also observe for larger fields.
Nevertheless a deeper analysis is necessary to confirm this observation. 

Another test was performed for a  $4 \times 4$ field, varying the initial condition (all zero, all one, random) and the update order
 (Table \ref{TableRuleA4x4}).
100 runs were performed for averaging. 
Starting with a initial all one or a random configuration, random sequence order  
and pure random show about the same performance.
We have to notice that the performance results statistically vary if another set of runs is analyzed. 
So these results could be evaluated more accurately by increasing the number of runs (e.g., up to 10 000).
Nevertheless we have observed that all theses different sequential updating orders worked well, where strict index order 
worked fastest.

Now, by simulation of the Cellular Automata  Rule A we want to get an impression of the patterns that can evolve for increasing larger fields.

\subsubsection{4 $\times$ 4 Patterns}
\begin{figure}[htb] 
\centering
\includegraphics[width=12cm]{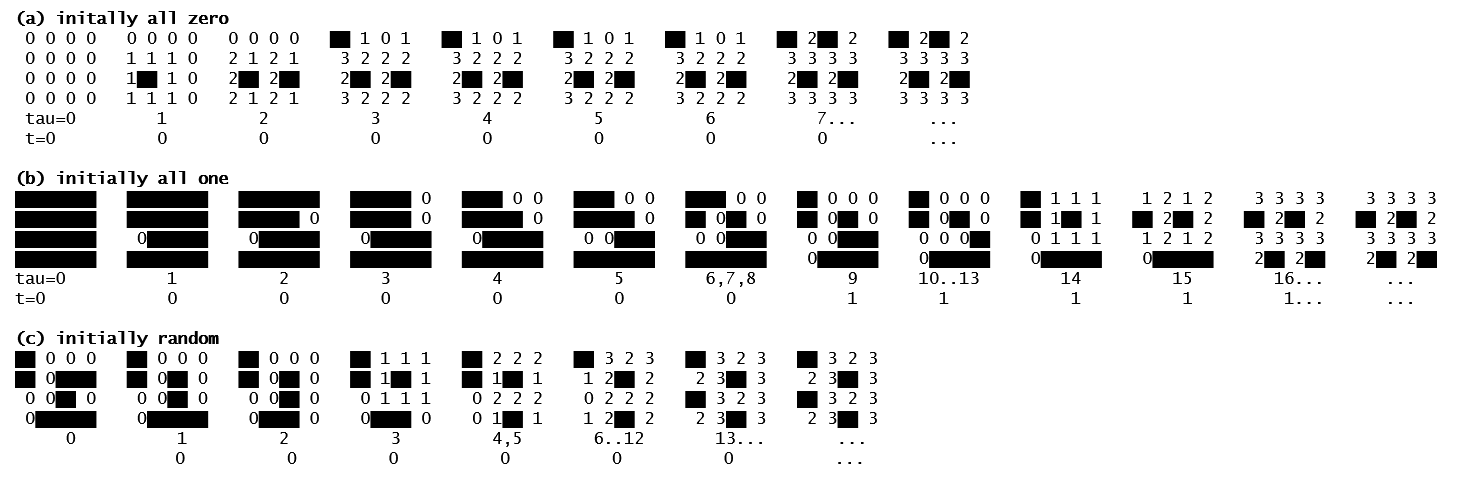}		
\caption{
The evolution of a $4 \times 4$ field under different initial conditions and random sequence update order.
}
\label{SimulationRuleA4x4}
\end{figure}

Fig. \ref{SimulationRuleA4x4} shows 
the evolution of a $4 \times 4$ field under different initial conditions and random sequence update order.
Similar 4-point patterns evolved. Notice that also similar 
symmetric patterns under reflection, shift and rotation may evolve.

\COMMENT{
The following  pattern is also a valid outcome.

\begin{verbatim}
                           0 0 0 0
                           0 1 0 1
                           0 0 0 0
                           0 1 0 1
\end{verbatim}
}

\subsubsection{5 $\times$ 5 Patterns}
\begin{figure}[htb] 
\centering
\includegraphics[width=12cm]{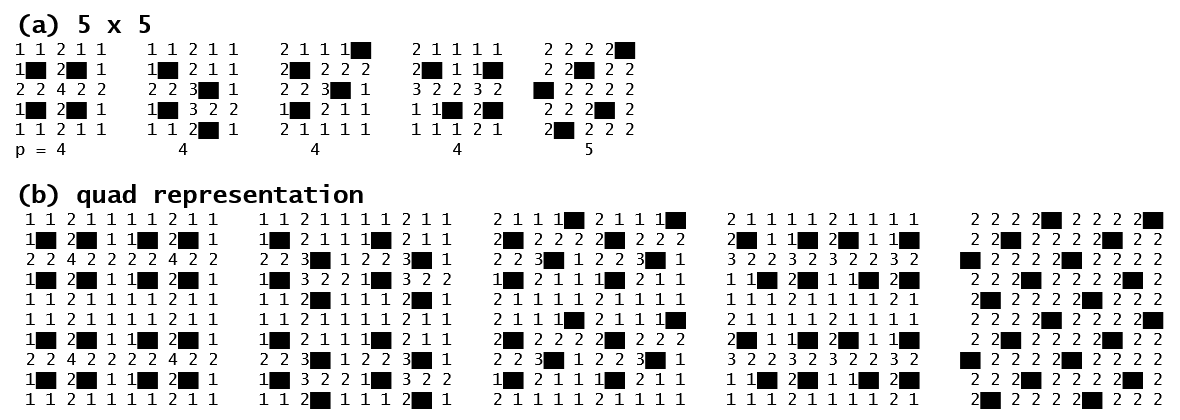}		
\caption{
Some valid $5 \times 5$ patterns evolved by Rule A.
(a) Patterns with $p=4,5$ points.
(b) Quad pattern representation: the (a) patterns are doubled in horizontal and vertical direction. 
Quad patterns allow to  observe  the regularities better.
}
\label{RuleA5x5}
\end{figure}

Now we can observe from 
Fig. \ref{RuleA5x5}
that patterns with 4 or 5 points may evolve, 
whereas for the size $4 \times 4$ only patterns with 4 points evolve.
We call valid patterns with a minimum number of points \emph{min patterns}
and  with a maximum number of points \emph{max patterns}.
The average cover level is 
$v_{avrg}(p,n)=9p/n^2$, ~where $p$ is the number of points.
So we get $v_{avrg}(4)=36/25=1.44$ for 4 points
and  $v_{avrg}(5)=1.8$ for 5 points.
There are different 4-point patterns with a different layout of the points. 

The point layouts of 
Fig. \ref{RuleA5x5}a
show a different distribution of the cover level.
The first 4-point pattern has the occurrence/frequency $f(v)$ of 
$(4=p)+12, 8, 0, 1$ for $v=1,2,3,4$.
We may use the equivalent notation $1^{4+12} 2^8 3^0 4^1$.

Equivalent pattern under symmetries have the same cover level distribution.

The second 4-point pattern has the distribution $1^{4+12} 2^7 3^2 4^0$. 

The third 4-point pattern has the distribution $1^{4+12} 2^9 3^1 4^0$.

The forth 4-point pattern has the distribution $1^{4+12} 2^7 3^2 4^0$.

The second and forth pattern have the same distribution but are not similar under symmetries.

The fifth pattern has 5 points and the distribution $1^{5+0} 2^{20} 3^0 4^0$.
It is the most regular, the distance from any point to its four nearest neighbors is equal ($\Delta x=1, \Delta y=2$).

In order to recognize better the inherent periodic pattern structure we can use the ``quad representation'' (Fig. \ref{RuleA5x5}b),
the pattern is repeated 4 times, twice in horizontal and twice in vertical direction. 
This representation allows also to detect equivalent patterns under symmetries. 
In general we may repeat the pattern horizontally \emph{U} times and vertically \emph{V} times
because of the periodic boundary condition. 
In the following section (Fig. \ref{RuleA6x6}b) we can see a $6 \times 6$ pattern repeated 9 times $(U=V=3)$.

\subsubsection{6 $\times$ 6 Patterns}
\begin{figure}[htb] 
\centering
\includegraphics[width=8cm]{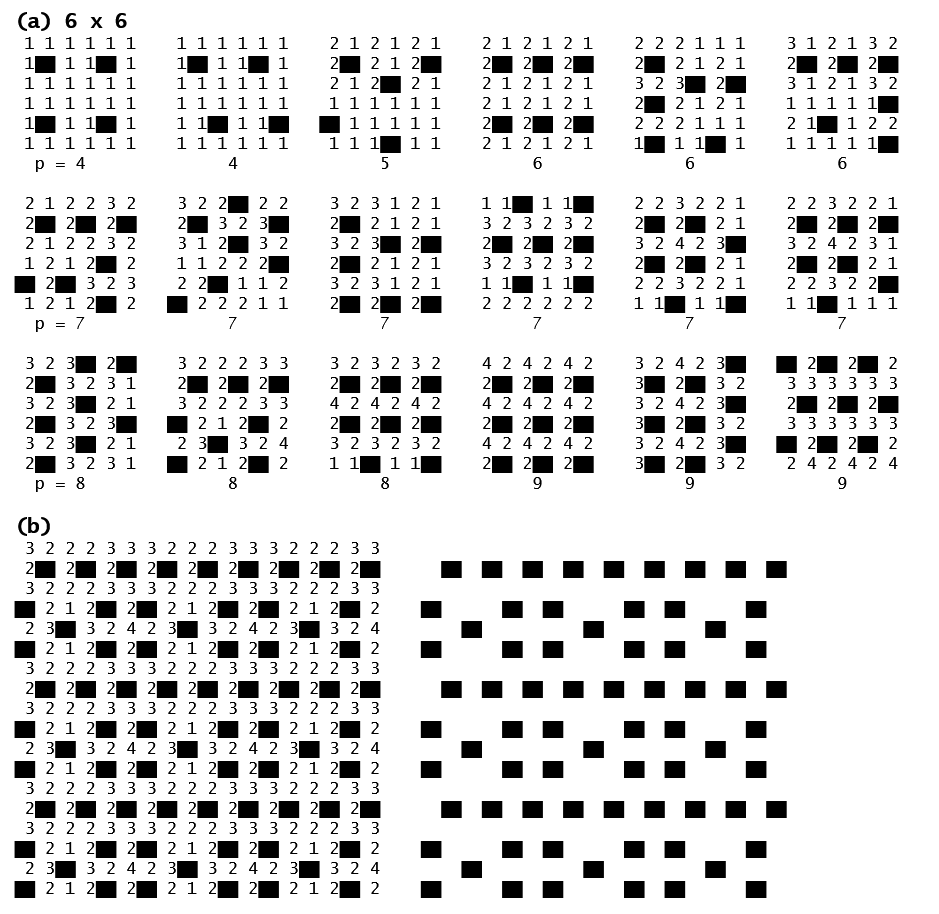}		
\caption{
Some valid $6 \times 6$ patterns evolved by Rule A.
(a) Patterns with $p=4 ~\dots~ 9$ points.
(b) An 8-point pattern repeated 3 times in horizontal and in vertical direction.
}
\label{RuleA6x6}
\end{figure}

Several evolved $6 \times 6$  patterns are shown in 
Fig. \ref{RuleA6x6}a.
The range of points lies between 4 and 9. 
4-point valid patterns consist of tiles that nowhere overlap, the cover level is $v=1$ everywhere,
and the horizontal and vertical distance between points is 3. 
In the most regular 9-point pattern the distance between points is 2, and the cover level distribution is
$1^{9+0} 2^{18} 3^0 4^9$.
The last two 9-point patterns are similar under rotation and shift, and their cover level distribution both is
$1^{9+0} 2^{12} 3^{12} 4^3$.
 No cover levels are shown in  Fig. \ref{RuleA6x6}b (right) in order to expose the pattern's structure.

\begin{figure}[t!] 
\centering
\includegraphics[width=9cm]{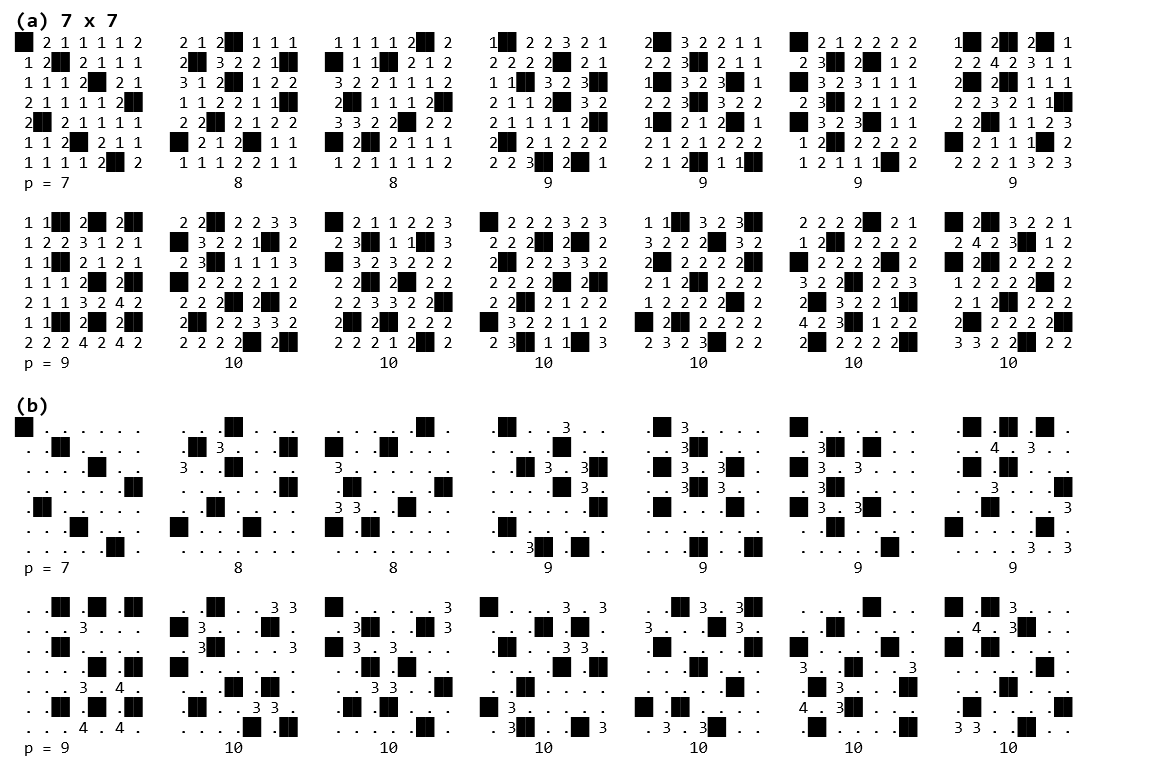}		
\caption{
Some valid $7 \times 7$ patterns evolved by Rule A.
(a) Patterns with $p=7 ~\dots~ 10$ points.
(b) The cover levels $v=1$ and $v=2$ were replaced by dots.
This representation emphasizes the sites with cover levels 3 or 4 allowing to detect easier equivalent patterns, 
e.g. the last two patterns are similar under symmetry.   
}
\label{RuleA7x7}
\end{figure}

\subsubsection{7 $\times$ 7 Patterns}

\begin{figure}[t!] 
\centering
\includegraphics[width=9cm]{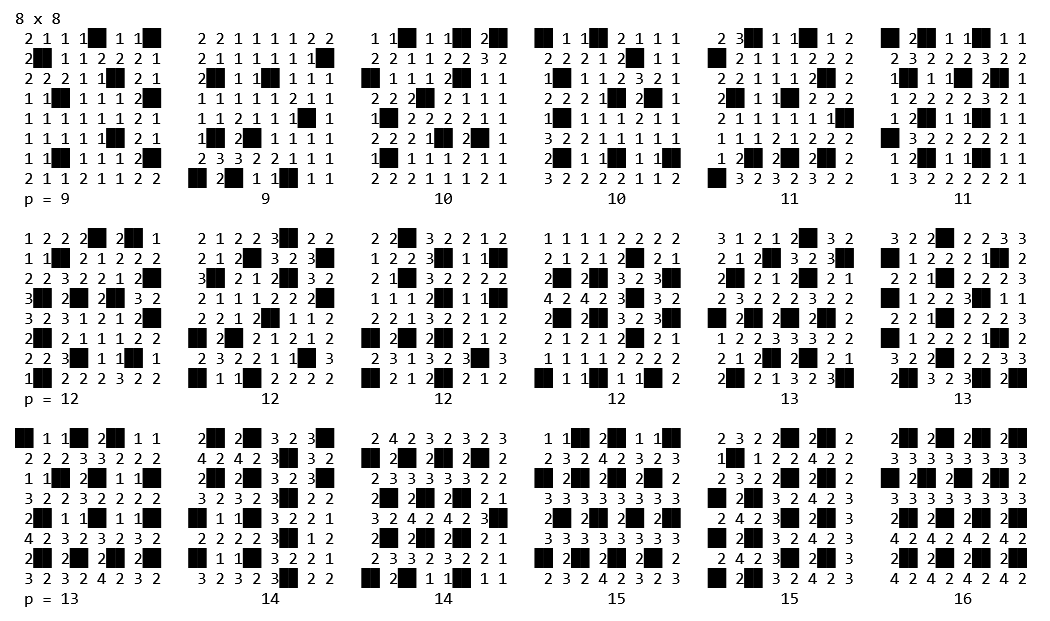}		
\caption{
Some valid $8 \times 8$ patterns evolved by Rule A
with $p=9 ~\dots~ 16$ points.  
}
\label{RuleA8x8}
\end{figure}

\begin{figure}[htb] 
\centering
\includegraphics[width=10cm]{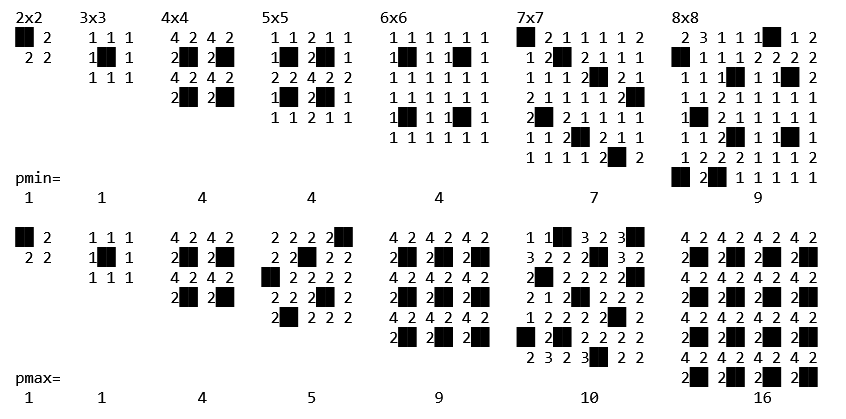}		
\caption{
Some patterns evolved by Rule A with a \emph{minimal} or \emph{maximal} number of points, \textit{min patterns} resp. \emph{max patterns} for $n=2 \dots 8$.
}
\label{minmax}
\end{figure}

Some evolved $7 \times 7$  patterns are shown in 
Fig. \ref{RuleA7x7}a.
The number of points range from 7 to 10.
The first one is the most regular showing the cover level distribution 
$1^{7+28} 2^{14} 3^{0} 4^{0} = 1^{p+(n^2-3p)} 2^{2p} 3^{0} 4^{0}$ with $p=n=7$.
In Fig. \ref{RuleA7x7}b only cover levels 3 and 4 are shown.
This representation can help to detect similar patterns more easily, e.g. the last two ones are similar. 
In addition sites or local regions with high levels of overlap (3 and 4) are highlighted.
For instance in the five 9-point patterns the overlap levels are varying: 
$3^4, 3^6, 3^6, 3^5 4^1, 3^2 4^3$
and in the six 10-point max patterns we find
$3^7, 3^7, 3^7, 3^6,   3^4 4^1, 3^4 4^1$.

\subsubsection{8 $\times$ 8 Patterns}

Fig. \ref{RuleA8x8} depicts evolved $8 \times 8$  patterns with 9 to 16 points.
Some patterns can easily be constructed.
The last one can be  composed of 4 rectangles (double lines) of size $2 \times 8$ containing 4 points each. 
Patterns with points  $p=16-M,(M=0...4)$ can be composed of $M$ rectangles with 3 points and $4-M$ rectangles with 4 points. 
Such patterns with a high regularity of construction by a human designer are more difficult to find by the CA rule.
Instead, most probably more irregular patterns are evolved which usually are difficult to construct by hand. 

\subsubsection{Min and Max Patterns}

Min and max patterns for fields up to  $8 \times 8$ are shown in 
Fig. \ref{minmax}.
The min patterns are unique for $n=2,3,4,7$,
and  the max patterns are unique for $n=2,3,4$. 

\subsection{Additional Rule B: Deleting Gaps}

Rule A works also fine for other tile shapes, like dominoes 
\cite{2019-pact-domino,Hoffmann-2021,Hoffmann-2021-MinimalDominoPact}. 
Depending on the shape, Rule A may produce patterns with uncovered cells (gaps).
It was not observed during many simulations that the here used point tile can produce gaps.
So the following rule shall be used in the case where gaps may appear, like for dominoes.
The rule injects additional noise at sites where no hits (matching templates) occur. 

In an implementation it is useful to use an additional temporary variable $hit$ that stores the number of hits
 (template matches).
So we use the cell state $(s, hit)$ where $s\in\{0,1\}$ is the cell's pattern state,
and $hit\in\{0,1,2,3,4\}$.
In our case using the point tile, the hit value needs not to be stored between generations, but for 
other tiles and objectives (e.g. minimization the number of tiles) it seems to be necessary that
the cell rule can use the hits in its neighborhood, either the hit just updated during the actual  computation
of the new generation, or the hit from the previous generation.

During the process of CA rule application the pattern may not be completely valid but only partially.
In partially valid pattern some tiles are detected 
but there exist some noisy or uncovered cells. 
The hits approximate the cover level.
The hits will converge to the cover level when the pattern gets valid and stable. 

The additional Rule B is

\[
s''(x,y) =
\left
\{
\begin{array}{lll} 

   random\in\{0,1\}  &with ~\textit{prob.} ~\pi_0 ~~\textbf{if} ~hit(x,y)=0     &(c)\\
                
   s'(x,y)           &\textbf{otherwise} \textit{~no change}                    &(d) \\
\end{array}
\right. .
\]

For our simple point tile problem, the influence of the probability $\pi_0$ on  the convergence speed and the 
resulting patterns is negligible until about $\pi_0=0.20$.
We have used $\pi_0=0.01$ in the further experiments. 

\subsection{Additional Rule C1: Maximizing the Number of Points}

When Rule A is applied, valid point pattern are evolved with a point number 
that lies between the maximum and the minimum, with a peak approximately at $p=(p_{min}+p_{max})/2$.
How can pattern be evolved with a high probability for $p=p_{max}$.
The idea is to inject additional noise when the local cover level $v$ is low,
in expectation to yield a more dense solution.  
So noise is injected where $v=1$ and if the cell is a white/green border cell with state zero:

\[
s'''(x,y) =
\left
\{
\begin{array}{lll} 

   1  &with ~\textit{prob.} ~\pi_1 ~\textbf{if} ~hit(x,y)=1 \wedge s''(x,y) = 0    &(e)\\
                
   s''(x,y)           &\textbf{otherwise} \textit{~no change}                    &(f) \\
\end{array}
\right. .
\]

So we yield the  combined rule ABC1 
which means that rule A, rule B, and rule C1 are executed sequentially at the current site $(x,y)$.

\subsubsection{Simulation Results Using Rule ABC1}

\begin{table}[htb] 
\centering
\caption{
Optimal or near-optimal min point patterns are evolved by rule ABC1.
The time $t_{stable}$ to reach a stable pattern is short. 
}
\includegraphics[width=10cm]{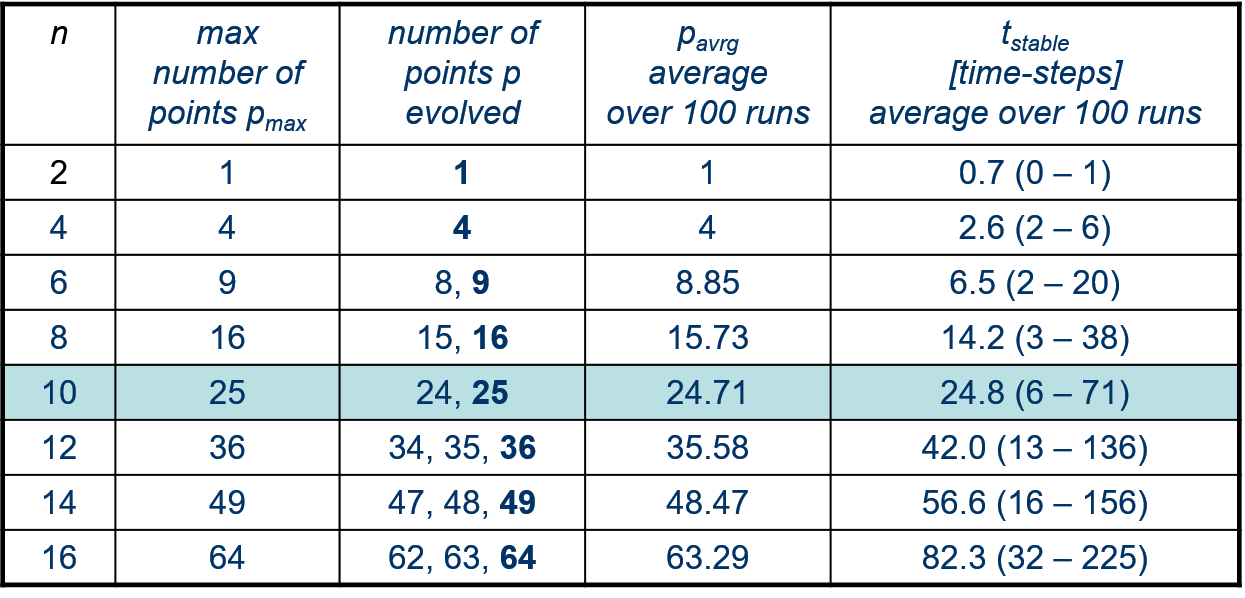}		
\label{TableRuleABC1}
\end{table}

\begin{figure}[htb] 
\centering
\includegraphics[width=10cm]{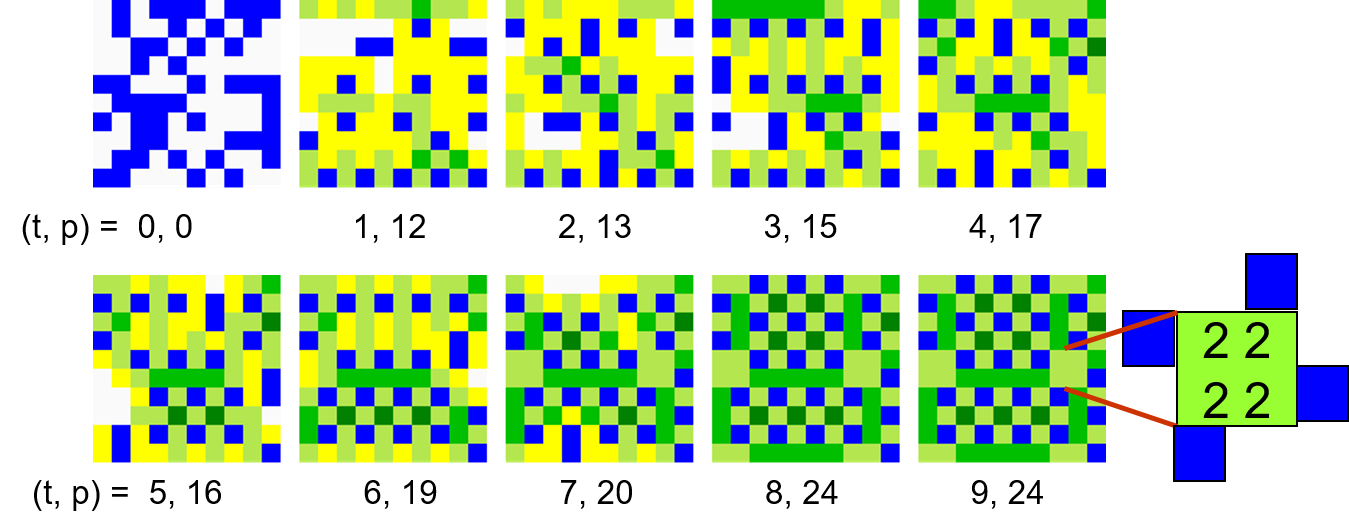}
\caption{
The evolution of a near-optimal pattern with 24 points.
The pattern is stable for $t\geq8$.
Squares 22/22 like the marked one have to be dissolved in order to yield max patterns.
Points are colored in blue, state = 1.
Colors for state = 0 and for cover levels: 
0 (white), 1 (yellow), 2 (light green), 3 (green), 4 (dark green).
}
\label{Simulation10x10-24points}
\end{figure}

The performance of rule ABC1  with $\pi_0=0.01$ and $\pi_1=0.25$ was evaluated
under 100 simulation runs, and averages were computed for $n$ even (Table \ref{TableRuleABC1}).

The evolved pattern show a high number of points, they are max patterns or close to them.
For instance, for $10 \times 10$ fields ($n=10$) the maximal number of points is $p_{max}=25$,
and the average number of evolved points is $p_{avrg}=24.71$, so in this case 71\% were max patterns. 
One can see that the speed of convergence is high, and $t_{stable}$ to find an optimal or near-optimal min pattern
is low.

Now it was analyzed for  $n$ even why not always optimal max point patterns evolve.
In   
Fig. \ref{Simulation10x10-24points}
we see the evolution of a near-optimal $10\times 10$ pattern with 24 points, not with the maximum of 25 points 
at which we aim at.
We can find in such near-optimal stable patterns $2\times 2$ squares with state 0 and cover level 2 surrounded by 4 
cells with state 1.
Such 22/22 squares have to be dissolved in order to yield max patterns.

\subsection{Additional Rule C2 for \textit{n} Even}

\noindent This additional rule is able to dissolve the mentioned 22/22-squares. 
Now the evolution converges to stable max patterns for $n$ even.
The rule C2 is:

\[
s''''(x,y) =
\left
\{
\begin{array}{lll} 

   1  &with ~\textit{prob.} ~\pi_2 ~~\\
                &\textbf{if} ~hit(x,y)=2   \wedge \exists !  s'''(x\pm1,y\pm1) = 1    &(g)\\
                
   s'''(x,y)           &\textbf{otherwise} \textit{~no change}                    &(h) \\
\end{array}
\right. .
\]

If the local situation says that the cover level is $v=2$ approximated by \textit{hit}, and
there exists exactly one NESW-neighbor with state $s=1$ (a point), then this situation
corresponds to the one mentioned before (22/22) which will be dissolved by noise injection.
For computing the condition in (g) the following equivalence can be used 

$\left( \exists !  s'''(x\pm1,y\pm1) = 1  \right) \equiv  \left(1= \sum_{i=-1,+1}\sum_{j=-1,+1}  s'''(x+ i,y+ j)\right)$.

\vspace{5pt}
The used rule probabilities are 
$\pi_0=0.01, \pi_1=0.25,  \pi_2=0.02$. 

\begin{figure}[htb] 
\centering
\includegraphics[width=10cm]{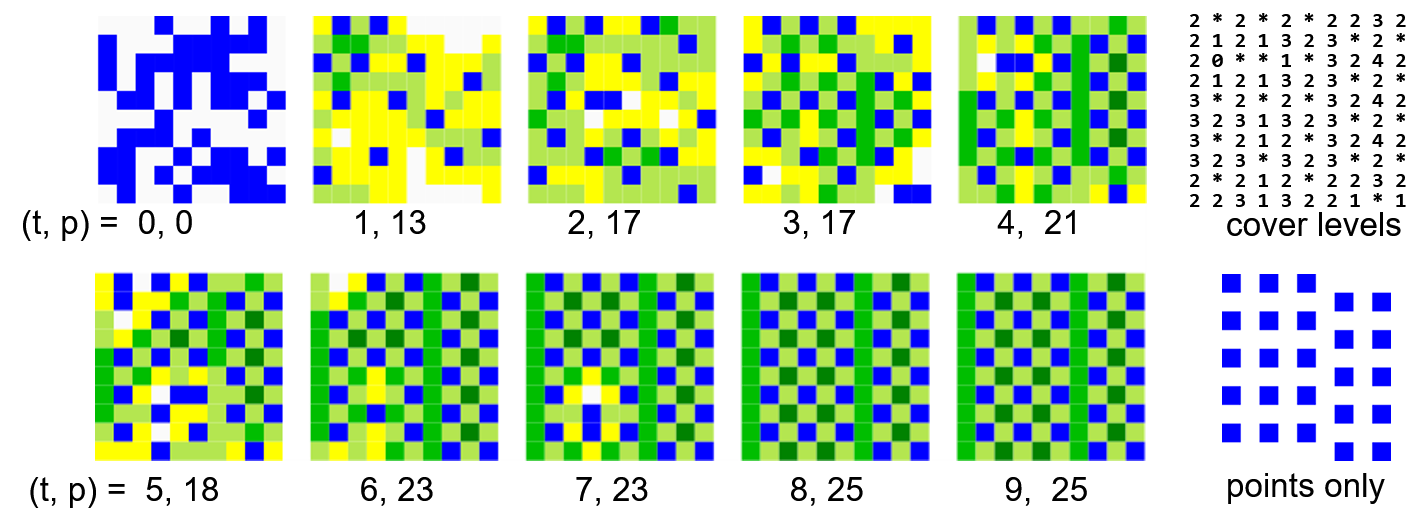}
\caption{
The evolution of a found $10\times10$  max pattern using rule ABC1C2.
}
\label{Evolution10x10Fast}
\end{figure}

\begin{figure}[htb] 
\centering
\includegraphics[width=10cm]{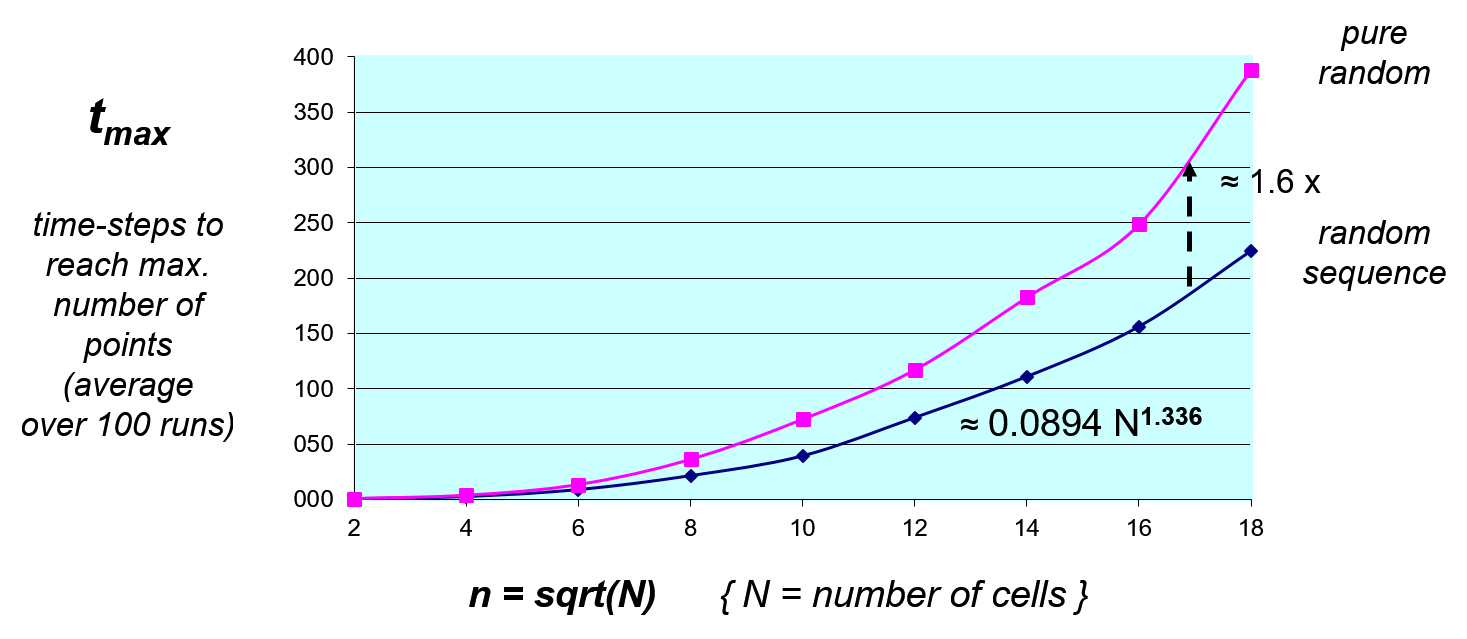}
\caption{
The average number of time-steps to reach a stable max pattern for $n$ even.
}
\label{TimeStepsMax}
\end{figure}
\begin{figure}[htb] 
\centering
\includegraphics[width=10cm]{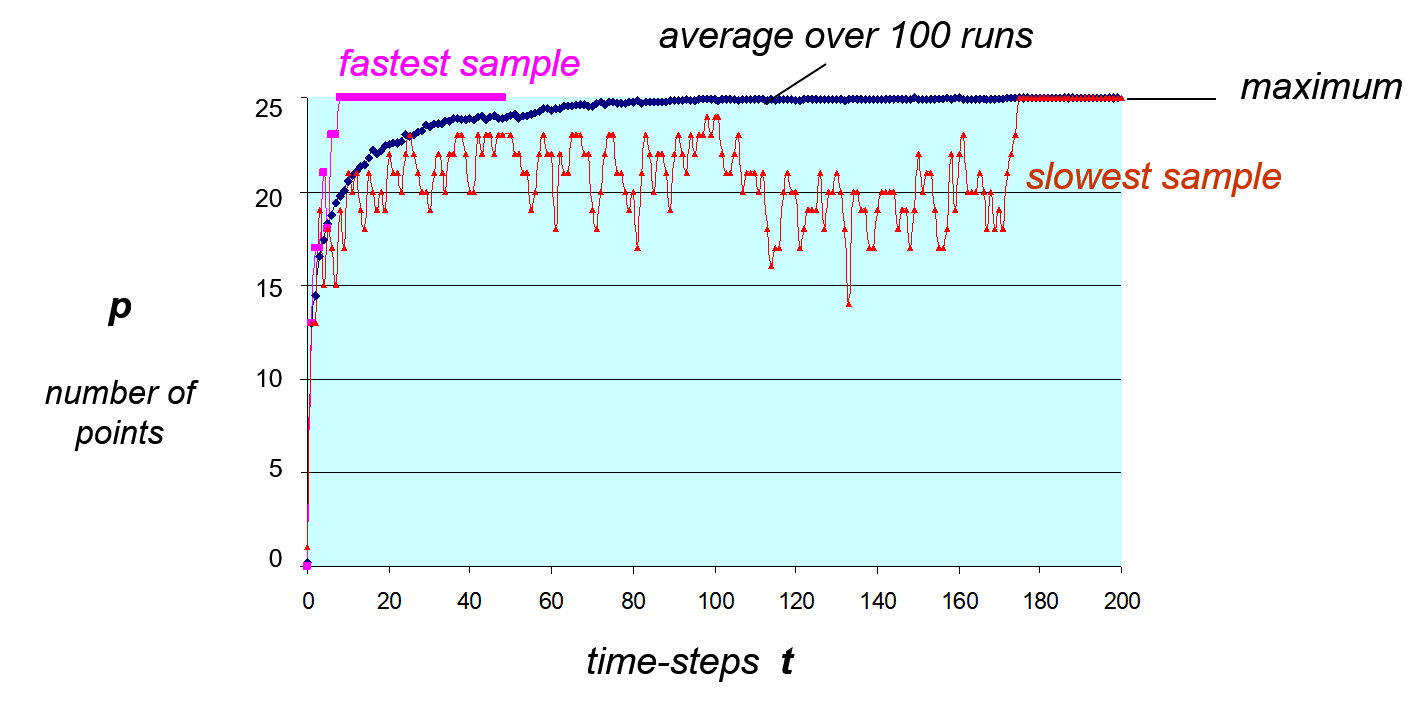}
\caption{
The average number of points vs time-steps for a $10\times10$ field.
The maximum is reached for time-steps in the range 8 ... 175 approximately.
}
\label{PointsVSTime}
\end{figure}

The evolution  of a stable $10\times10$ max pattern is shown in Fig. \ref{Evolution10x10Fast}.
Fig. \ref{TimeStepsMax} shows the needed number of time-steps (average over 100 runs) $t_{max}$ vs $n$.
The update method \textit{pure random} is about 1.6 times slower than \textit{random sequence}. 
As expected the computational effort increases exponentially,
for the given range we can find the approximation $t_{max} \approx 0.0894 N^{1.336}$.
Fig. \ref{PointsVSTime} 
shows the number of points vs the time-steps for a  $10\times10$ field.
We can see that the maximum of points is reached after about 80 time-steps on average.
The fastest observed evolution took only a few time-steps, whereas the slowest took about 180
time-steps with a lot of fluctuations.

\subsection{Does Rule C1 Work Also for \textit{n} Odd?}
\begin{figure}[htb] 
\centering
\includegraphics[width=10cm]{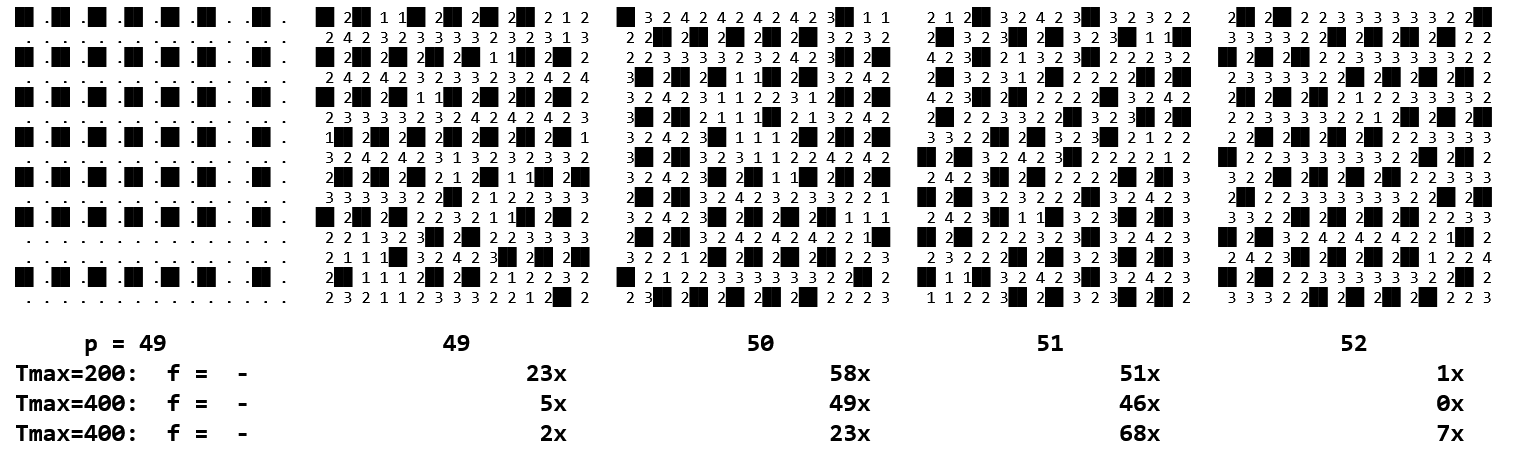}
\caption{
Found $15\times15$  pattern using rule ABC1. The patterns were not stable. 
The first pattern is a constructed one. 
The others are evolved samples for a certain number of points. 
}
\label{CaseOdd15}
\end{figure}

\begin{figure}[htb] 
\centering
\includegraphics[width=12cm]{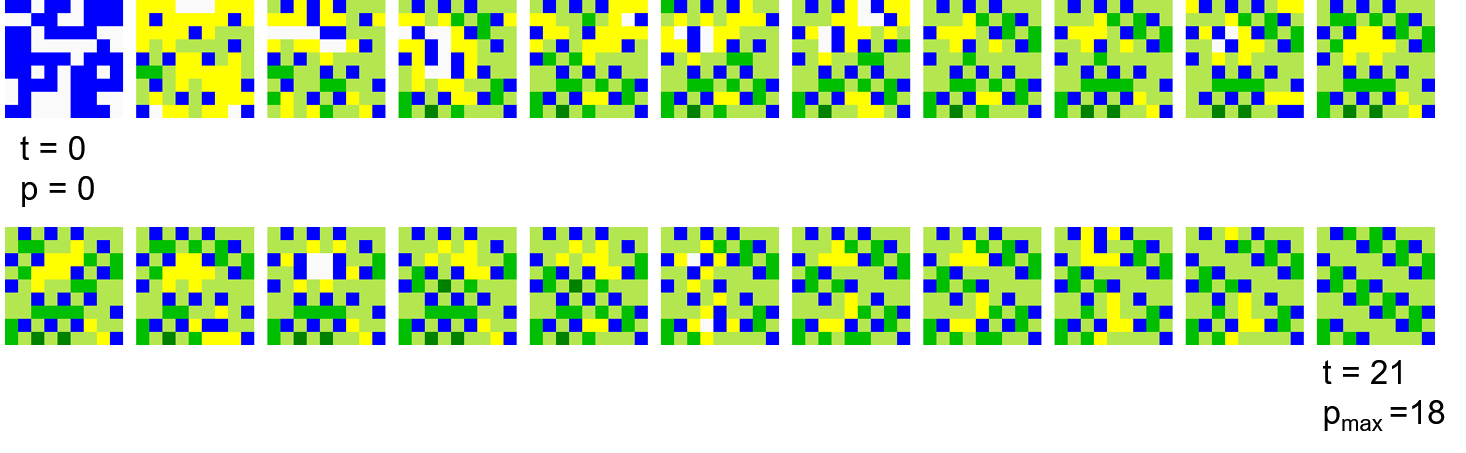}
\caption{
Evolving an optimal $9\times9$ 18-point pattern
using rule ABC1, random sequence updating, $\pi_0=0.01, \pi_1=0.025$.
}
\label{Sequence9x9}
\end{figure}

Now the question arises whether this rule ABC1C2 works also for $n$ odd.
The problem is that optimal max patterns may contain sites with cover levels 1 and 2 besides 3 and 4.
For instance for $n=7$ (Fig. \ref{RuleA7x7}) there is an optimal pattern with a cover level distribution of
$1^{10+5 } 2^{27} 3^7$ and we find there also local 22/22 squares.
Therefore optimal patterns may appear but they are transients (not stable) because noise is injected for $v=1$
by rule C1 and for $v=2$ by rule C2.
Rule ABC1C2 was checked against ABC1 with respect to achieve a higher number of points. 
It turned out that C2 is not useful for $n$ odd because many sites with cover levels 2 
may be part of optimal patterns and therefore should not be destroyed. 
Therefore for $n$ odd it is better to use ABC1 only. 
We have to note that max patterns that may appear during the evolution are usually not stable
for $n$ odd.
In Fig. \ref{CaseOdd15} we see some $15\times15$ patterns with 49, 50, 51, 52 points that were evolved using rule ABC1. 
100 simulation runs were performed with the time limit $T^{max}=200, 400, 800$.
The frequency $f$ of (49--52)--point pattern is also documented in Fig. \ref{CaseOdd15}.
One can see that the probability to find a max pattern during the evolution increases with the maximal number of simulated time-steps $T^{max}$.
Fig \ref{Sequence9x9} depicts a simulation sequence that evolves an optimal $9\times9$ pattern.
The applied rule is again ABC1.

\section{The Number of Points in Max Patterns and More Patterns}

\begin{figure}[htb] 
\centering
\includegraphics[width=8cm]{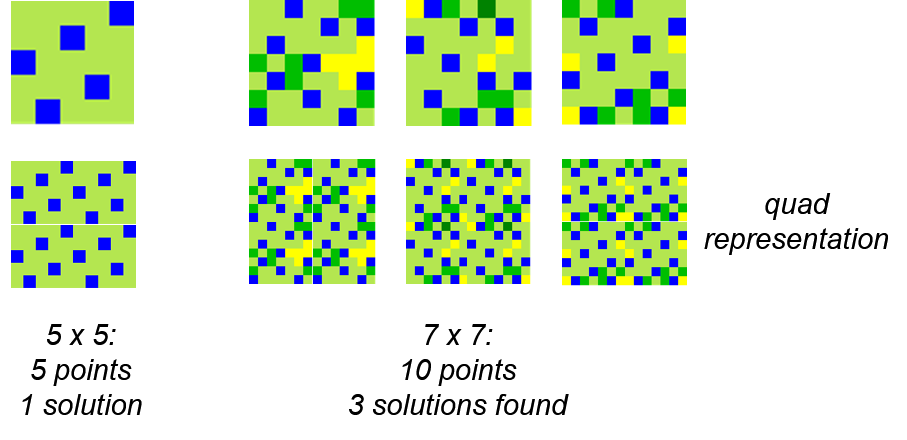}
\caption{
Evolved max patterns of size $5\times5$ and $7\times7$. 
}
\label{Special5x5And7x7}
\end{figure}


\begin{figure}[htb] 
\centering
\includegraphics[width=10cm]{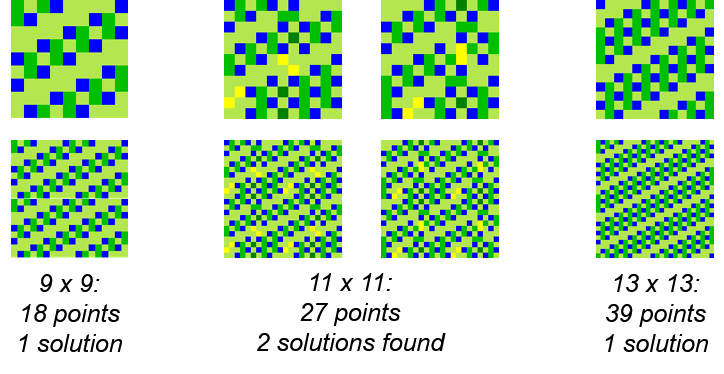}
\caption{
Solutions found for $n=9,11,13$.
}
\label{Special9x9And11x11And13x13}
\end{figure}

\begin{figure}[htb] 
\centering
\includegraphics[width=10cm]{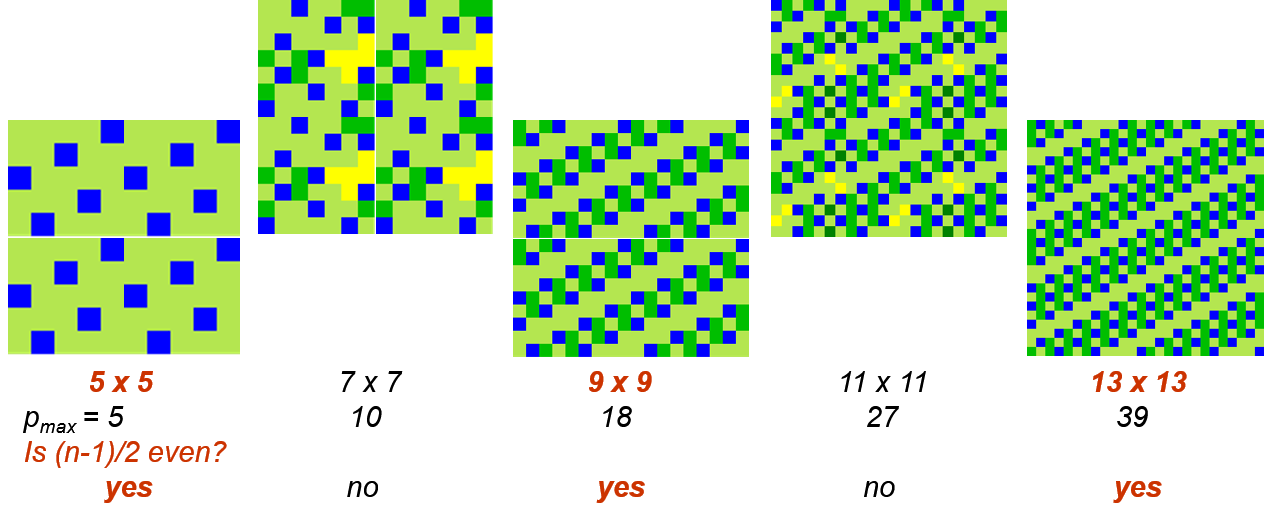}
\caption{
Solutions found for $n=5,7,9,11,13$. The solutions for
$n=5,9,13, ...$ seem to be unique and stable without cells having cover level 1.
}
\label{SpecialPatterns}
\end{figure}

\begin{figure}[htb] 
\centering
\includegraphics[width=10cm]{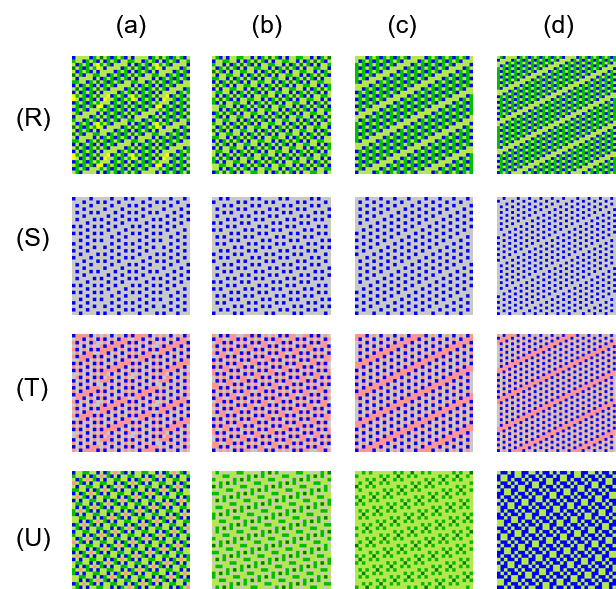}
\caption{
(R) Points are colored in blue, state = 1.
Colors for state = 0 and for cover levels: 
0 (white), 1 (yellow), 2 (light green), 3 (green), 4 (dark green).\\
R(a--c) Evolved $17\times17$ patterns by rule ABC1:
R(a) unstable with 67 points, R(b, c) stable with the maximum of 68 points. 
R(d) $21\times 21$ stable max pattern with 105 points.\\
(S) Only points are displayed. 
(T) Points and cover level 2 (in light red) are shown. \\
(U) R(b) is displayed using different colors for cover levels. 
}
\label{Special17x17}
\end{figure}

Analyzing the evolved max patterns for $n$ odd we constructed the following formula 
to be confirmed by a formal proof.

\[
p_{max}(n) =
\left
\{
\begin{array}{lll} 

(n(n-1)-2)/4             &\textbf{if} ~n=3, 7, 11, ... = 3+4k,  & k=0, 1, 2, 3 ... \\ 
                      
n(n-1)/4                 &\textbf{if} ~n=5, 9, 13, ... = 5+4k,  &k=0, 1, 2, 3 ...  \\

n^2/4                    &\textbf{if} ~n ~\textit{even}  \\
\end{array}
\right. .
\]

The following table displays $p_{max}(n)$ for odd $n=3, \ldots,23$.

\vspace{9pt}
\begin{minipage}[h]{.9\textwidth}
\begin{verbatim}
n      3  5  7  9 11 13  15  17  19  21
pmax   1    10    27     52      85          (n(n-1)-2)/4
          5    18    39      68     105       n(n-1)/4
\end{verbatim}
\end{minipage}
\vspace{9pt}

Let us have a look at more evolved patterns. 
Fig. \ref{Special5x5And7x7}: 
There is only one very regular solution for $5\times5$ with 5 points. 
Three solutions were found for $7\times7$ that are not very regular. 

Fig. \ref{Special9x9And11x11And13x13}: 
For $9\times9$ and $13\times13$  we observe a similar regular construction principle.
The found $11\times11$ patterns shows structures that lie between the $9\times9$ and $13\times13$ structures.

Fig. \ref{SpecialPatterns}: 
The patterns were arranged in this way in order to find the above formula for the number of points for $n$ odd. 
We observe a certain relation between  patterns for  $n=5,9,13 ...$ and for  $n=7, 11, ...$. 

Fig. \ref{Special17x17}:
R(a) shows and unstable sub-optimal, and R(b, c) optimal $17\times17$ patterns. 

Interesting is that the optimal pattern R(b) does not follow the structure scheme observed for $n=5,9,13$,
but R(c) does. 
The $21\times21$ max pattern R(d)  shows again the structure scheme of $n=5,9,13$.
Only the points are displayed in the patterns (S).
Points and cover level 2 (in light red) are shown in (T). 
The bottom line (U) of patterns show that different structures can be elucidated from R(b) if different colors for the
cover level are used. In fact we may derive different patterns by interpreting not only the state $s$ but also the 
cover level $v$.
Therefore we may use such CA rules of pattern generation also to produce intentionally certain ``cover level patterns''.
Patterns with $n=3+4k$ and near-optimal patterns may have some aesthetic value because there structure is not regular
in a simple fashion.

\newpage
\section{Conclusion and Future Work}

The aim was to find a cellular automata rule that can evolve a point pattern  with a maximal number of points (max patterns). 
The problem was considered as a tiling problem using overlapping point tiles. 
Nine templates are systematically derived form the point tile that are used for local matching.
The number of matches defines the number of hits.
Four Rules were designed that supplement each other. 
Rule A adjust the cell to the template's center value (reference value) if there is a hit, 
otherwise noise is injected.
Rule A is able to evolve stable point patterns very fast, but the number of points is seldom maximal.
Rule B injects noise if no tile covers the cell (cover level zero) in order to avoid gaps.
Rule C1 drives the evolution to a maximal number of points. 
Rule C2 drives the evolution to a stable optimum if the field length $n$ is even.  
A formula is presented that defines the maximal number of possible points. 
The patterns are more interesting if $n$ is odd.
Two different structures for  optimal patterns were found for $n=17$.
It would be interesting to find the number and structure of possibly different optimal patterns for $n>17$, especially if $n$ is odd. 
The introduced  method was already applied to other patterns (domino pattern, sensor point pattern) 
and offers a way to generate more complex patterns using more complex tiles. 
Future work can address the following problems
\begin{itemize}
\item parallel implementation,
\item  synchronous updating instead of asynchronous,
\item  using several tiles, more complex tiles, or more colors,
\item  using different grid dimensions, different regular grids (triangular, hexagonal, ...), or graphs in general,
\item  optimizing time efficiency using for instance a ``divide and conquer'' approach,
\item  relating this method to other methods, like constraints satisfaction,
\item  relating the problem to physical problems, like the  attraction and repulsion of particles,
\item  consider the problem in the context of combinatorics,
\item  classify the results into the theory and applications of tilings,
\item  find interesting applications, like the generation of artistic patterns. 
\end{itemize}

\DREIECK{3}

\newpage
\section{Appendix: Presentation 2019}

\begin{figure}[h!] 
\centering

\includegraphics[width=2.3cm]{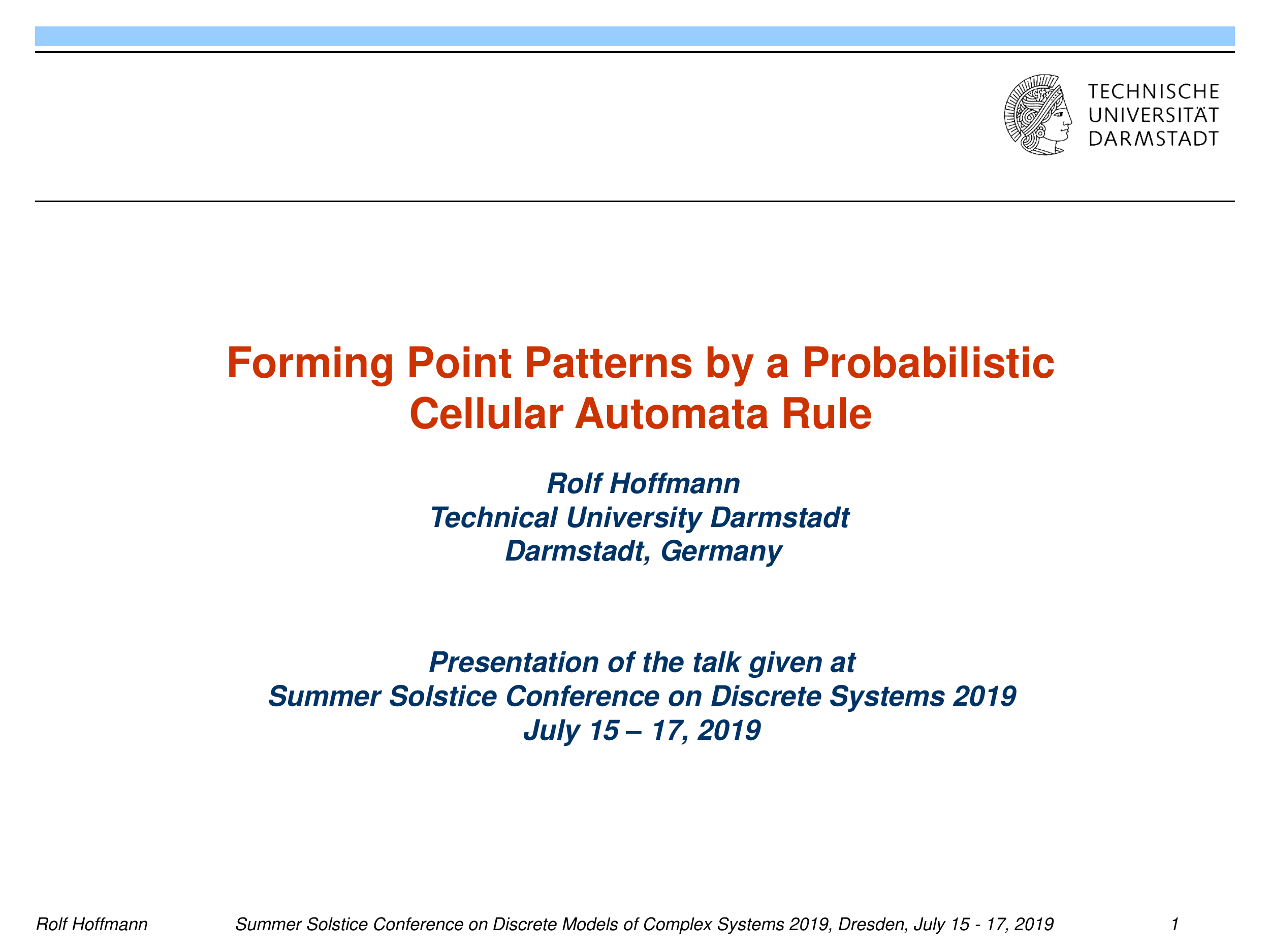}
\includegraphics[width=2.3cm]{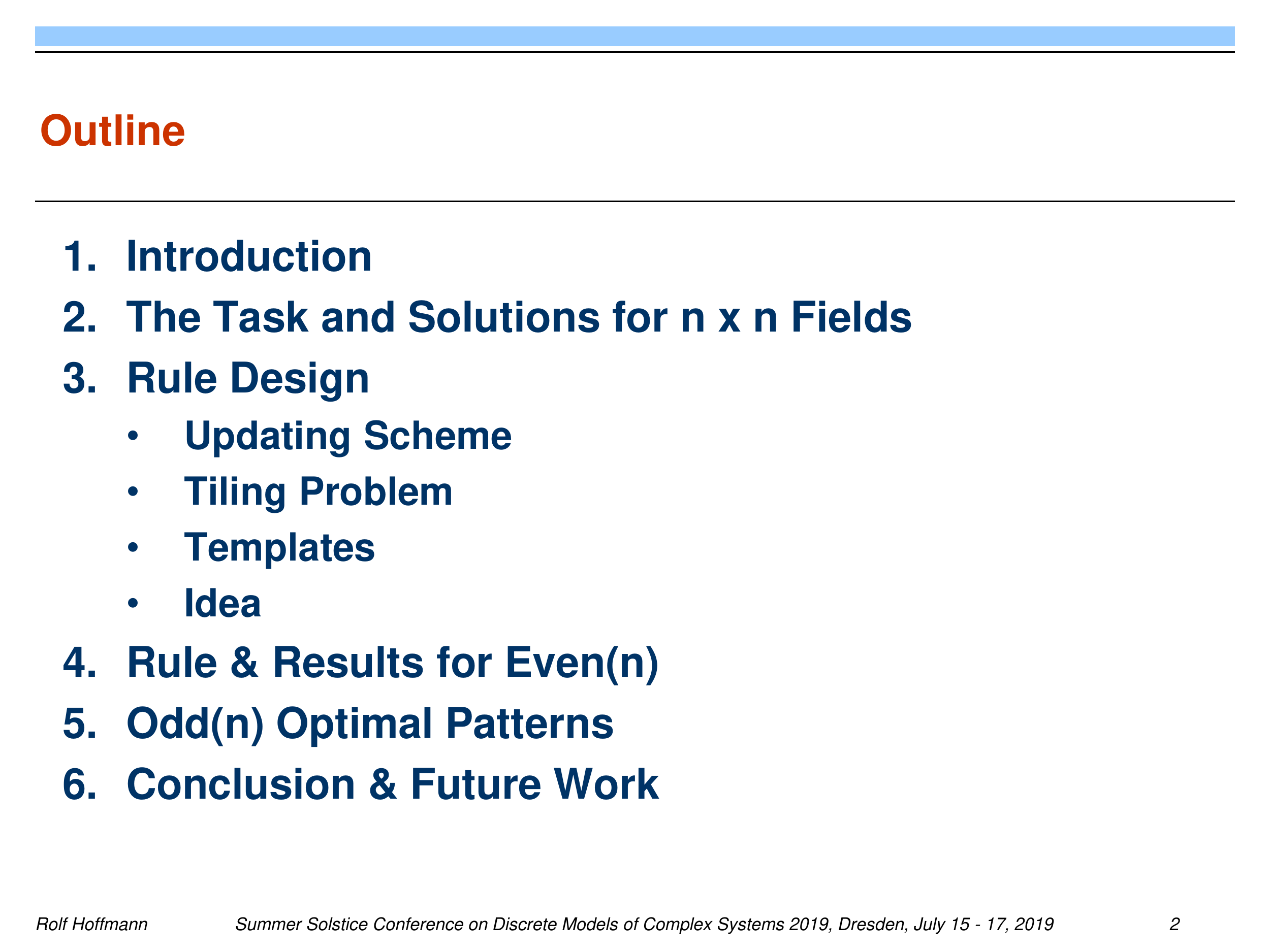}
\includegraphics[width=2.3cm]{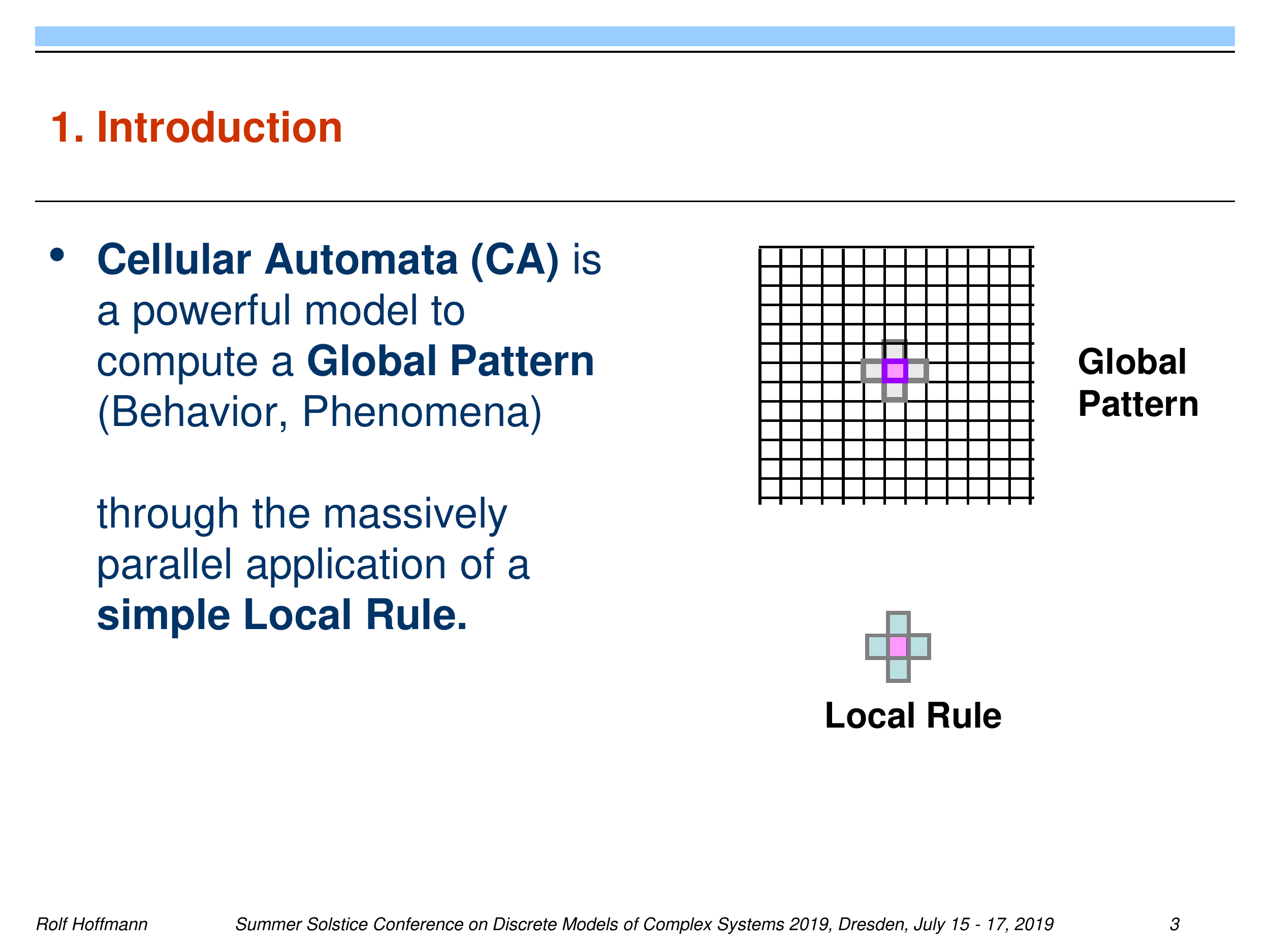}
\includegraphics[width=2.3cm]{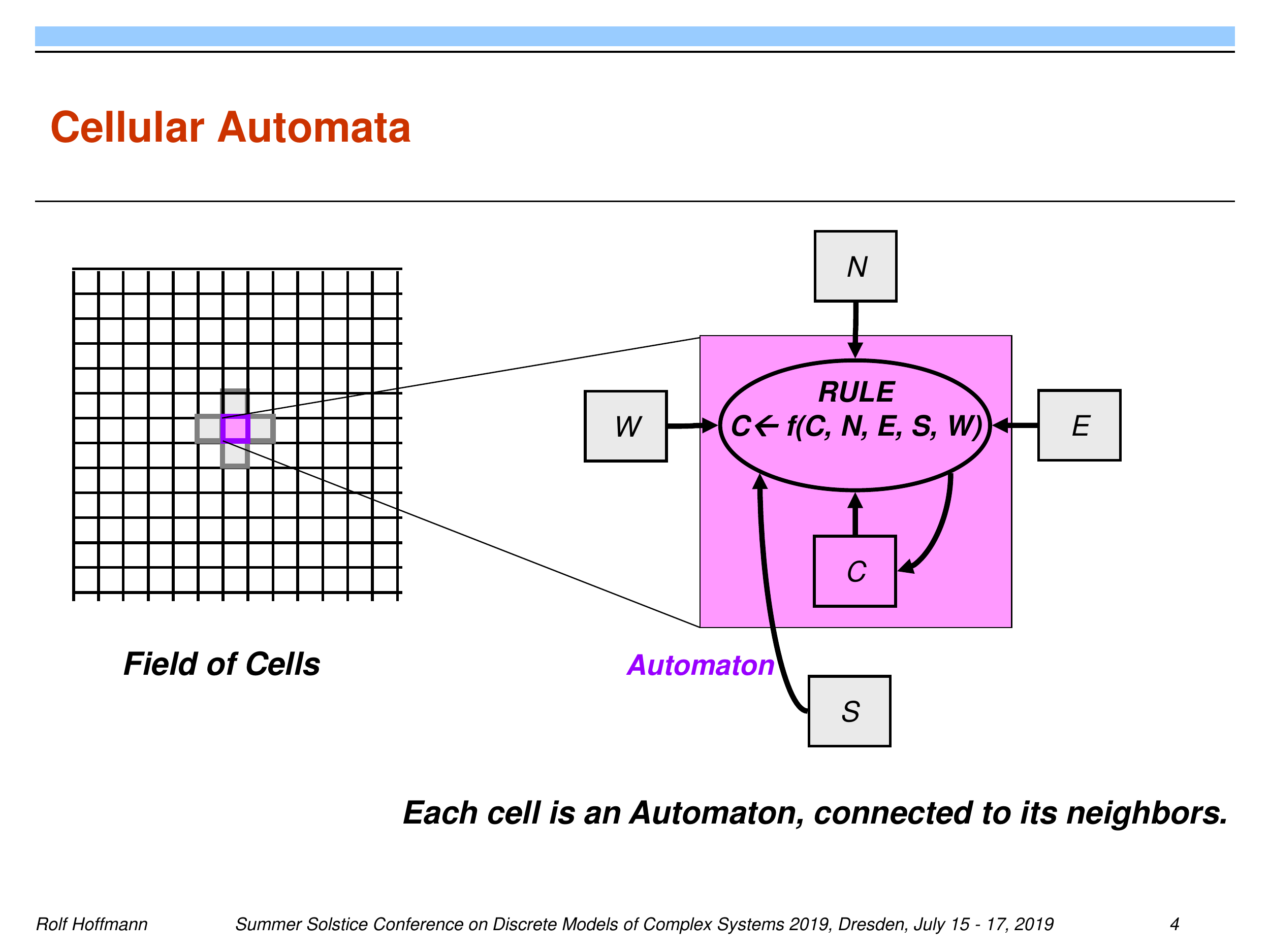}
\includegraphics[width=2.3cm]{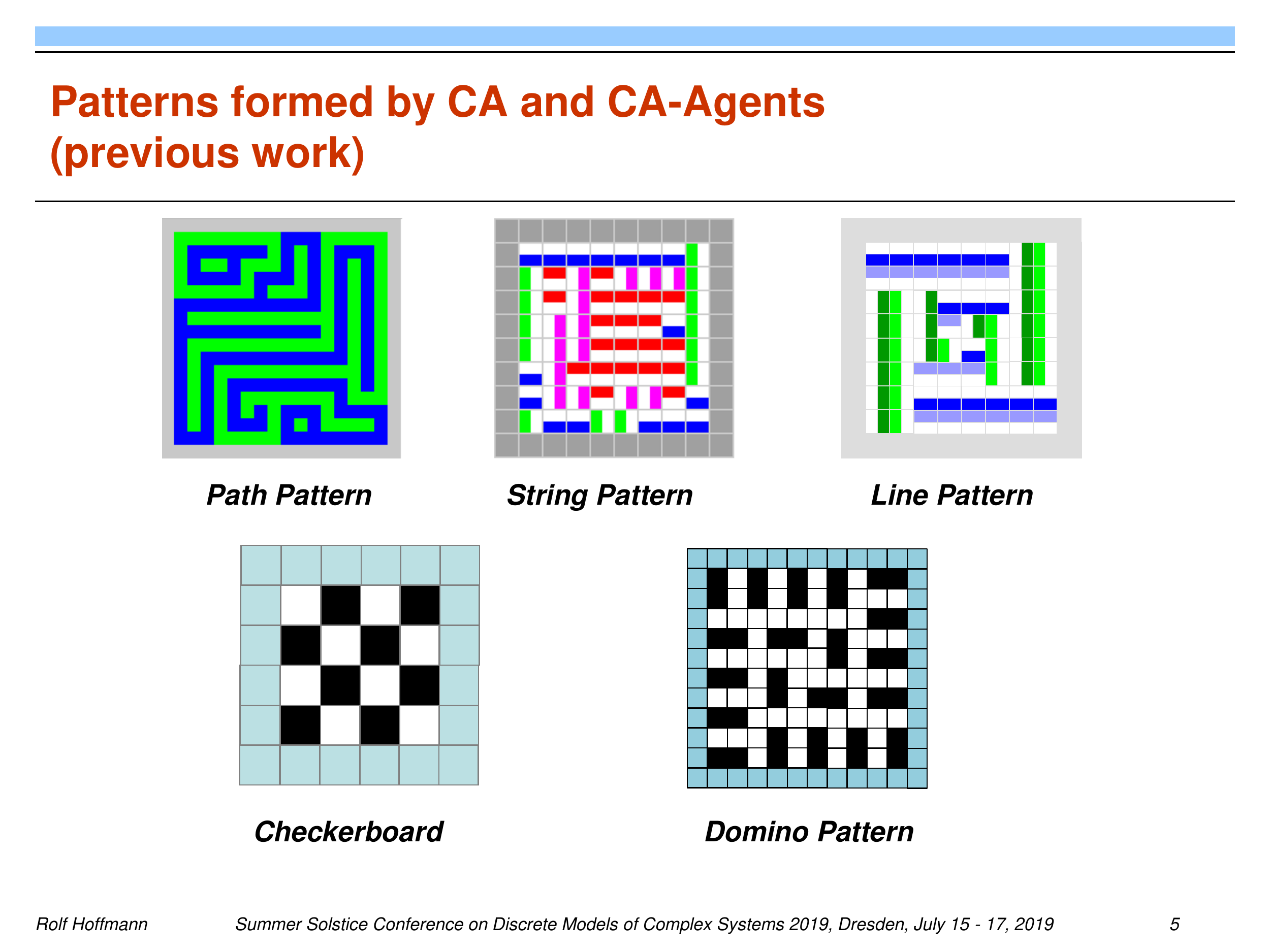}

\includegraphics[width=2.3cm]{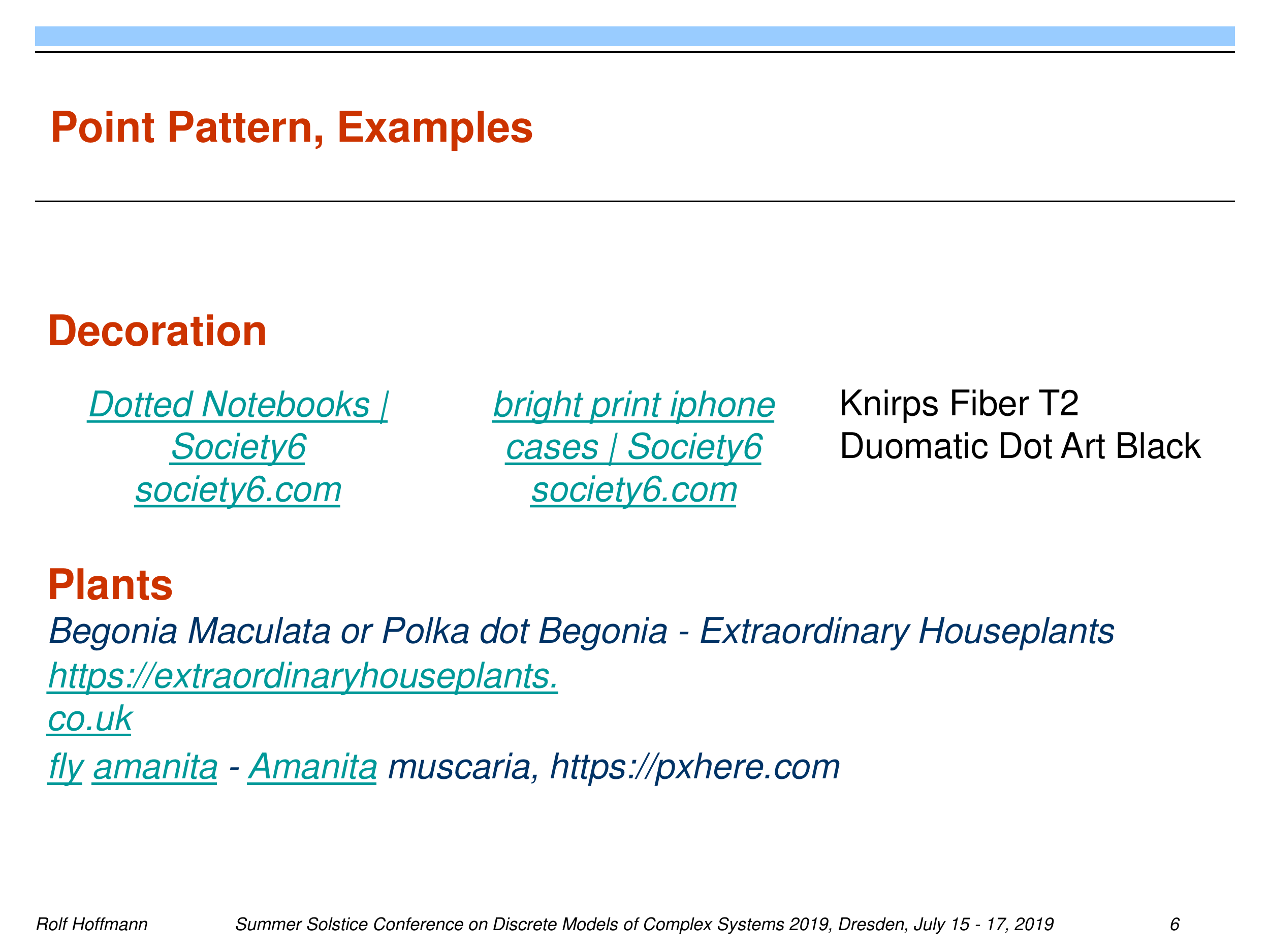}
\includegraphics[width=2.3cm]{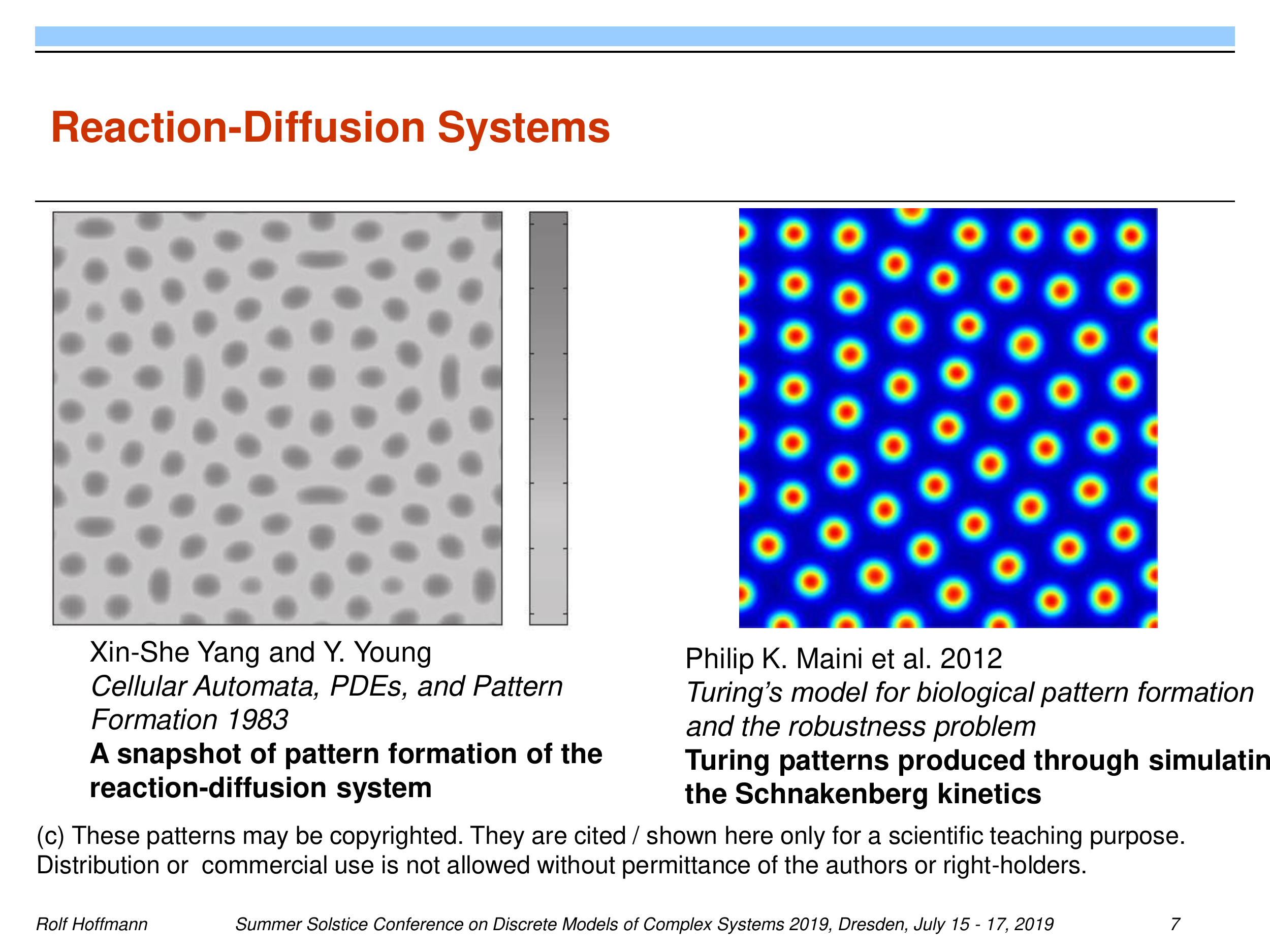}
\includegraphics[width=2.3cm]{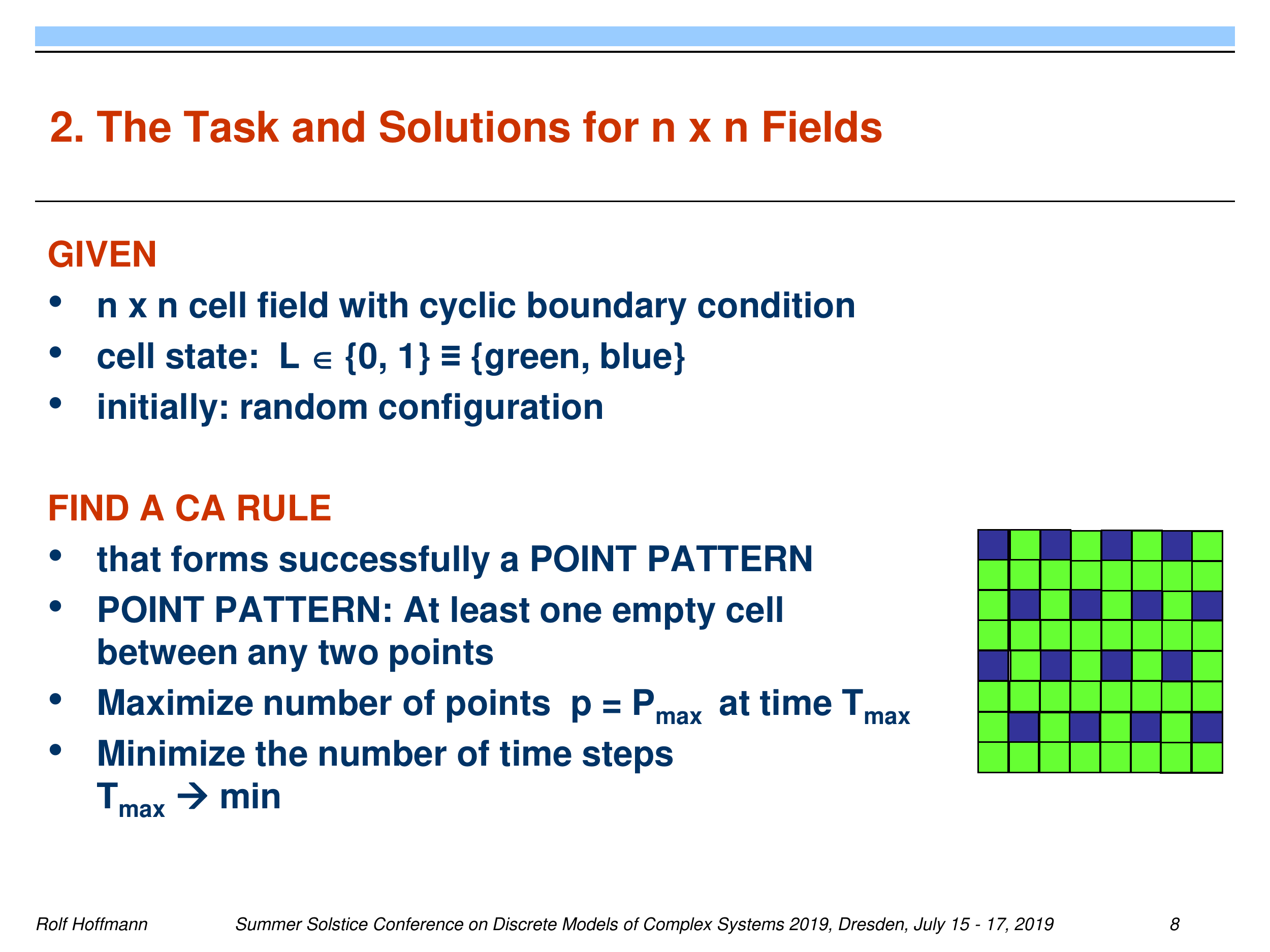}
\includegraphics[width=2.3cm]{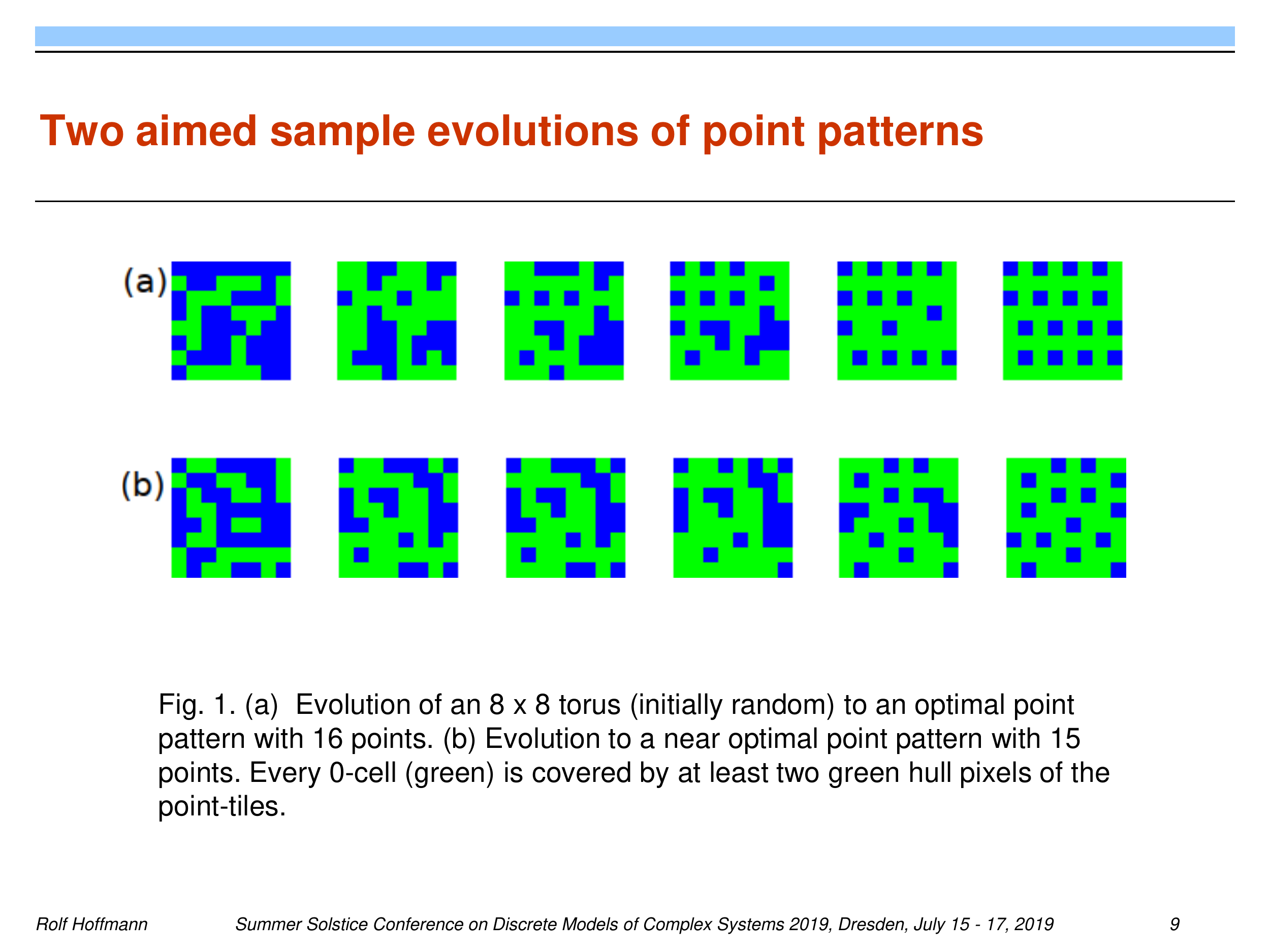}
\includegraphics[width=2.3cm]{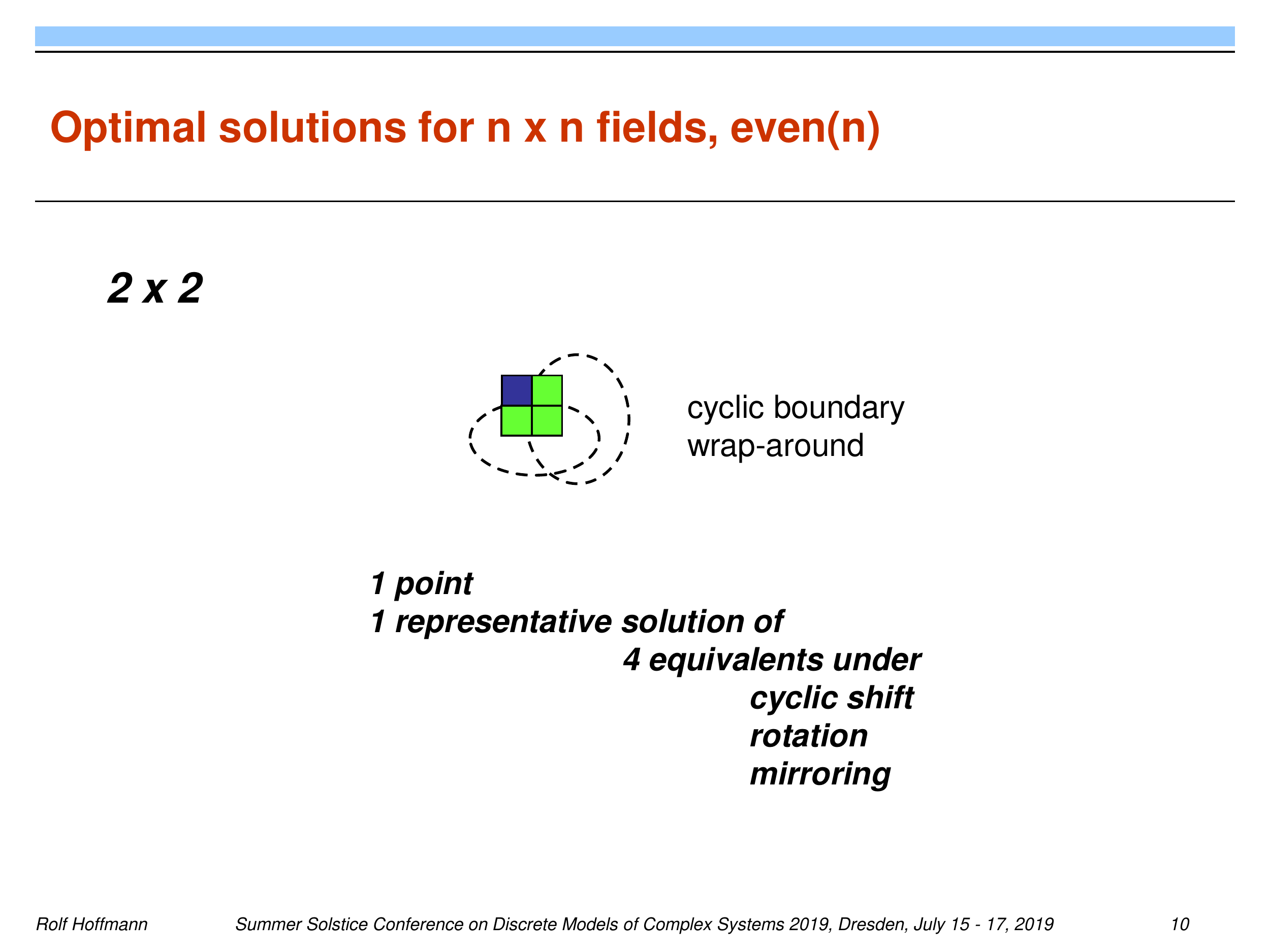}

\includegraphics[width=2.3cm]{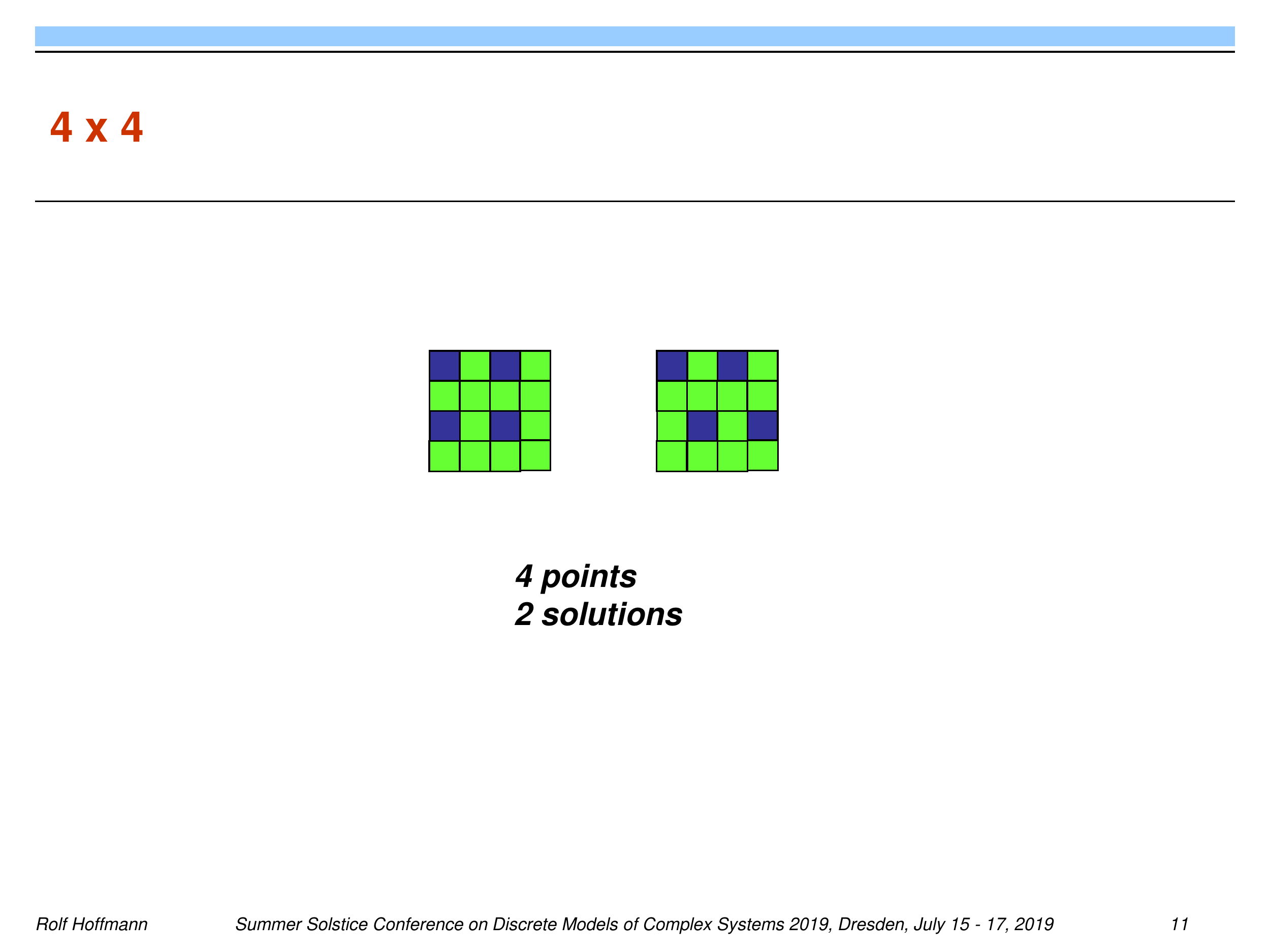}
\includegraphics[width=2.3cm]{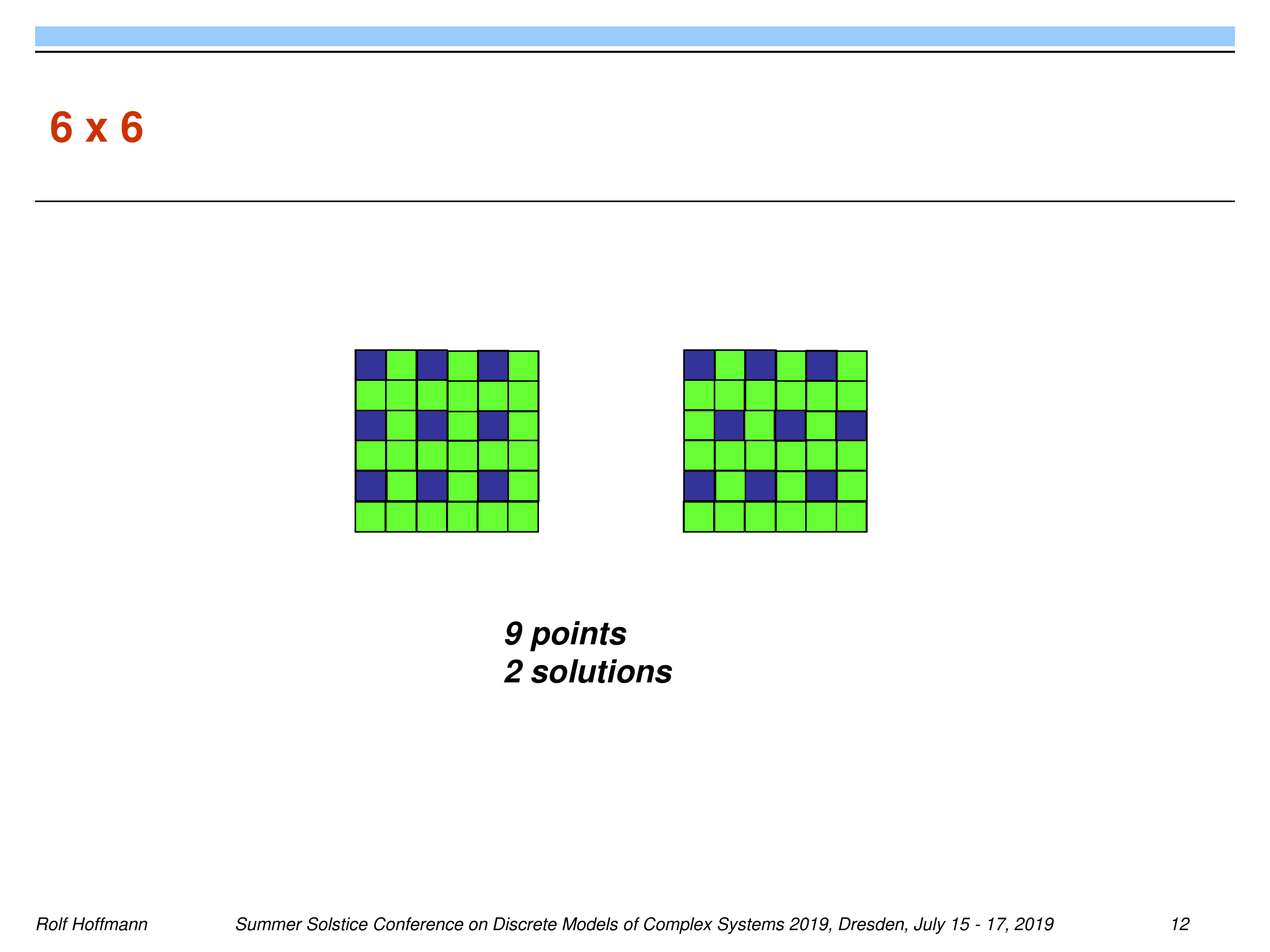}
\includegraphics[width=2.3cm]{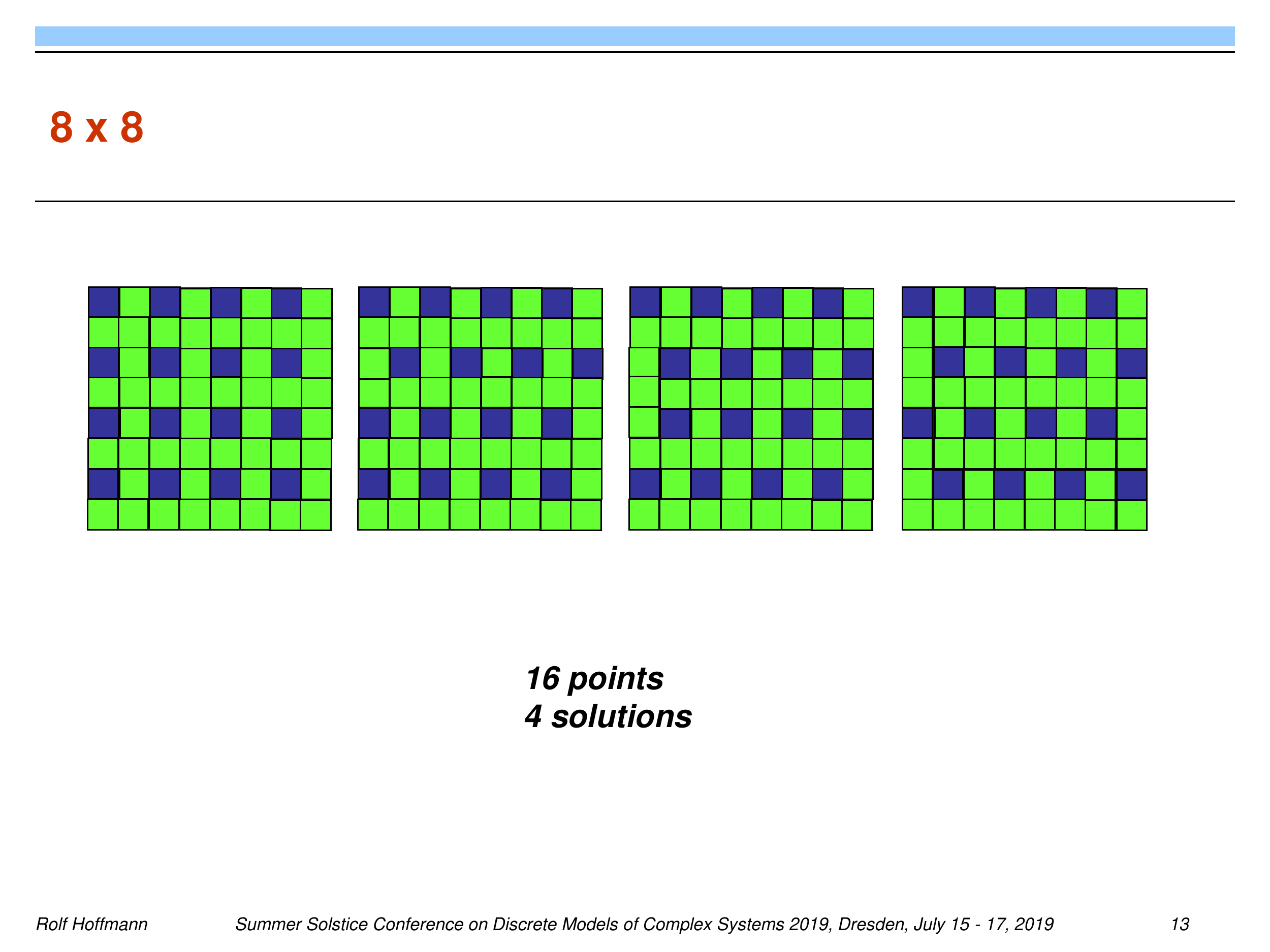}
\includegraphics[width=2.3cm]{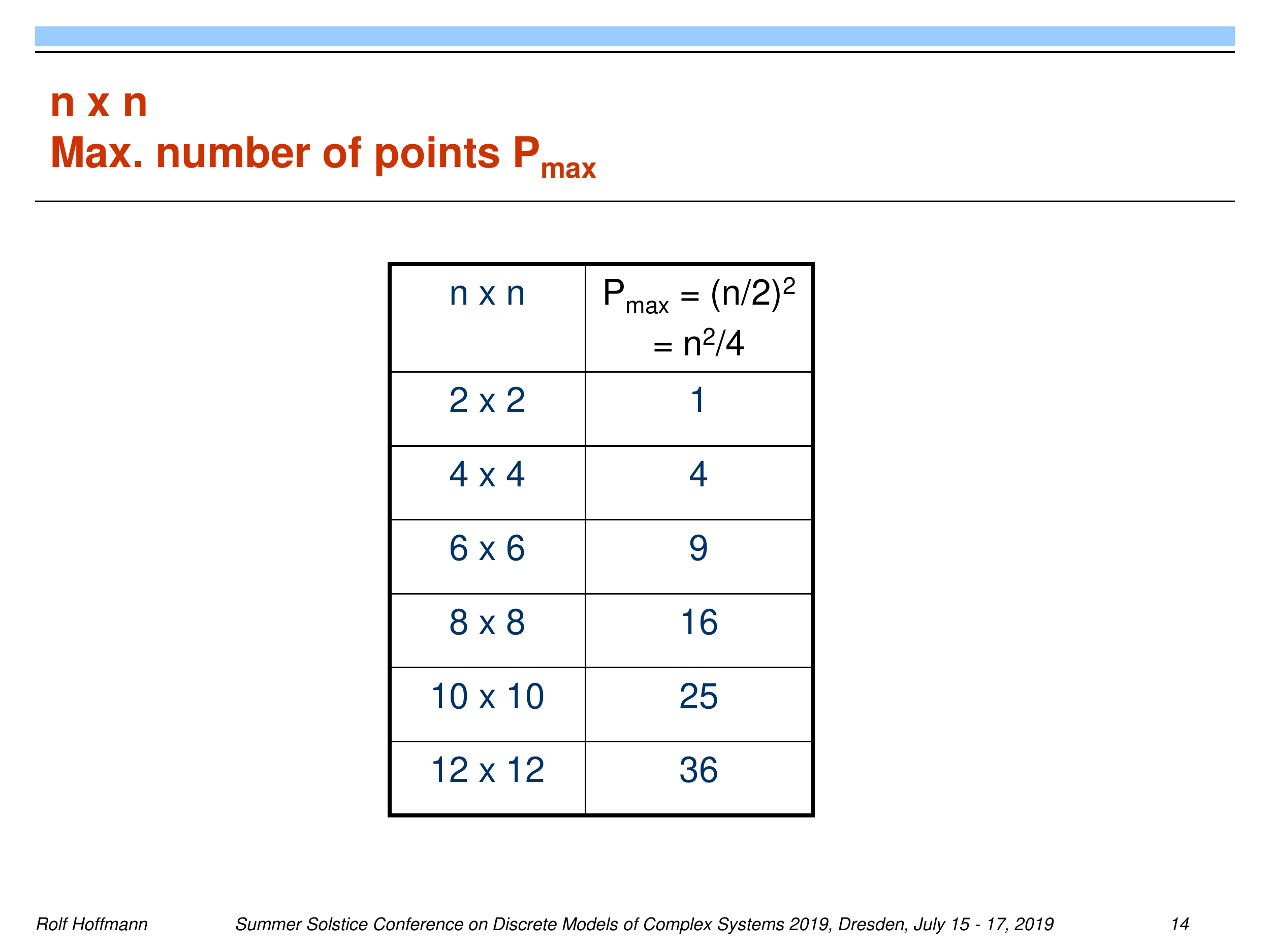}
\includegraphics[width=2.3cm]{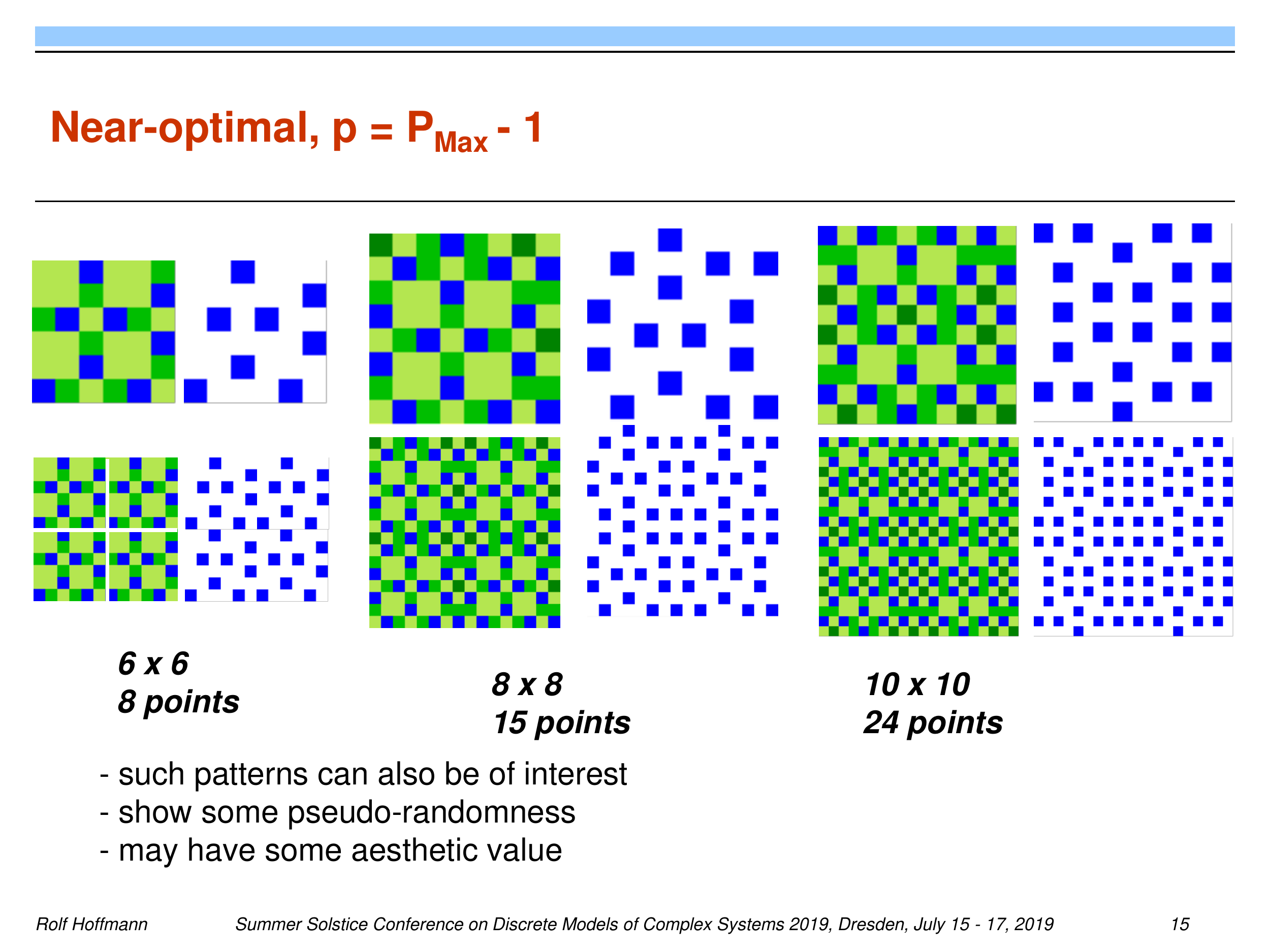}

\includegraphics[width=2.3cm]{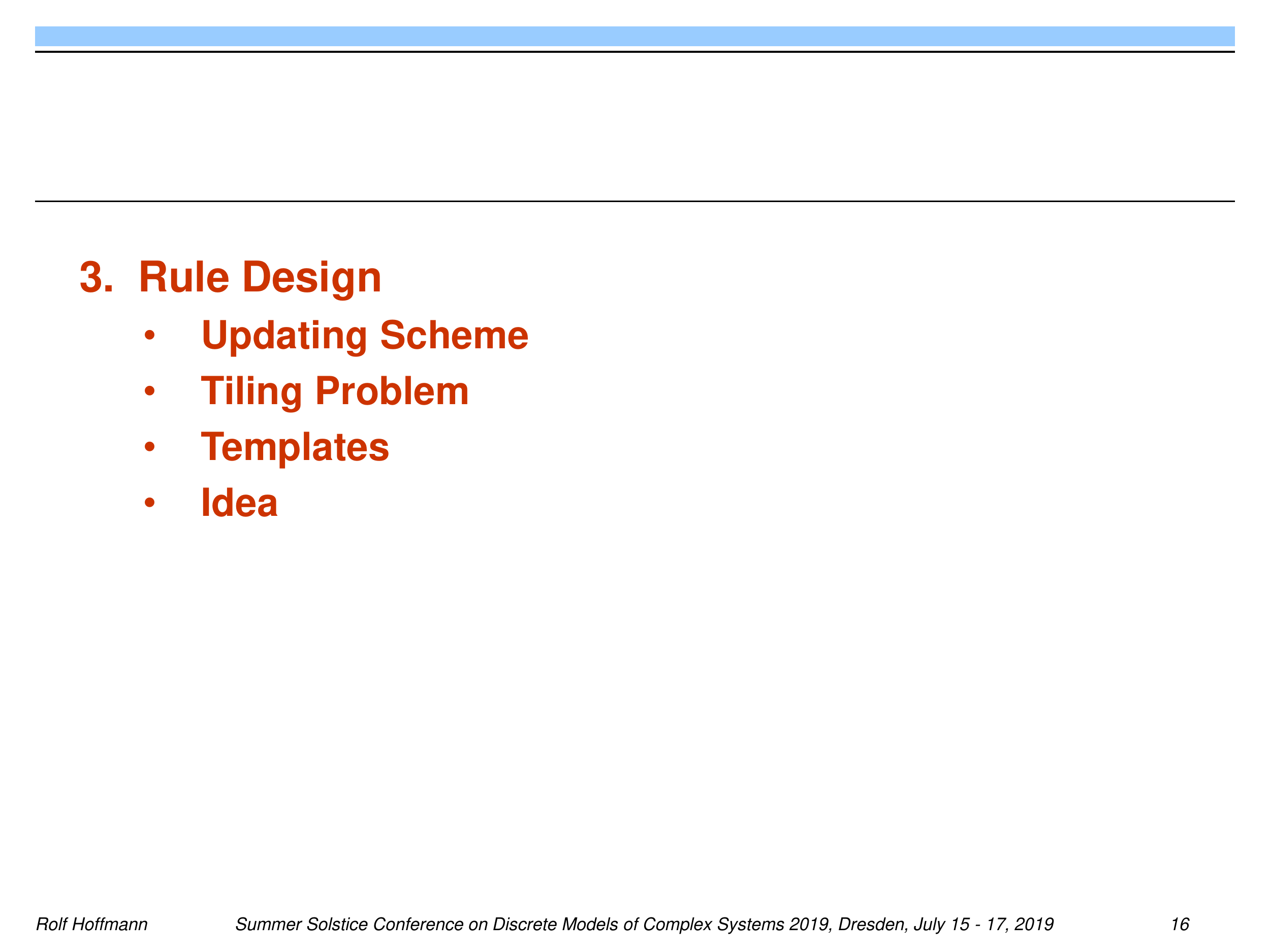}
\includegraphics[width=2.3cm]{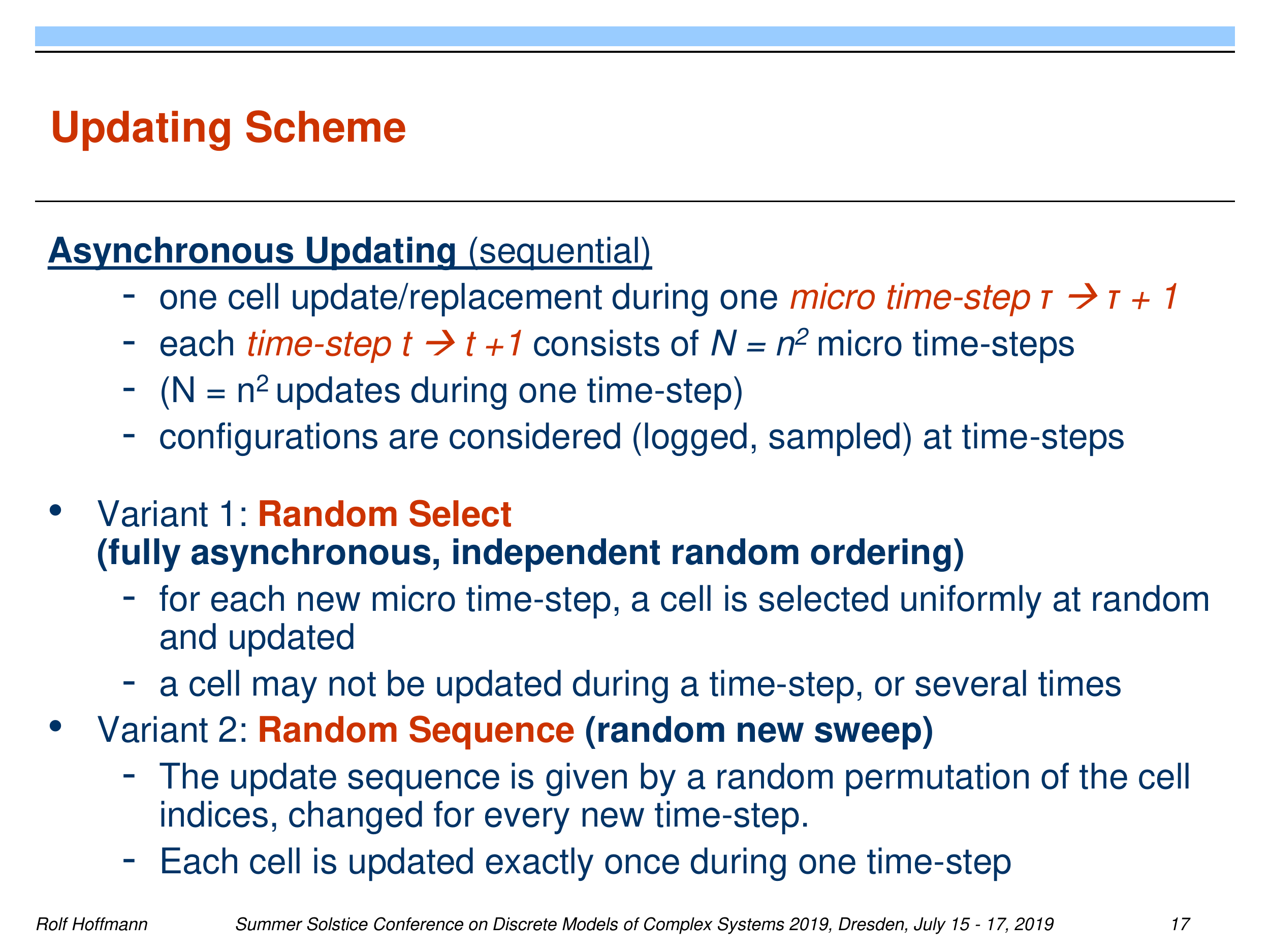}
\includegraphics[width=2.3cm]{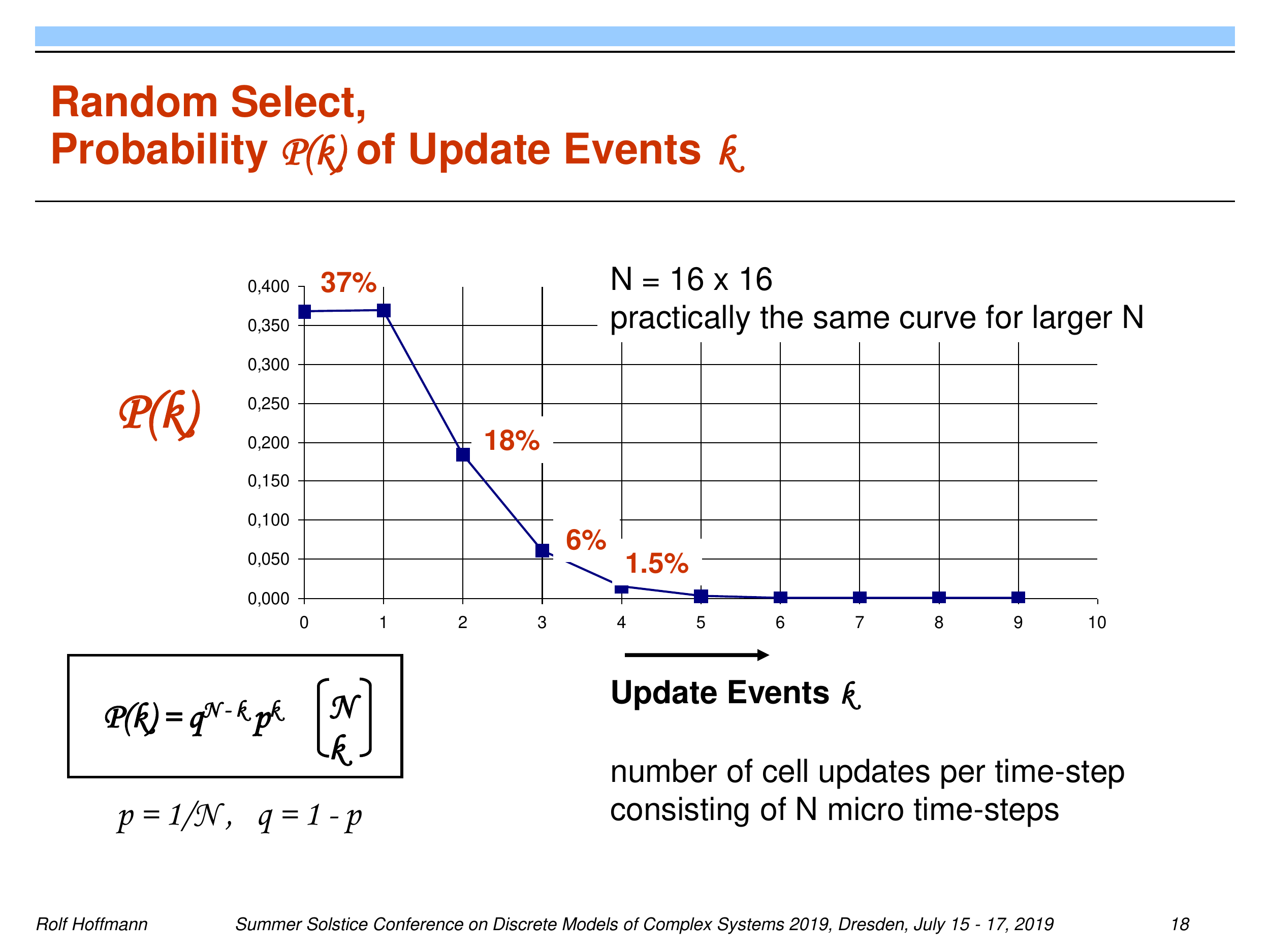}
\includegraphics[width=2.3cm]{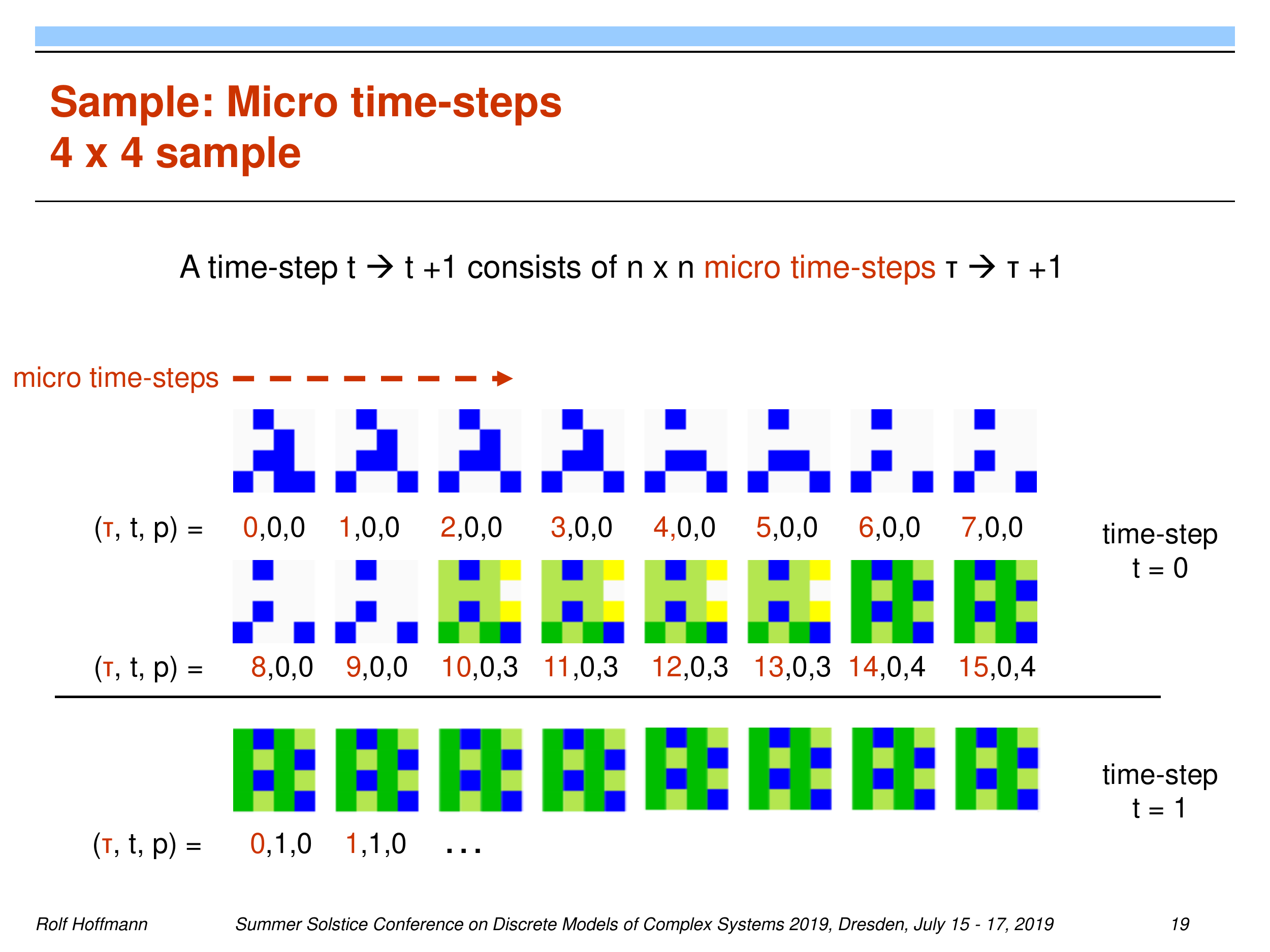}
\includegraphics[width=2.3cm]{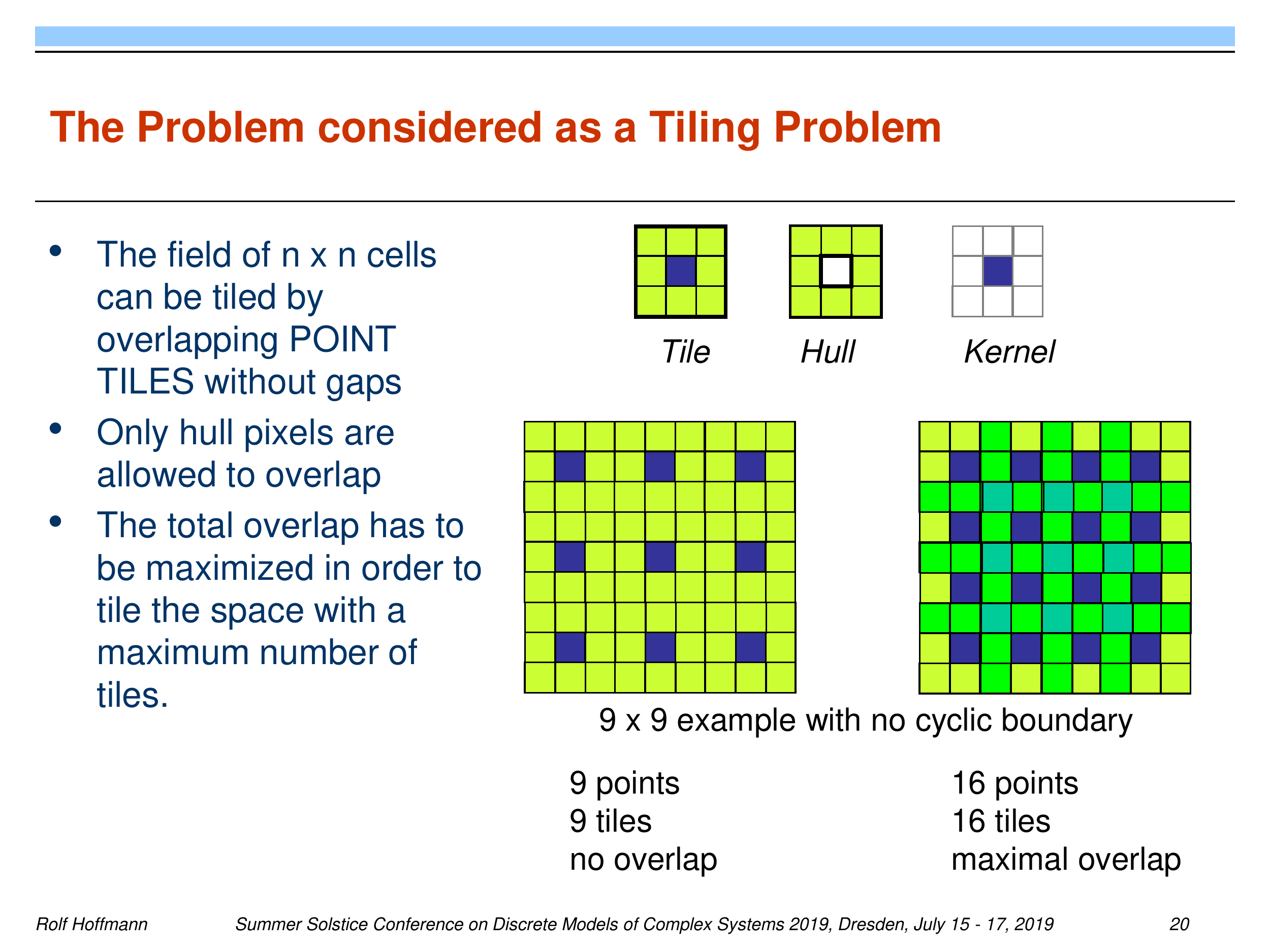}

\includegraphics[width=2.3cm]{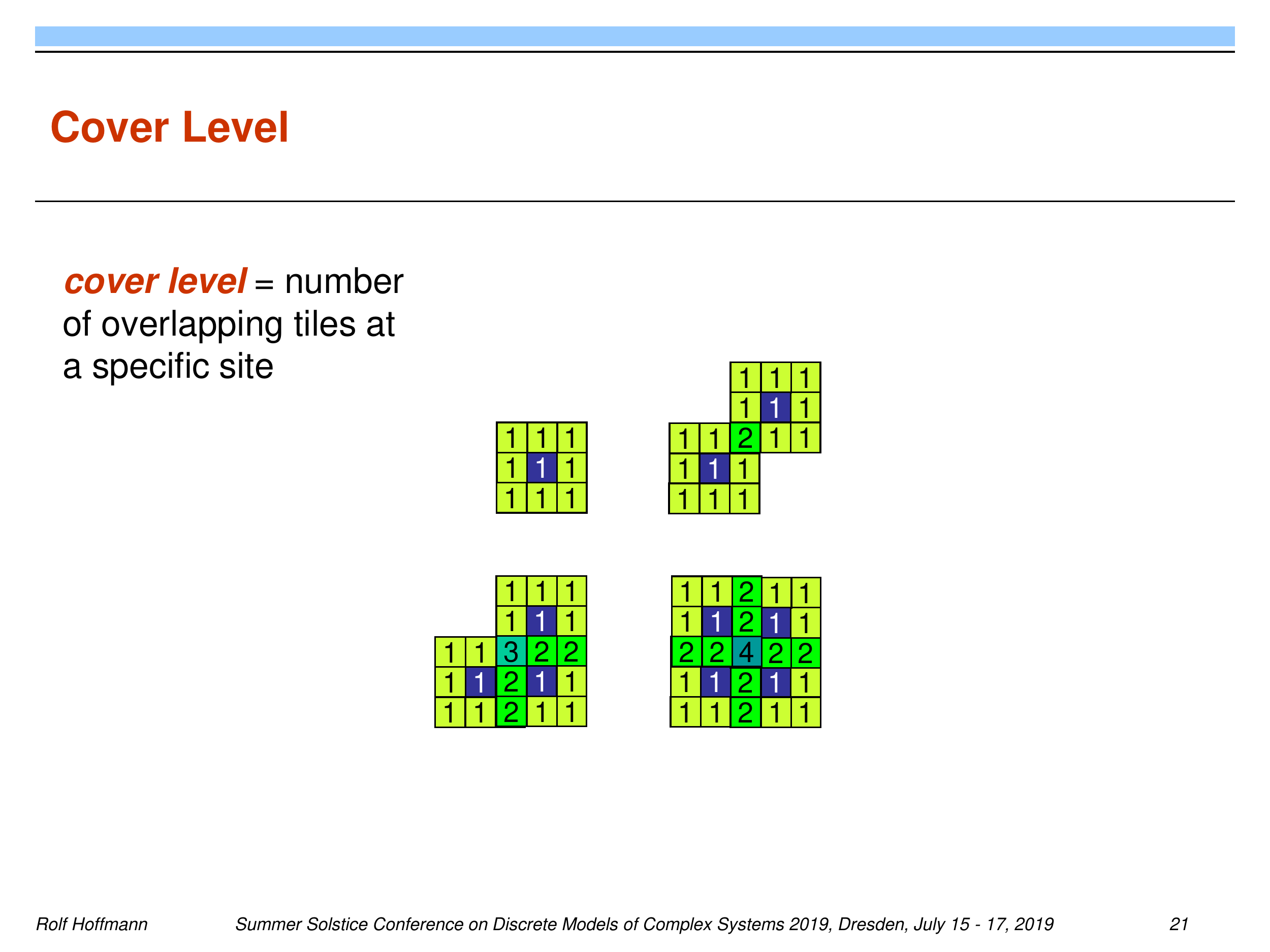}
\includegraphics[width=2.3cm]{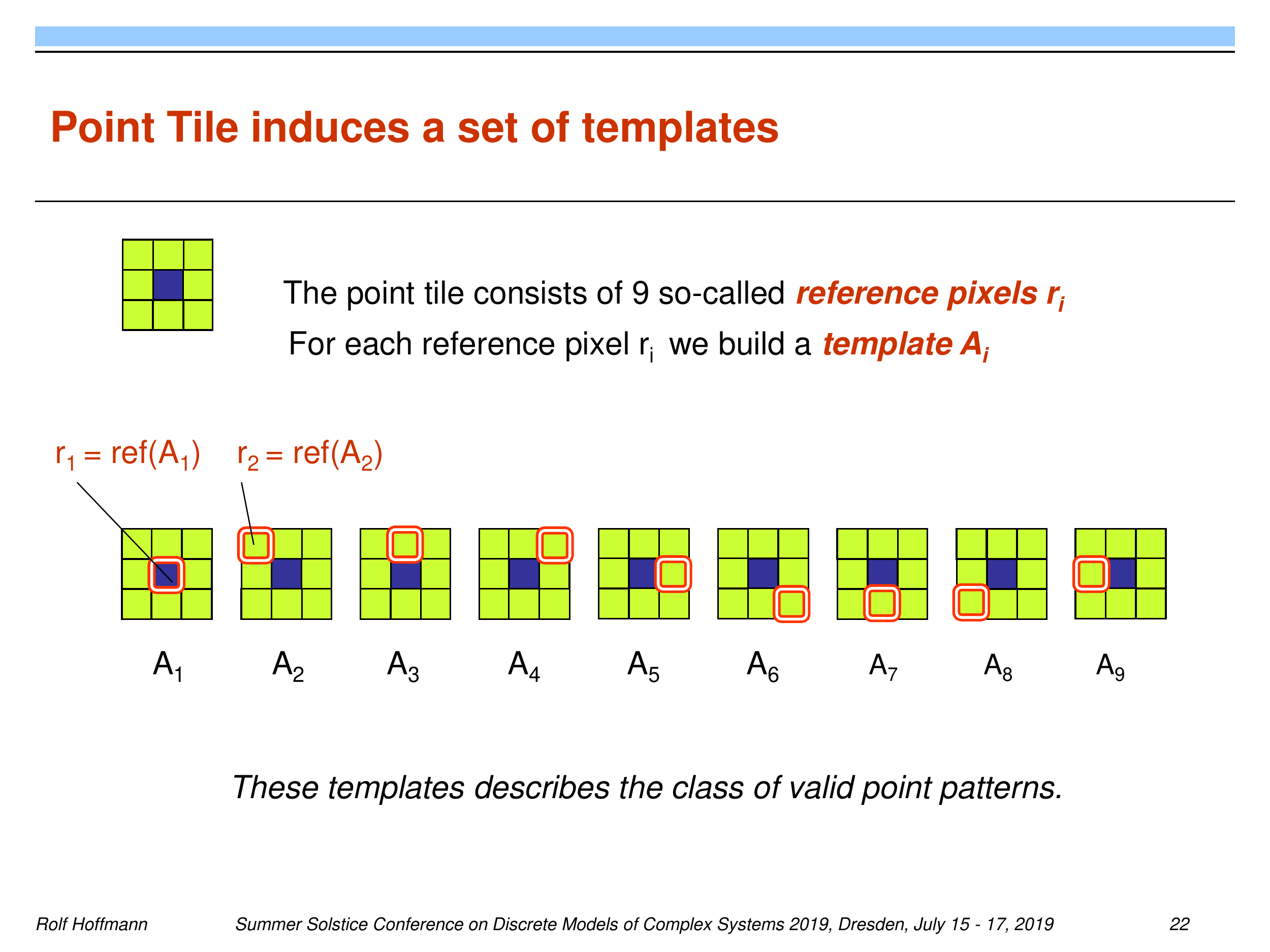}
\includegraphics[width=2.3cm]{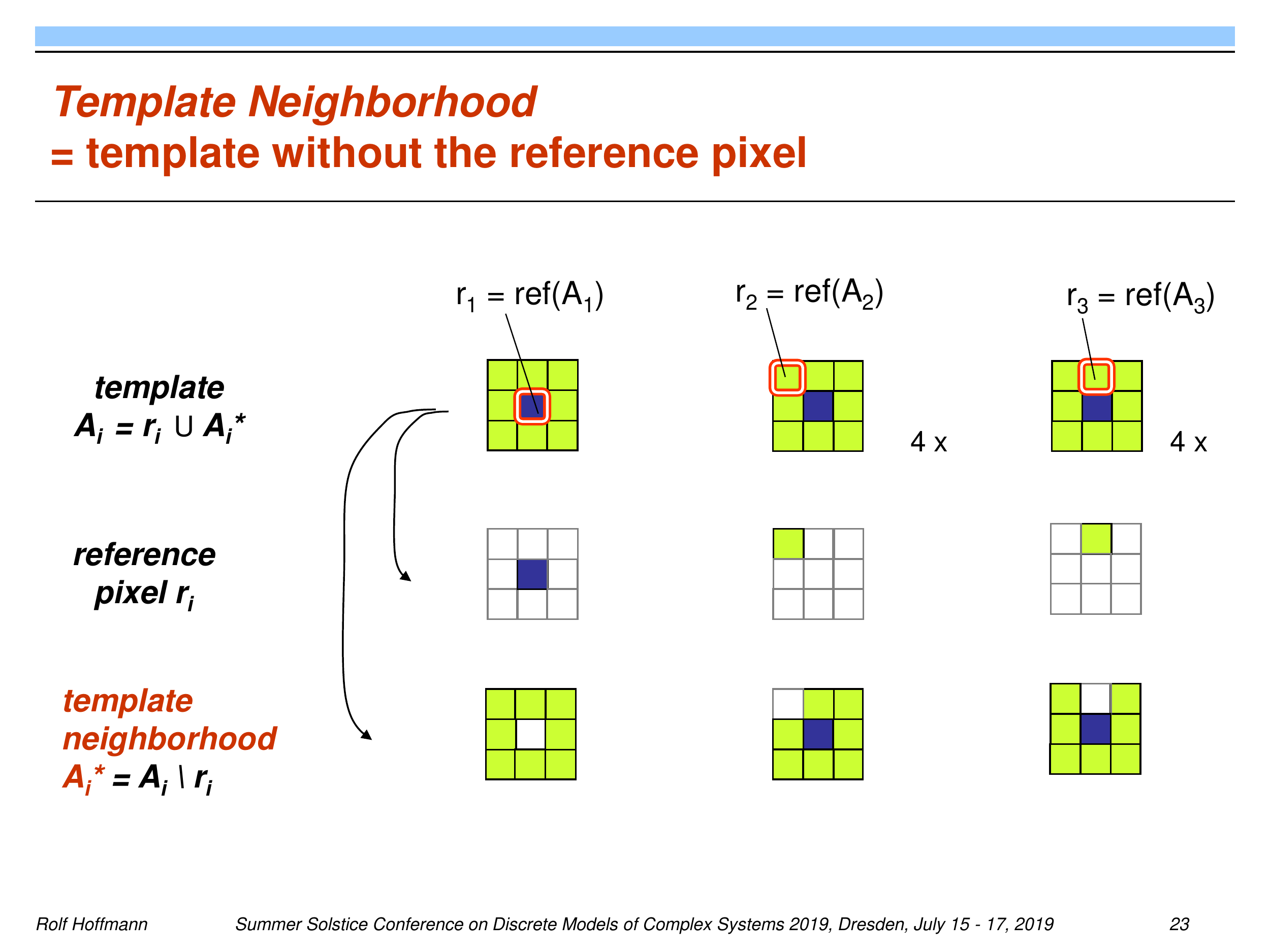}
\includegraphics[width=2.3cm]{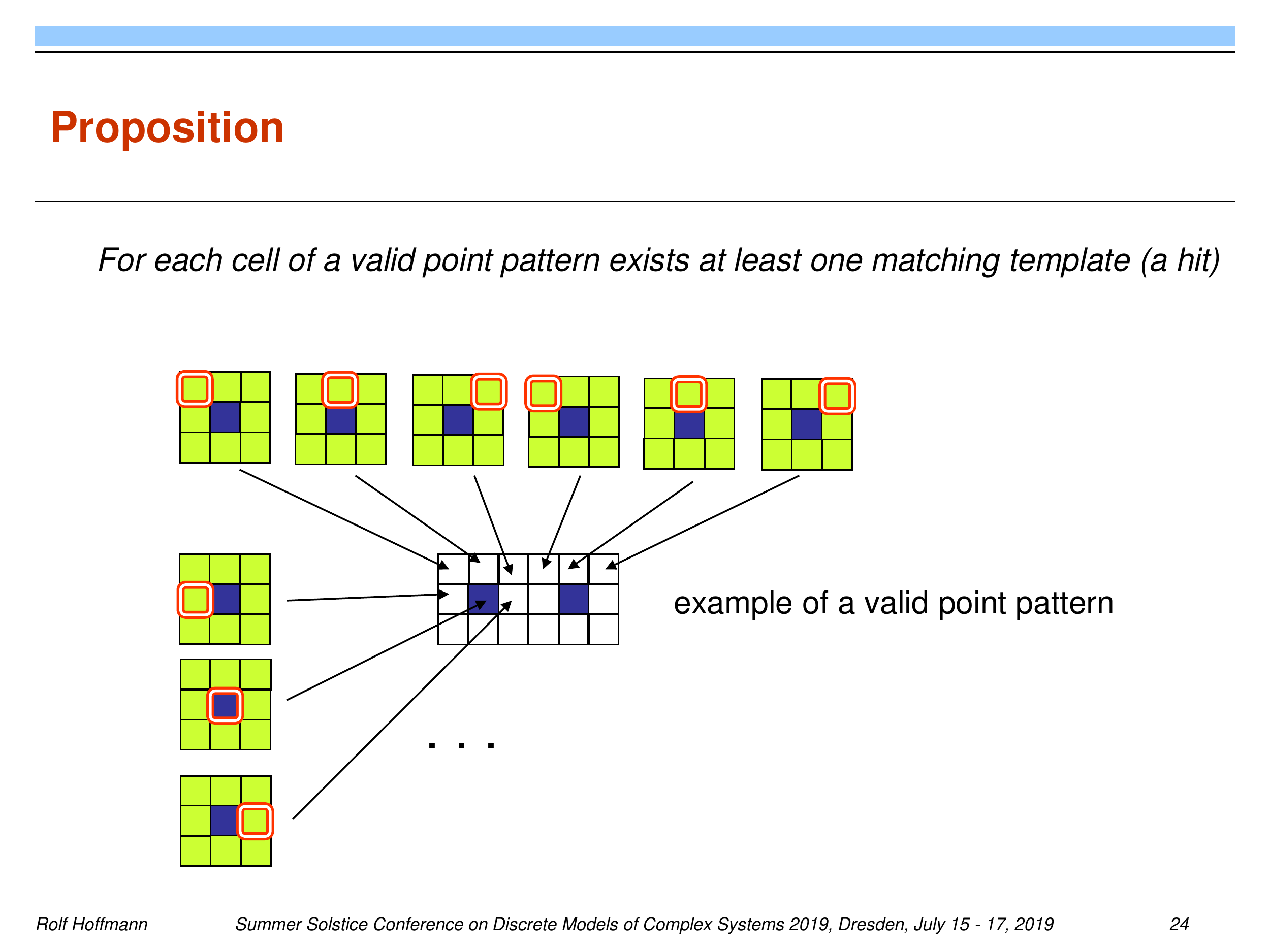}
\includegraphics[width=2.3cm]{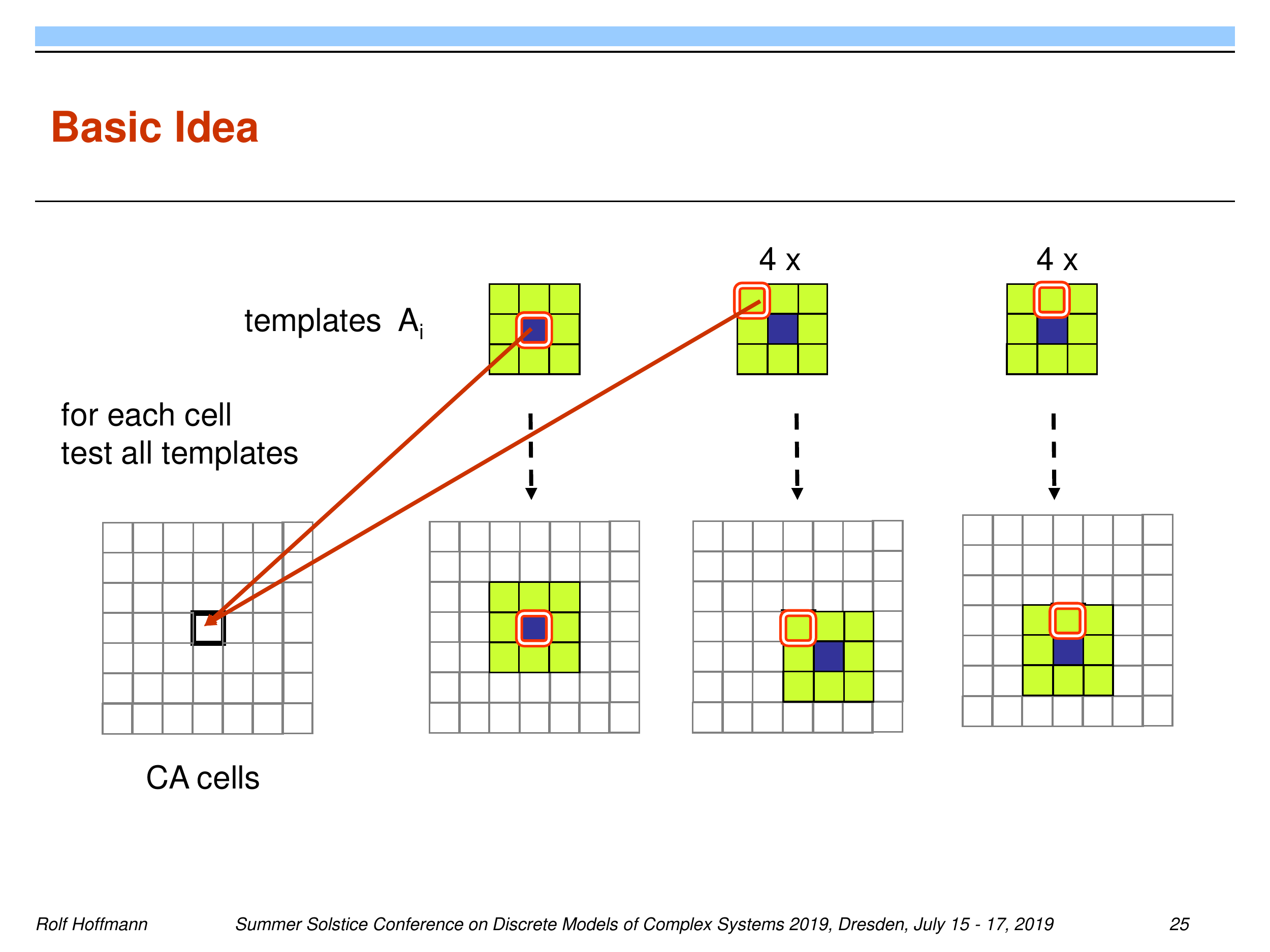}

\includegraphics[width=2.3cm]{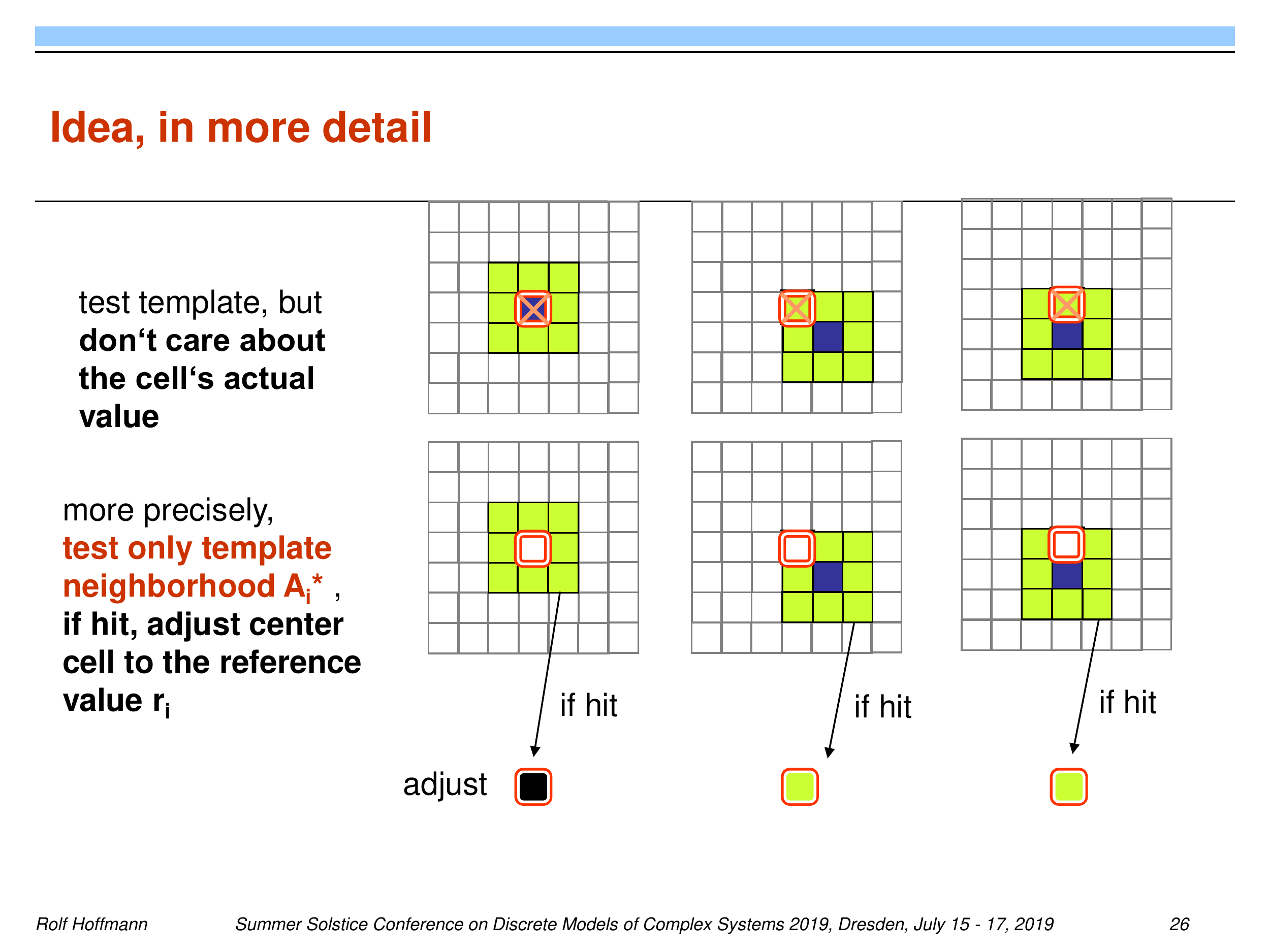}
\includegraphics[width=2.3cm]{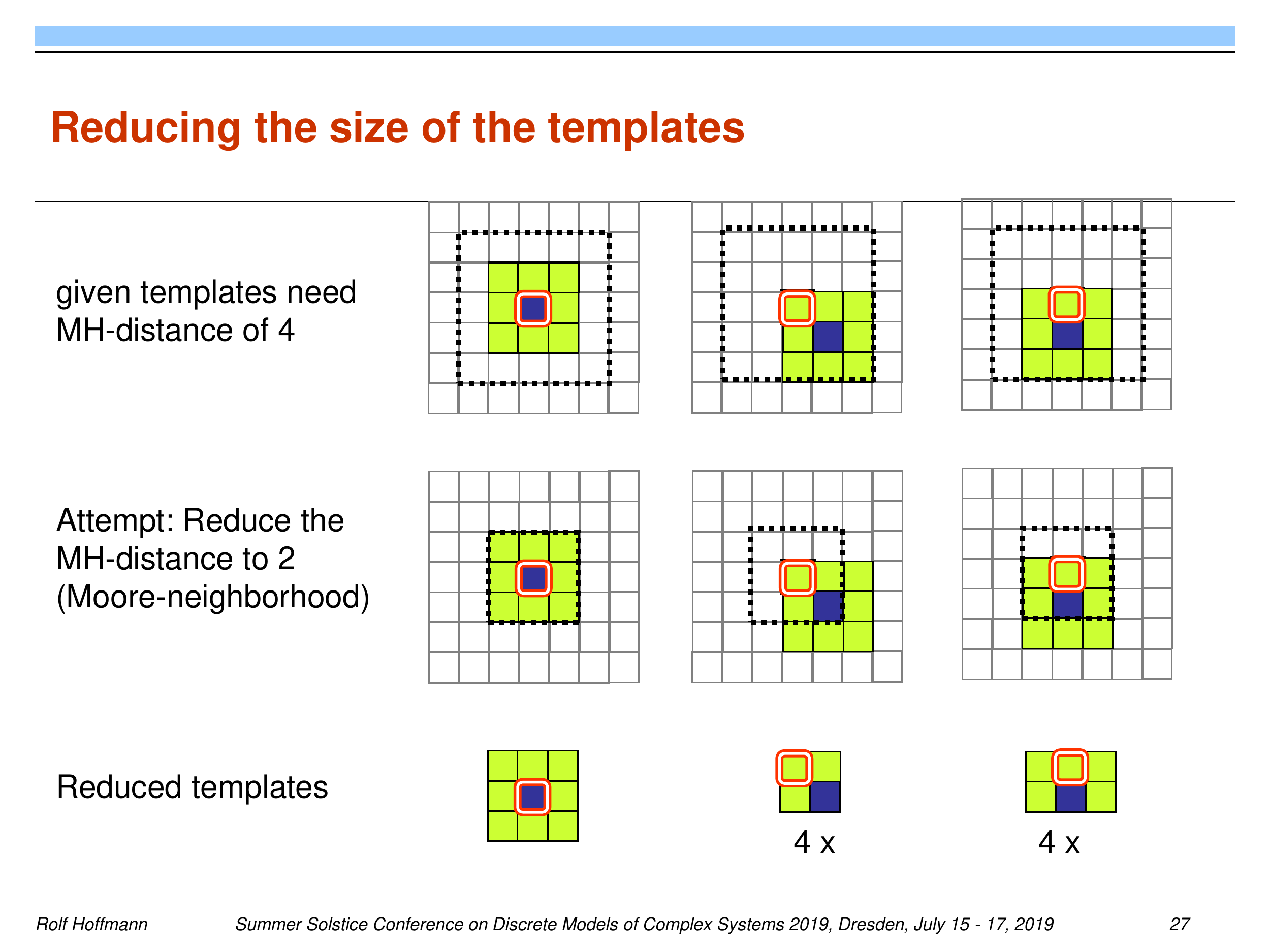}
\includegraphics[width=2.3cm]{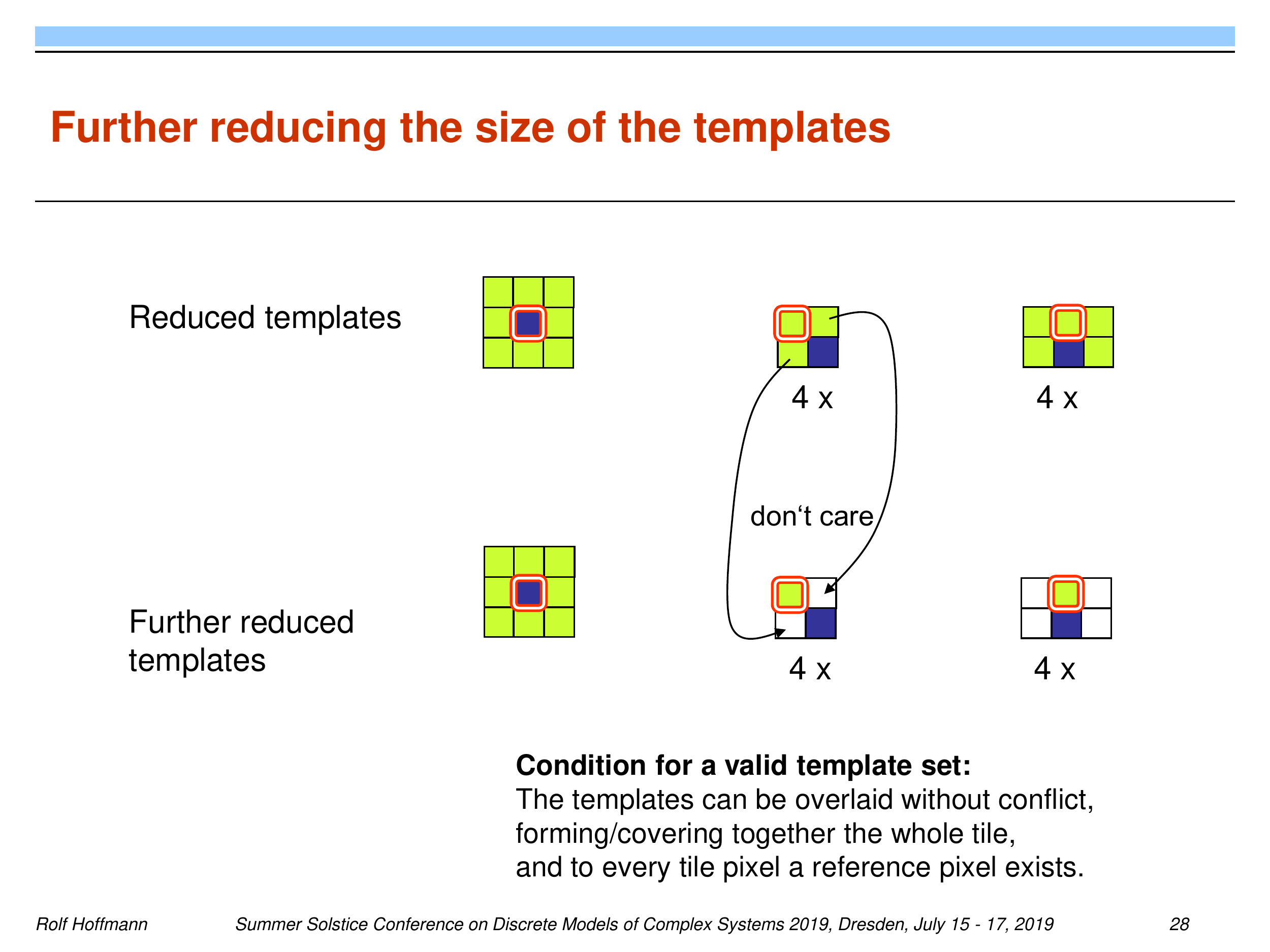}
\includegraphics[width=2.3cm]{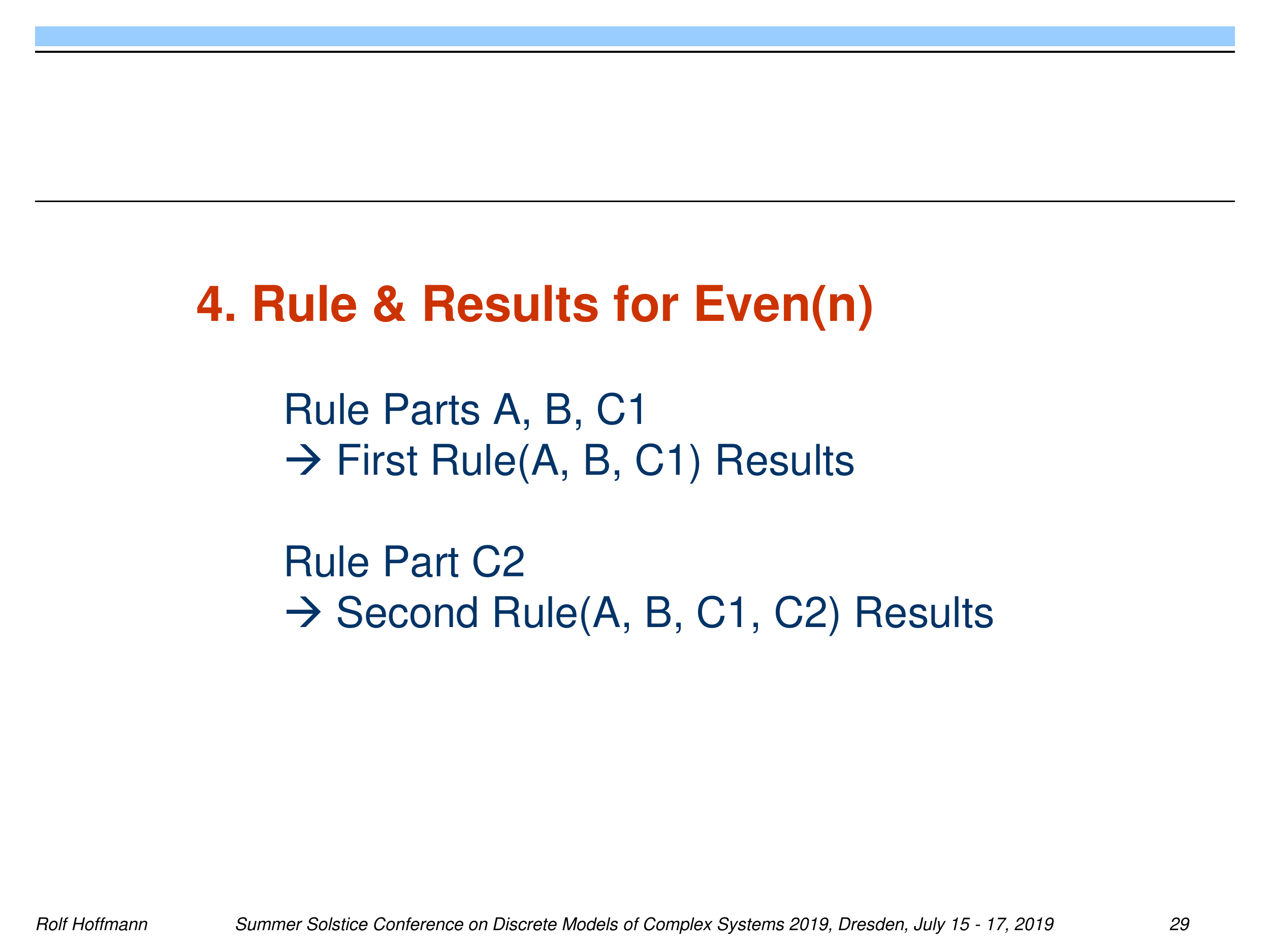}
\includegraphics[width=2.3cm]{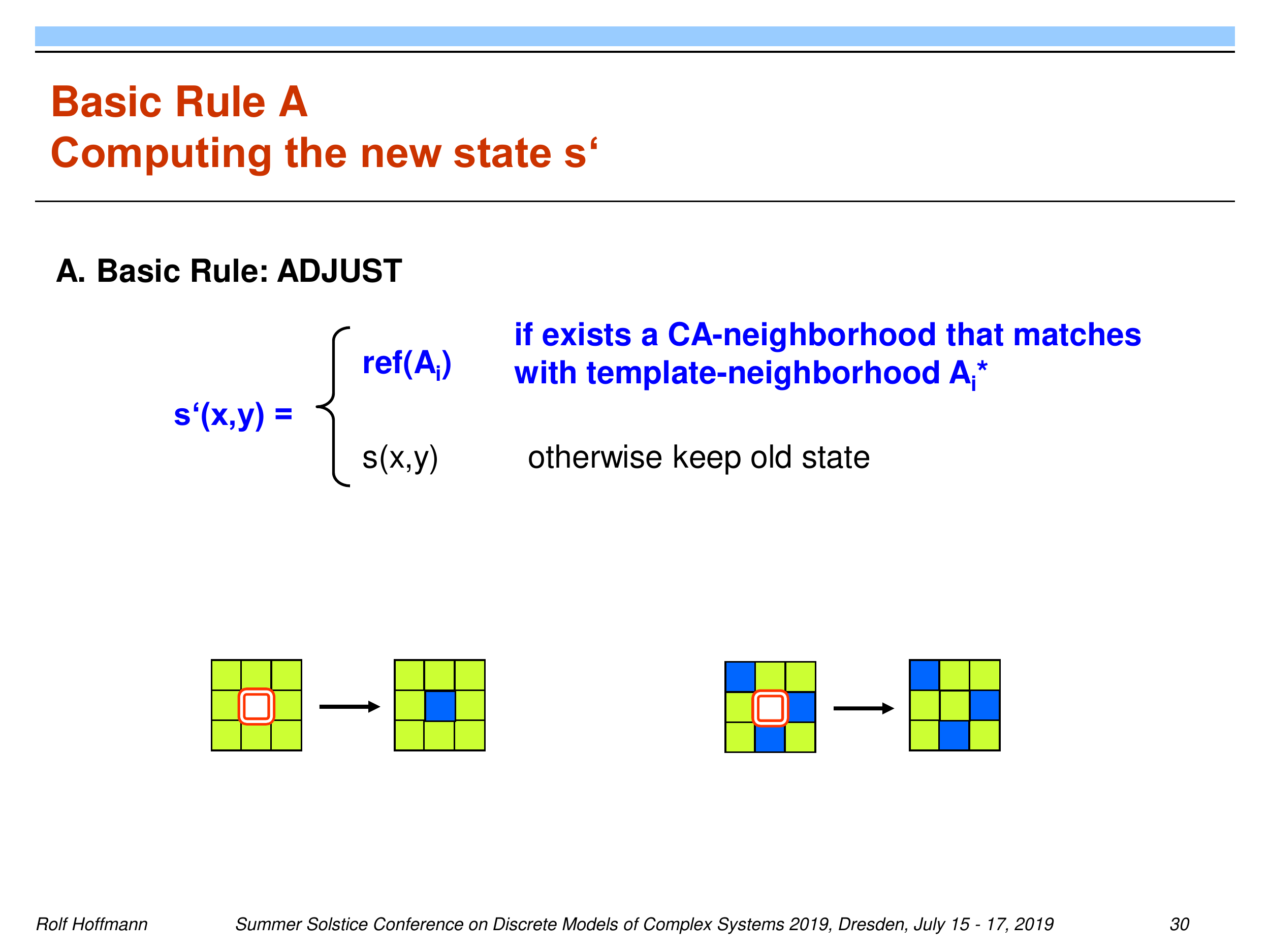}

\includegraphics[width=2.3cm]{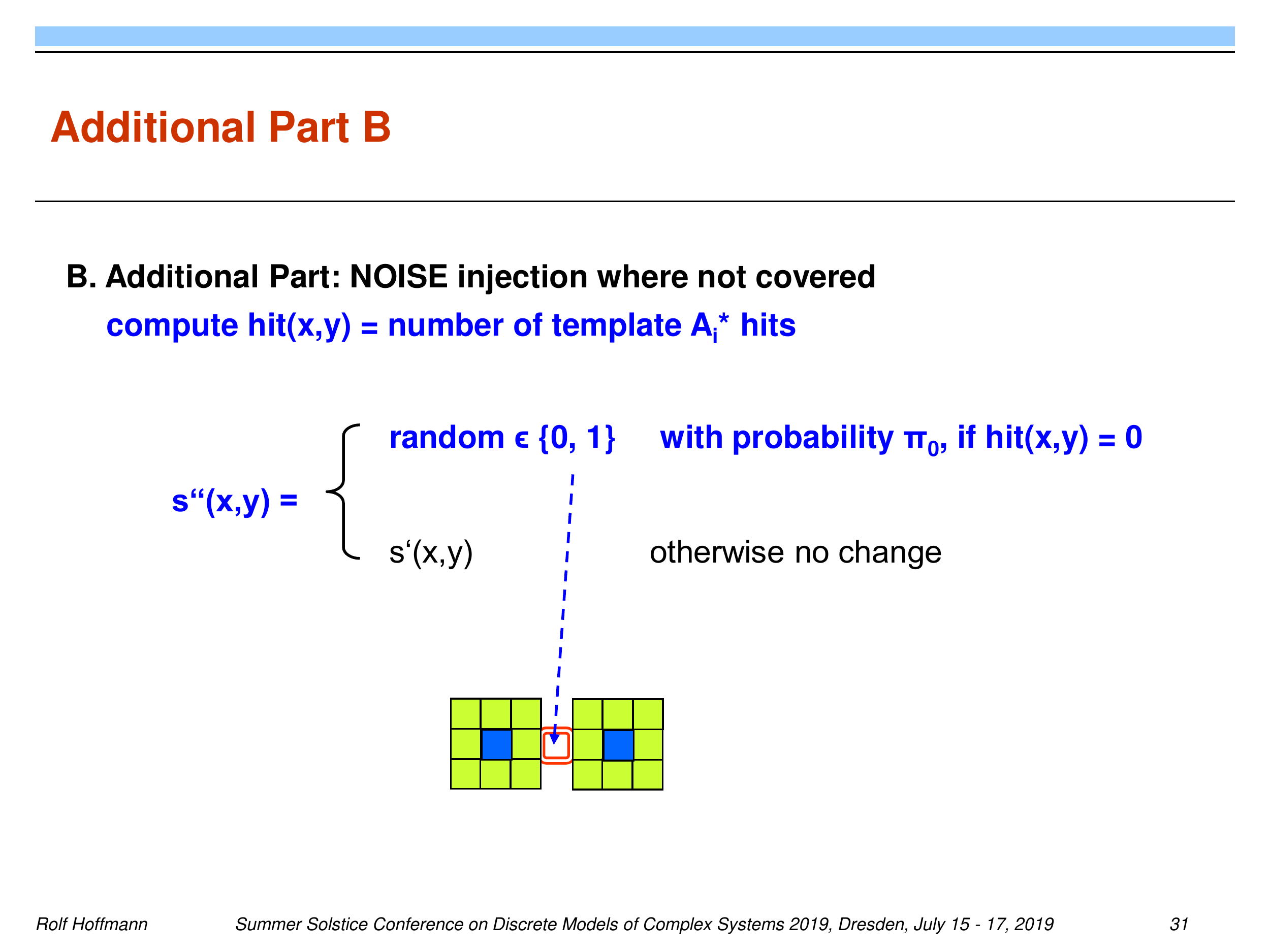}
\includegraphics[width=2.3cm]{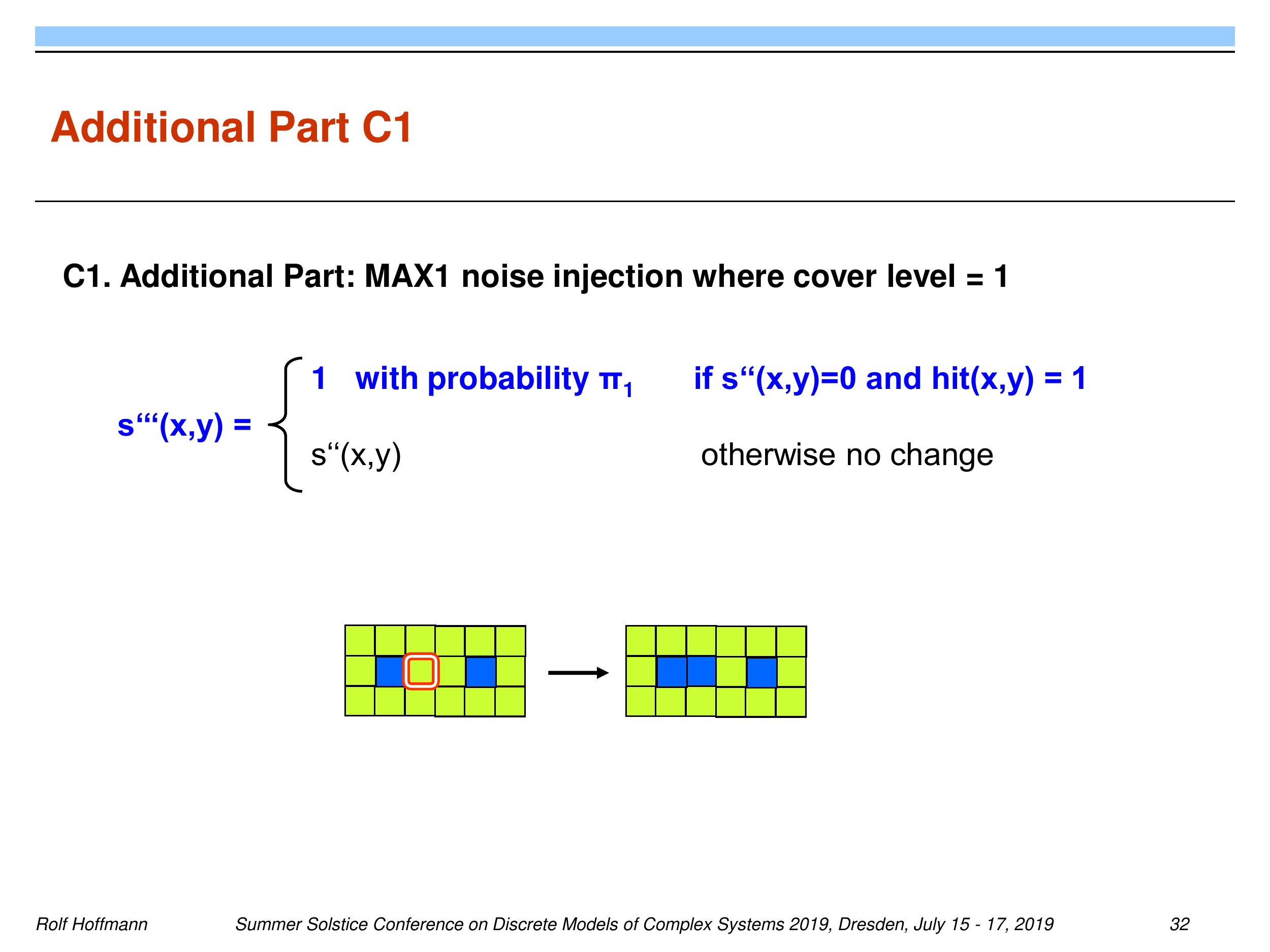}
\includegraphics[width=2.3cm]{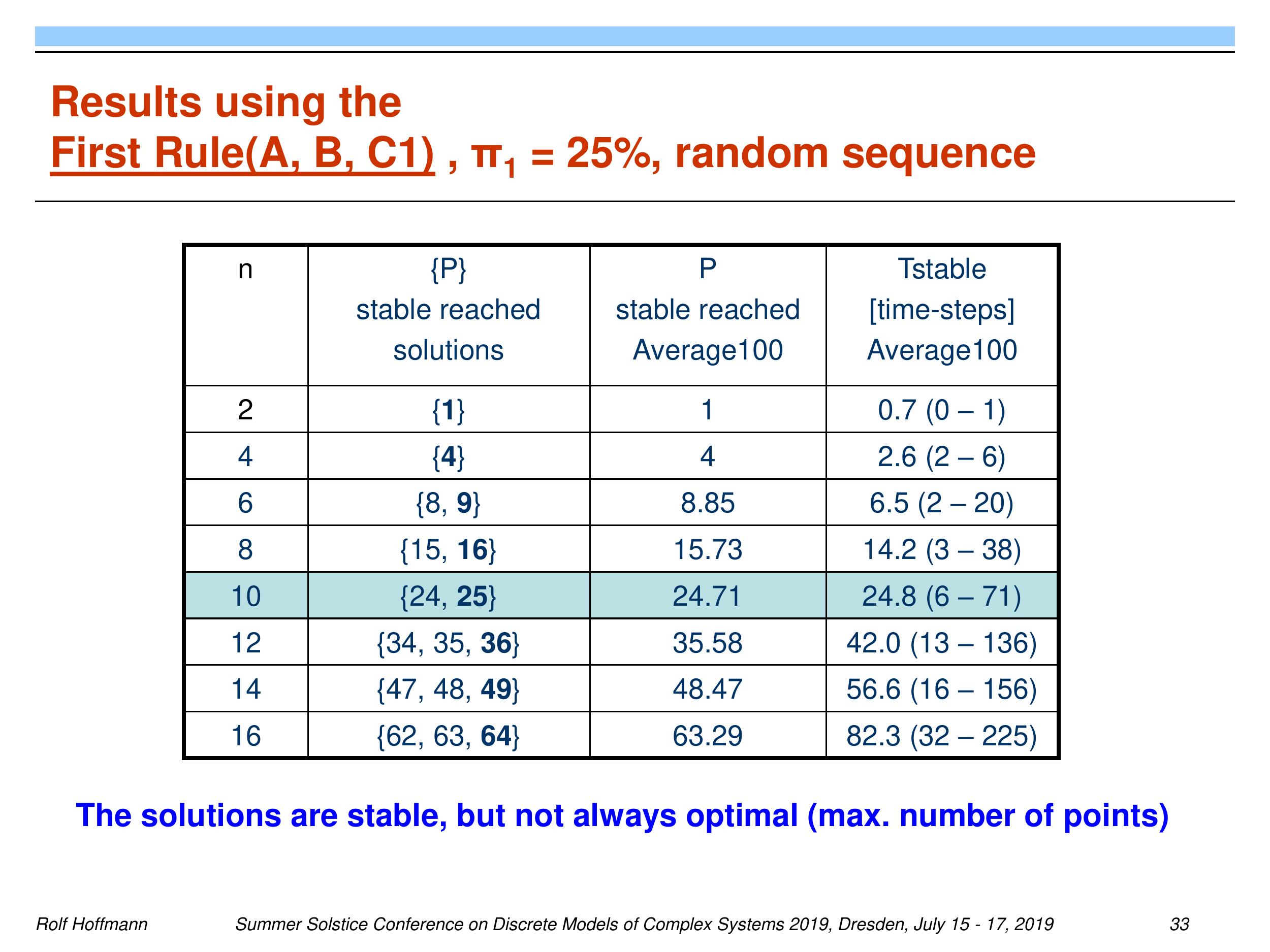}
\includegraphics[width=2.3cm]{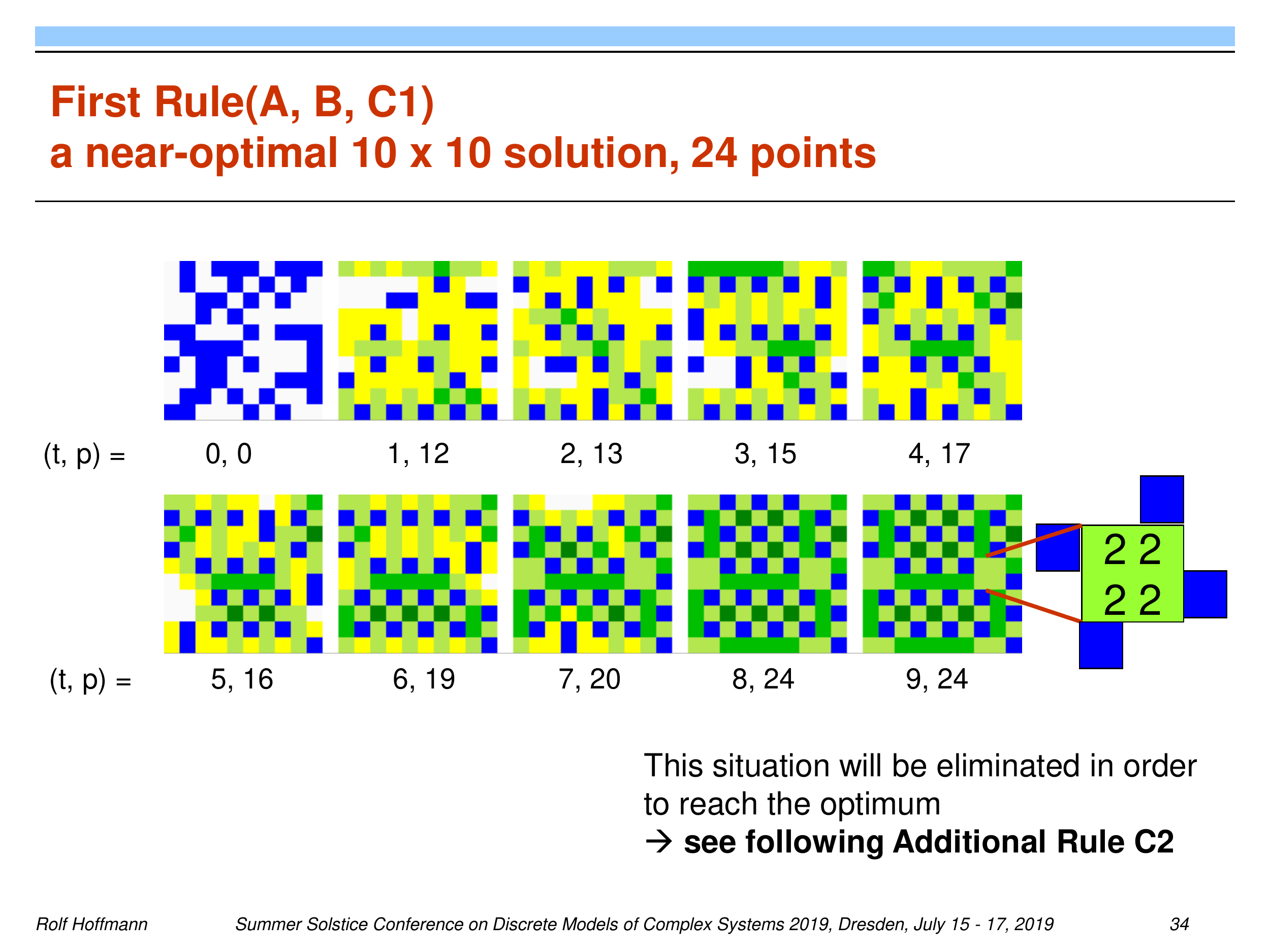}
\includegraphics[width=2.3cm]{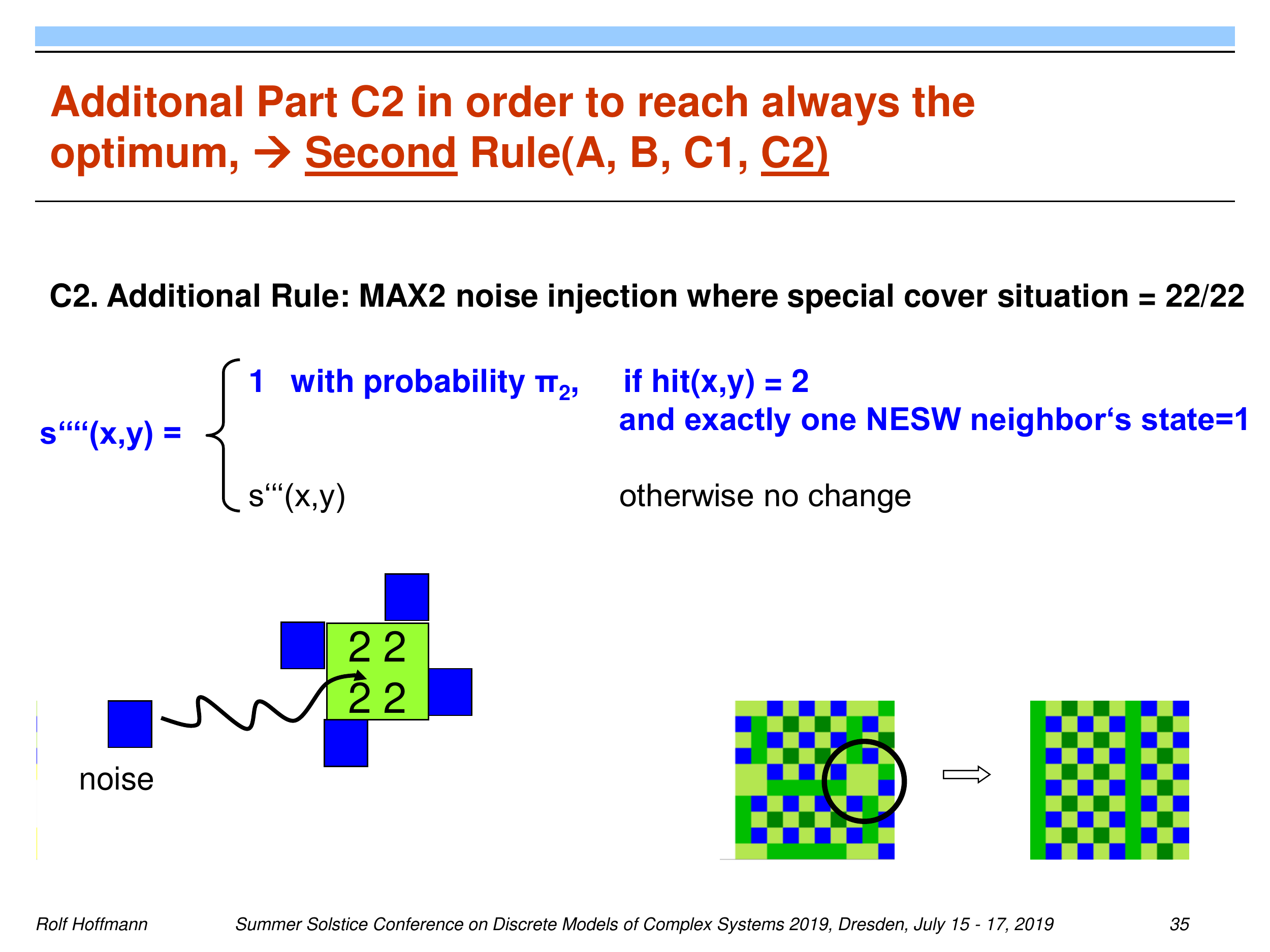}

\includegraphics[width=2.3cm]{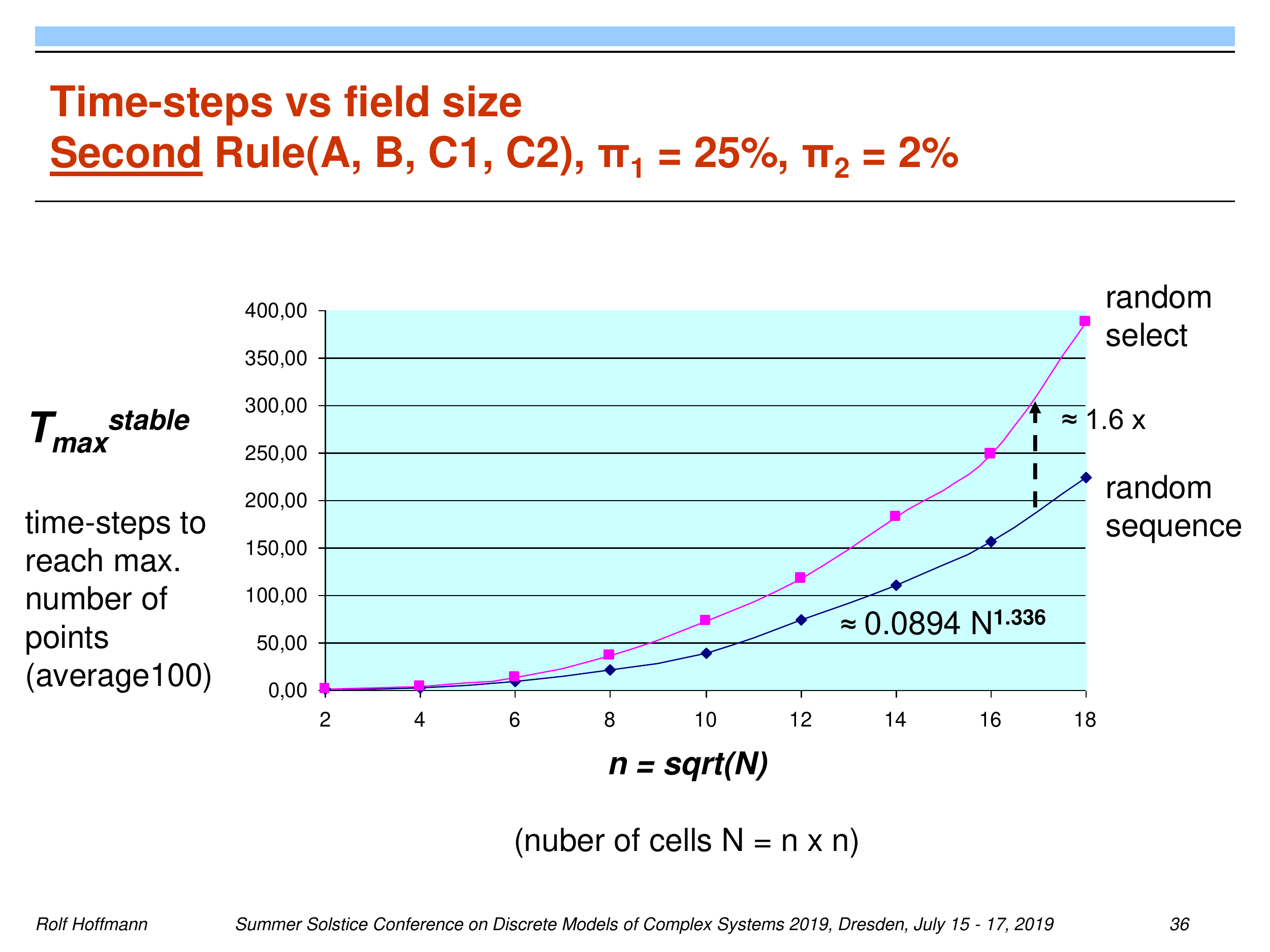}
\includegraphics[width=2.3cm]{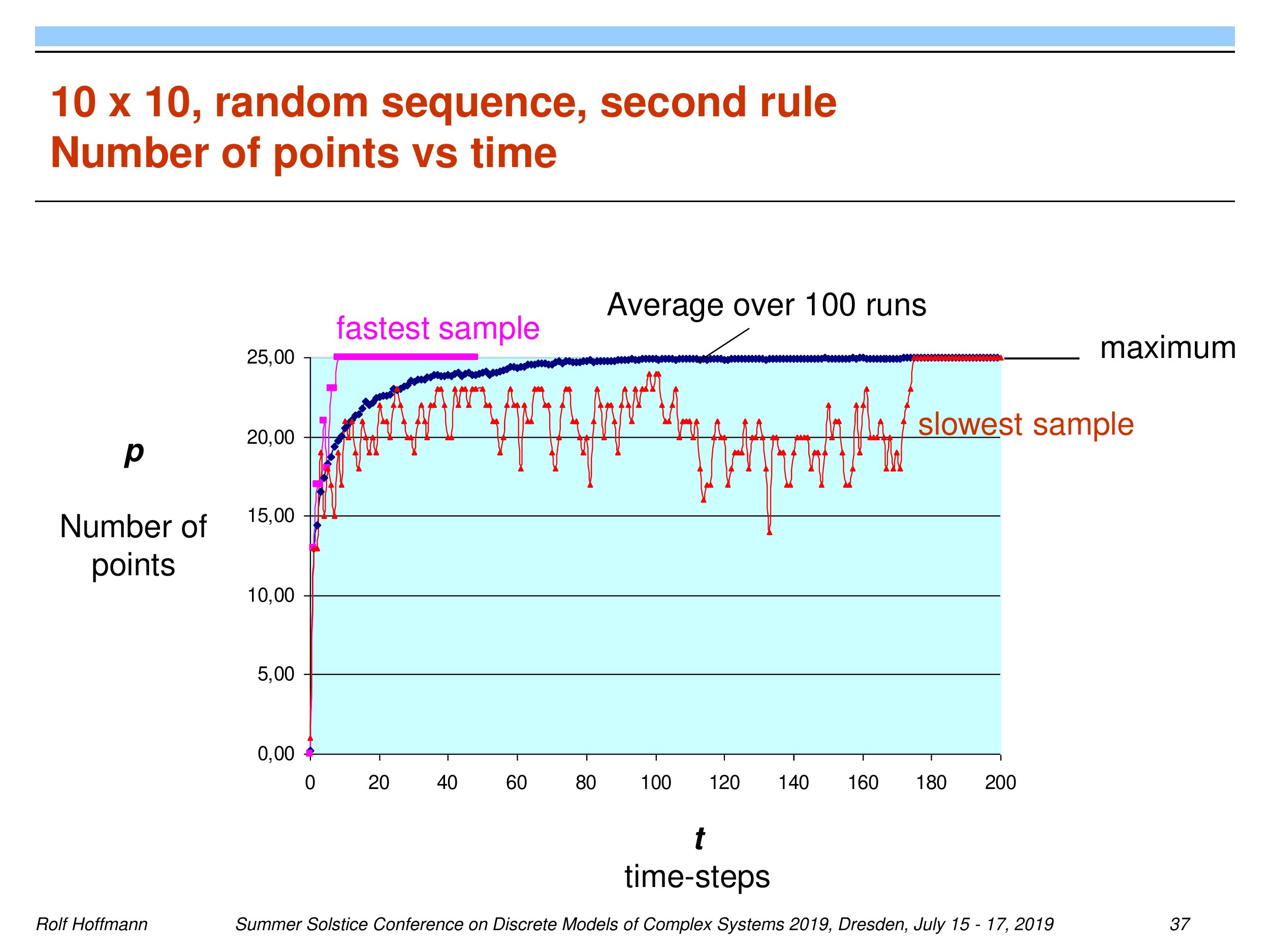}
\includegraphics[width=2.3cm]{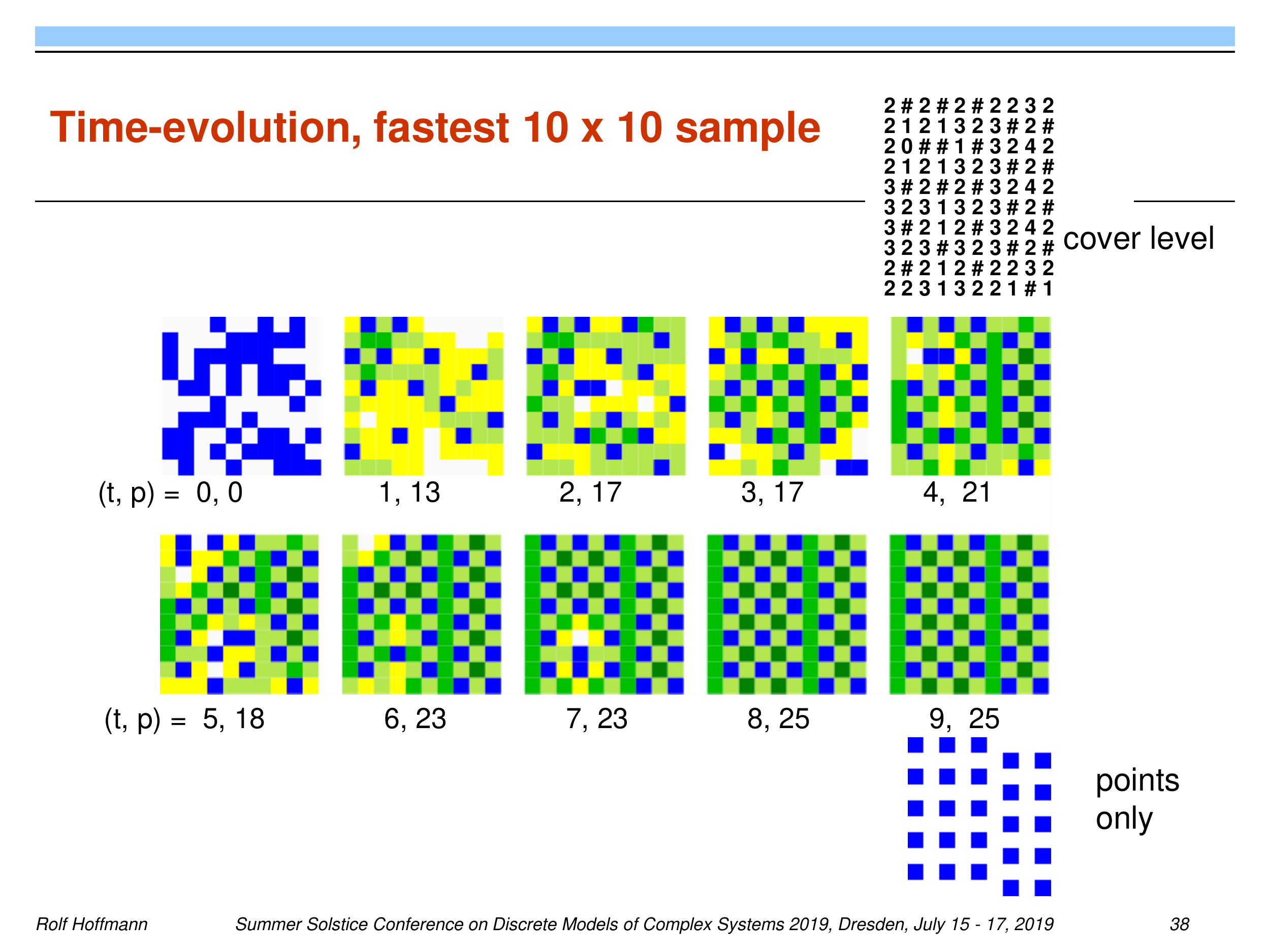}
\includegraphics[width=2.3cm]{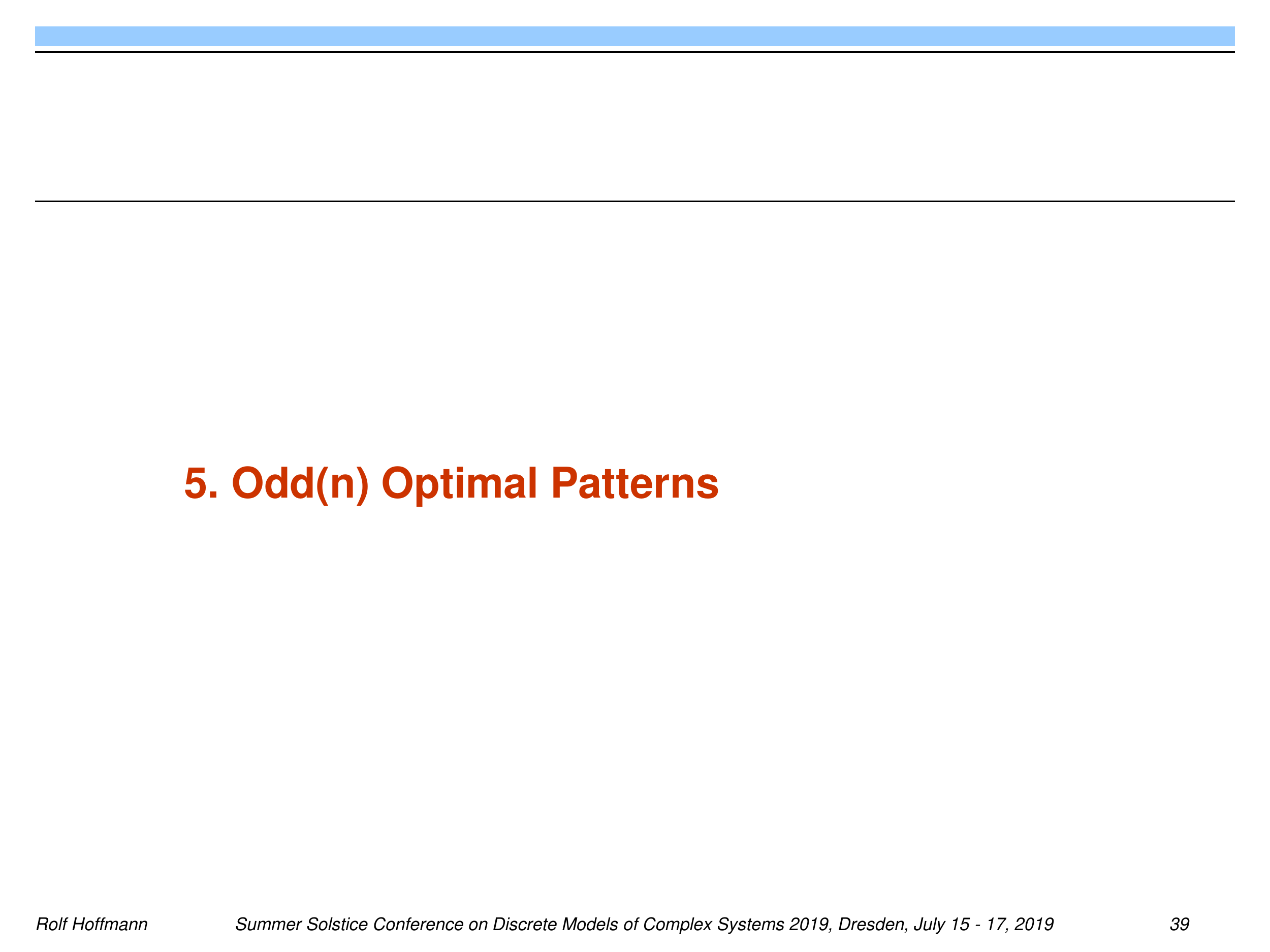}
\includegraphics[width=2.3cm]{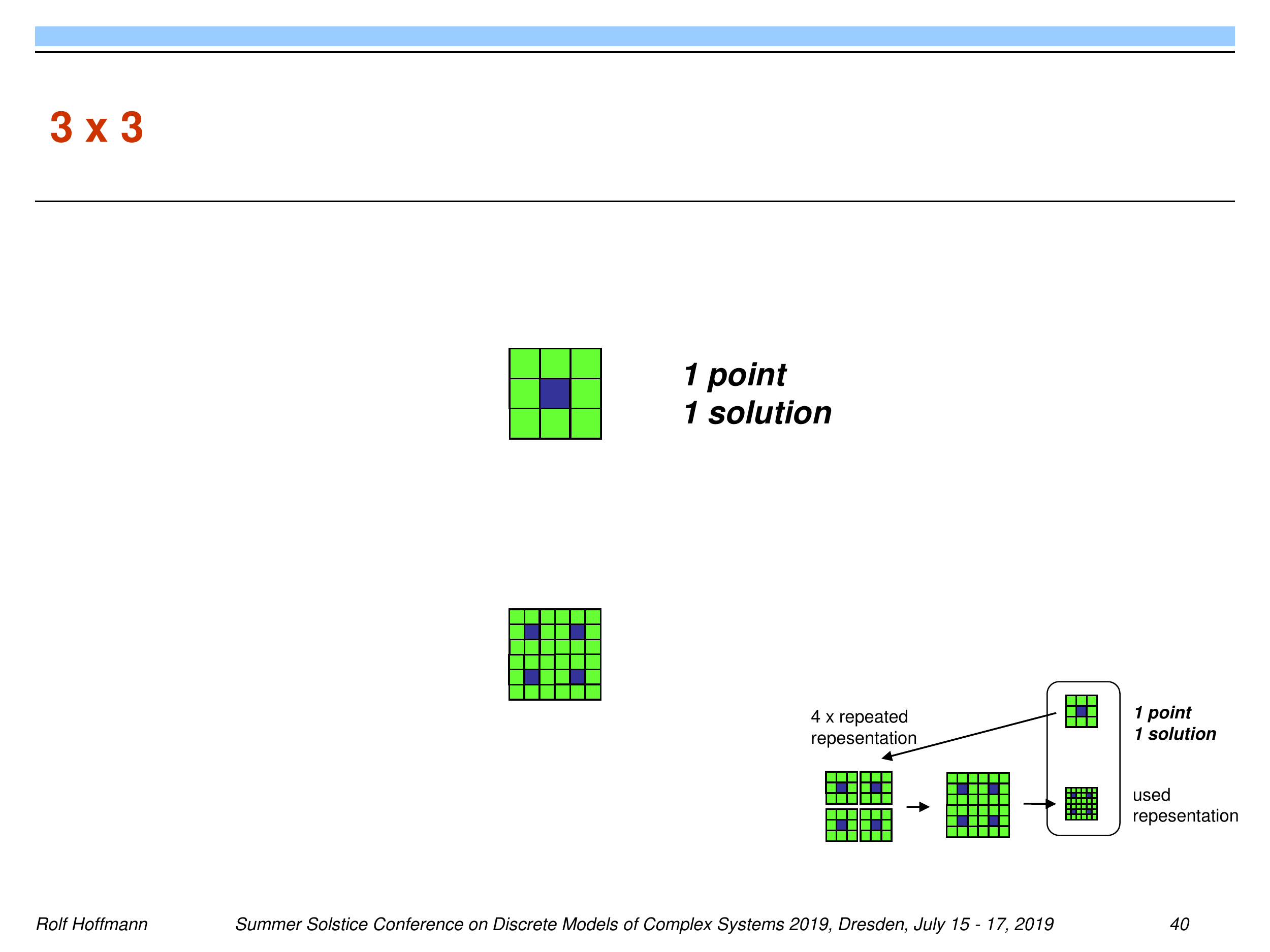}

\includegraphics[width=2.3cm]{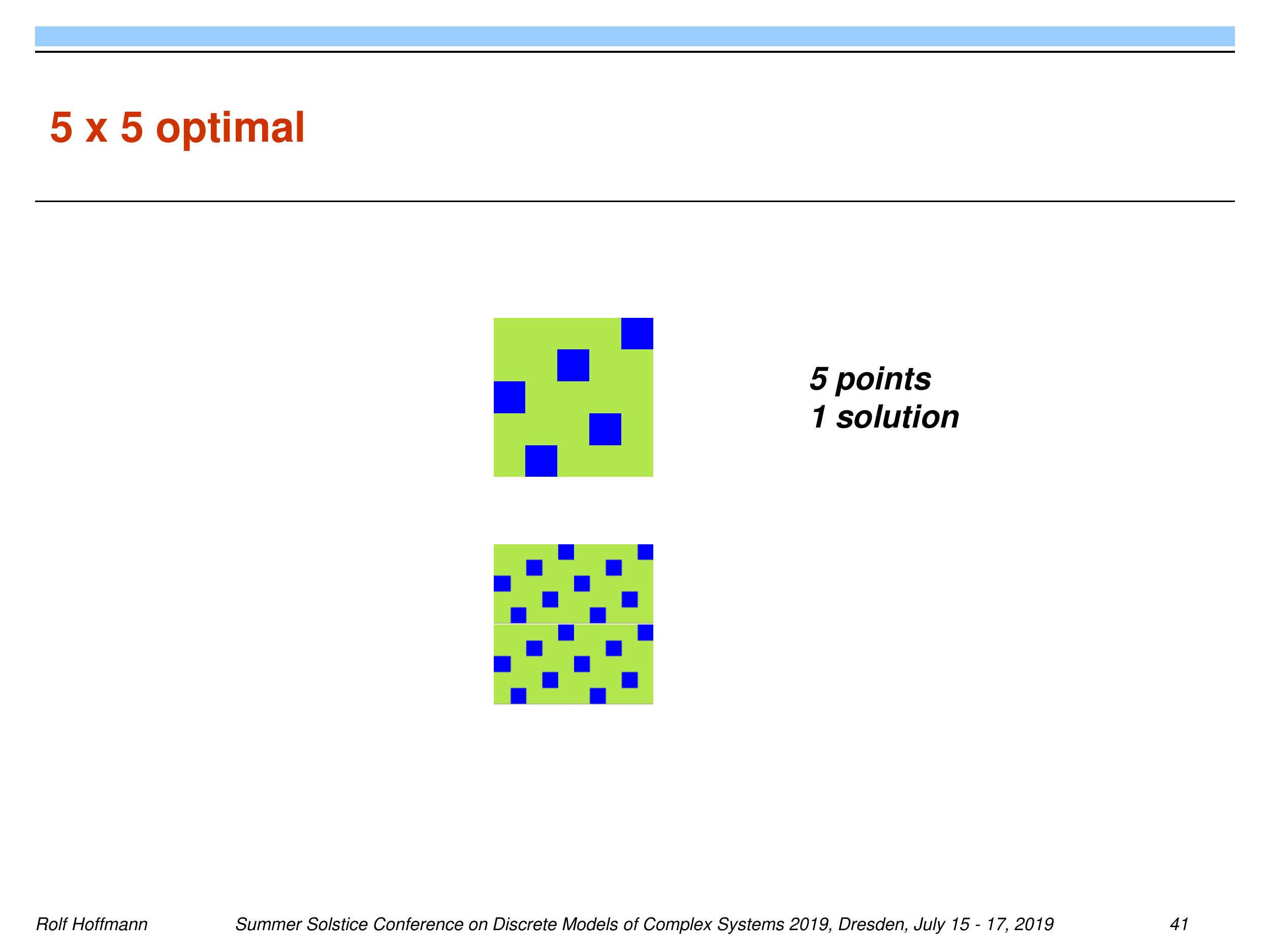}
\includegraphics[width=2.3cm]{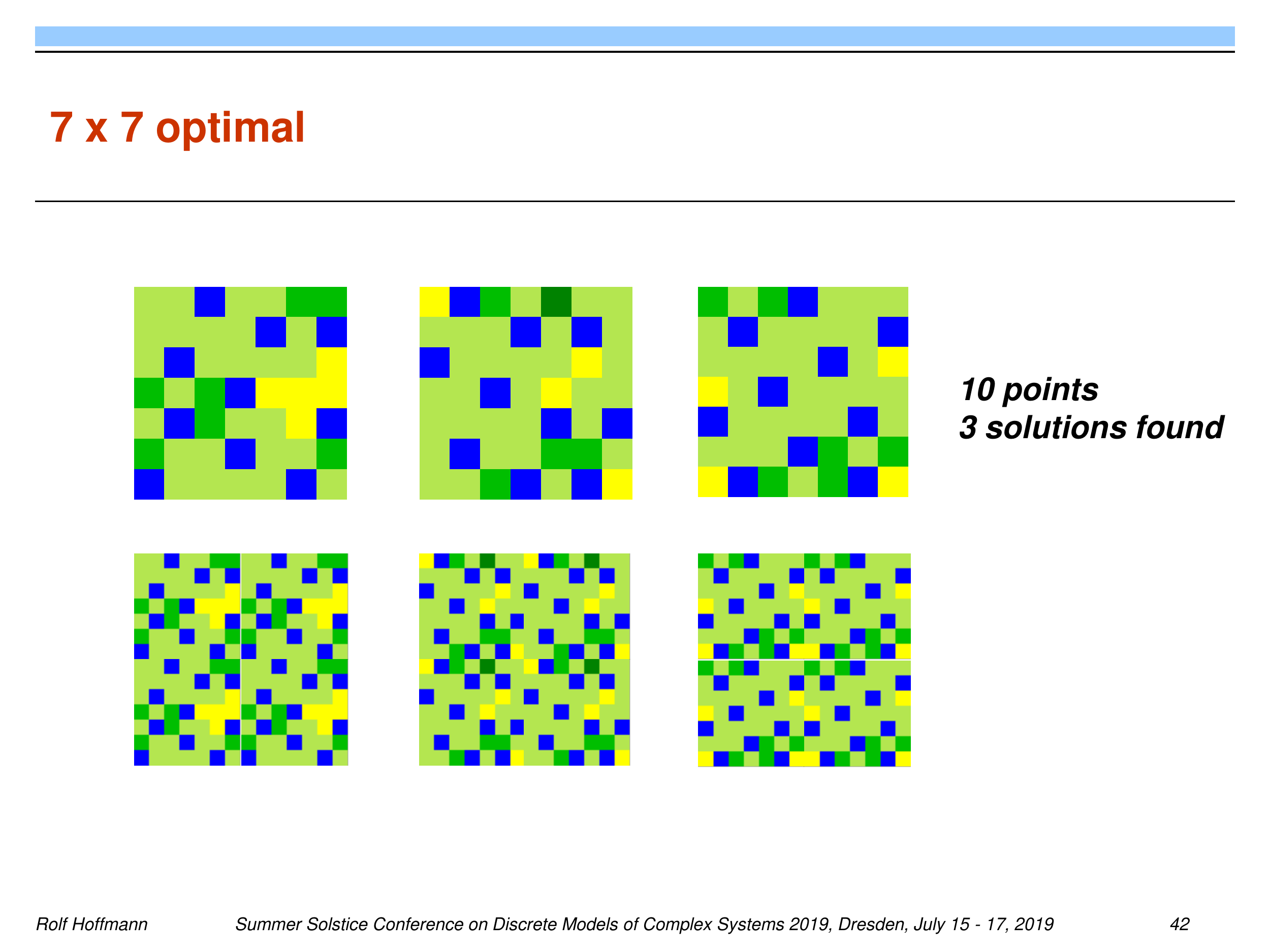}
\includegraphics[width=2.3cm]{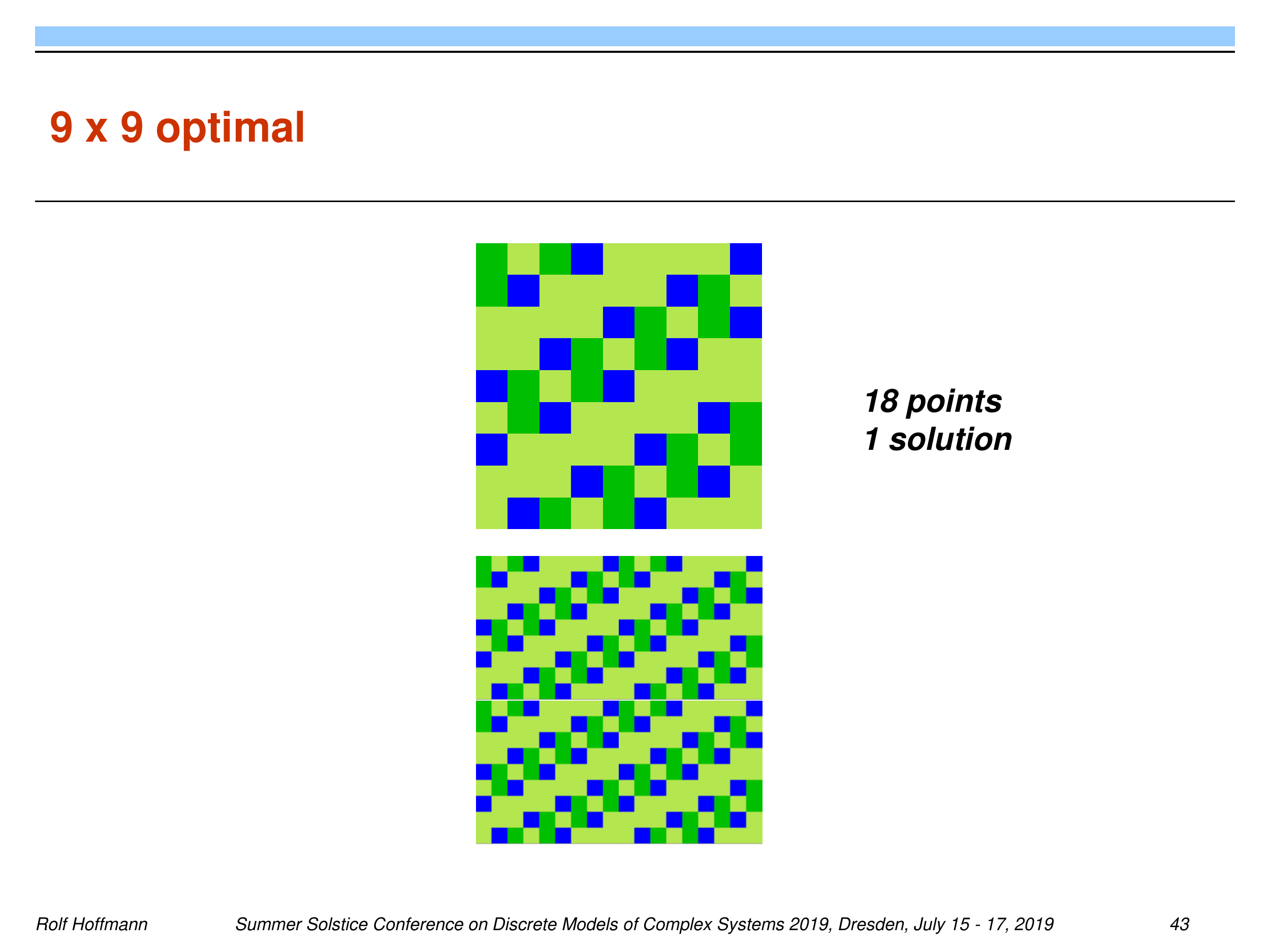}
\includegraphics[width=2.3cm]{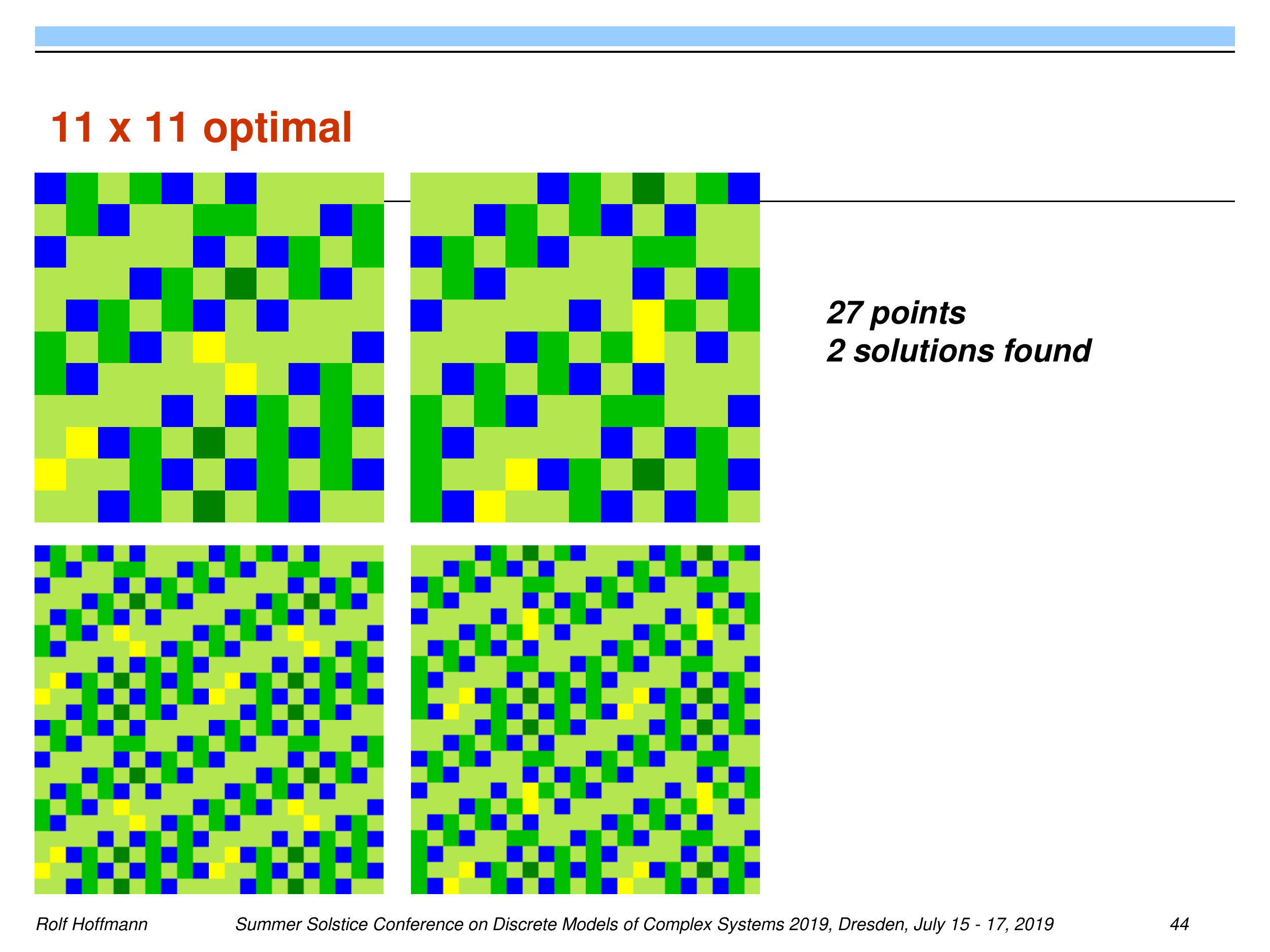}
\includegraphics[width=2.3cm]{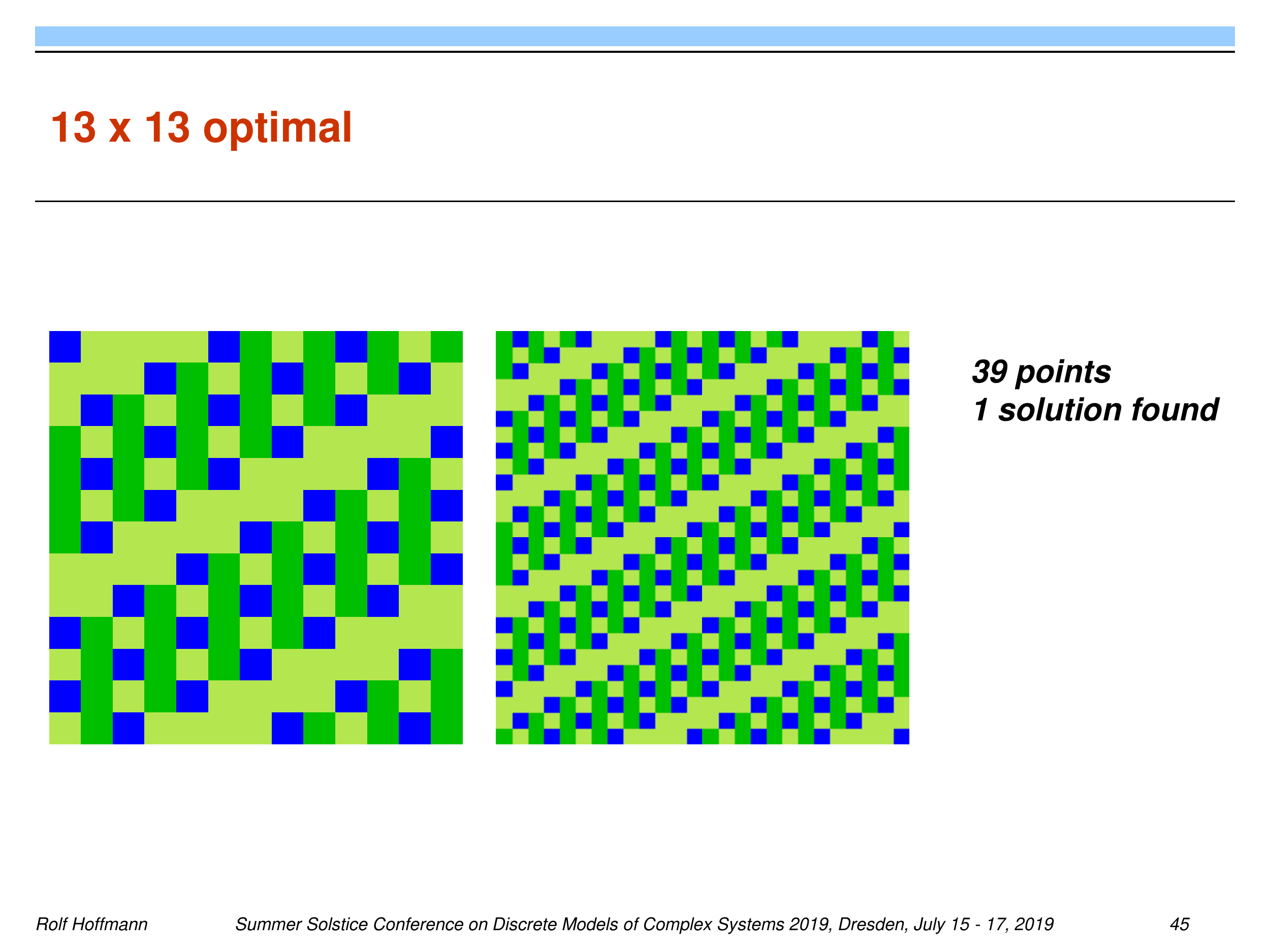}

\includegraphics[width=2.3cm]{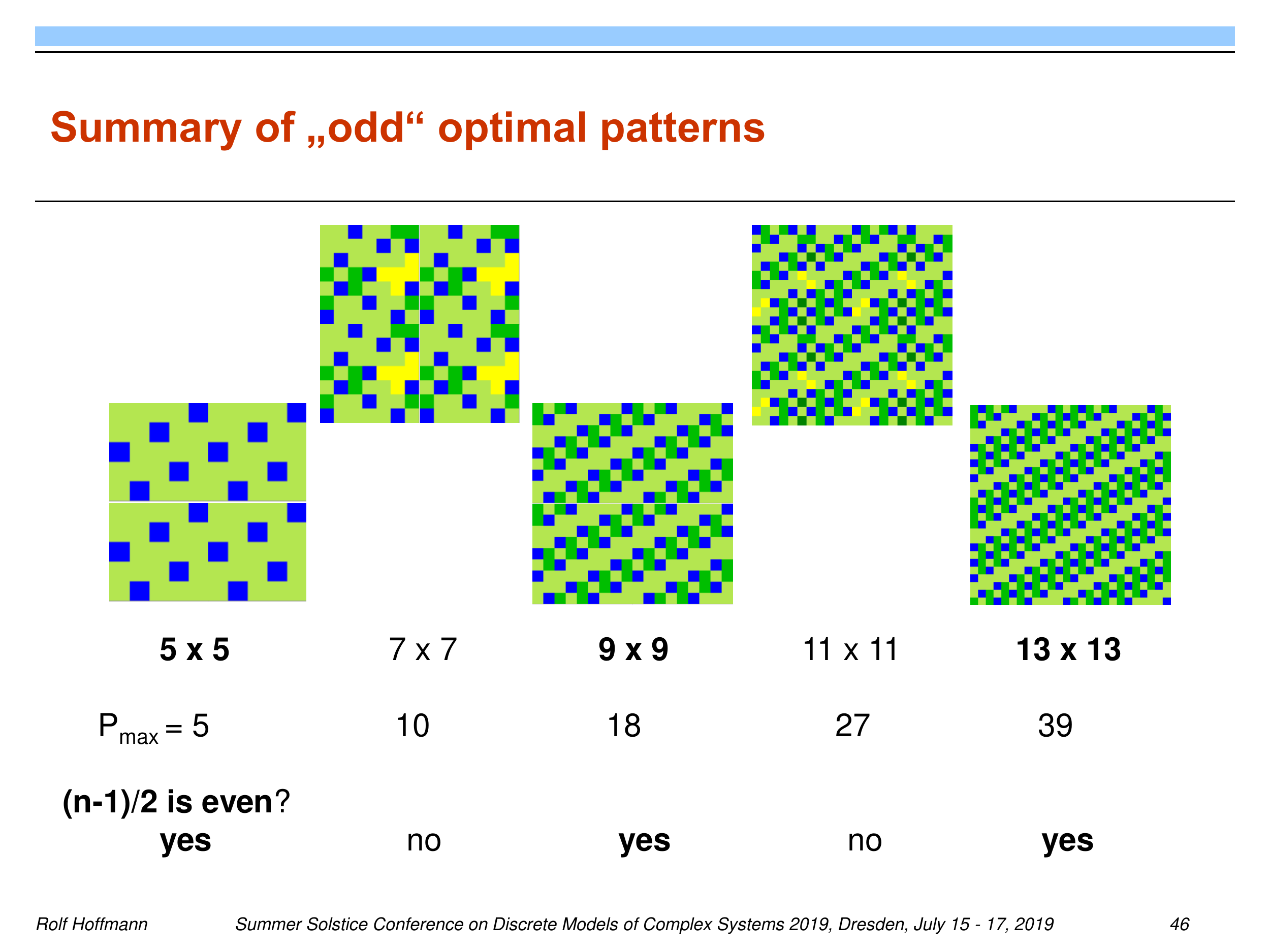}
\includegraphics[width=2.3cm]{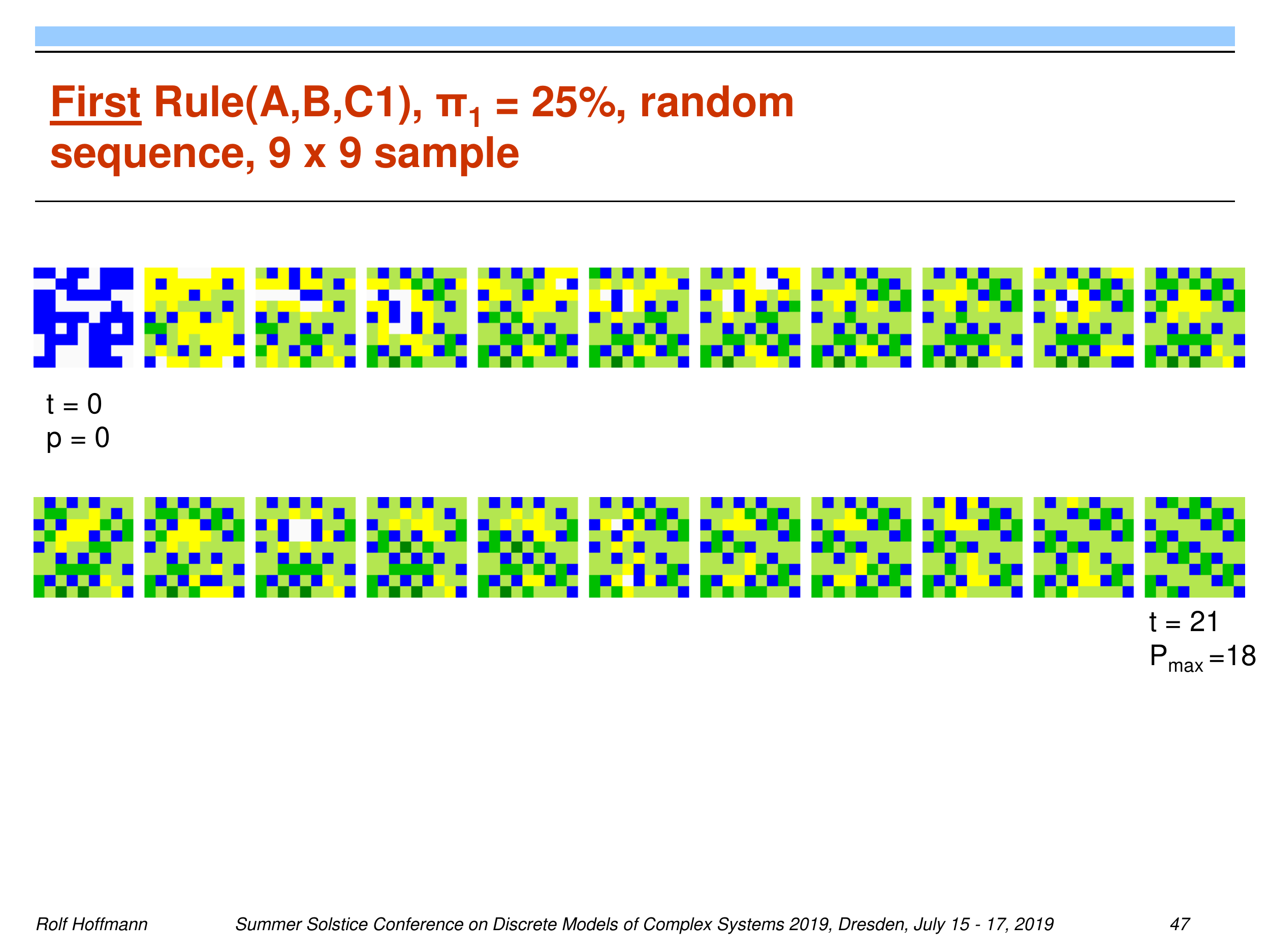}
\includegraphics[width=2.3cm]{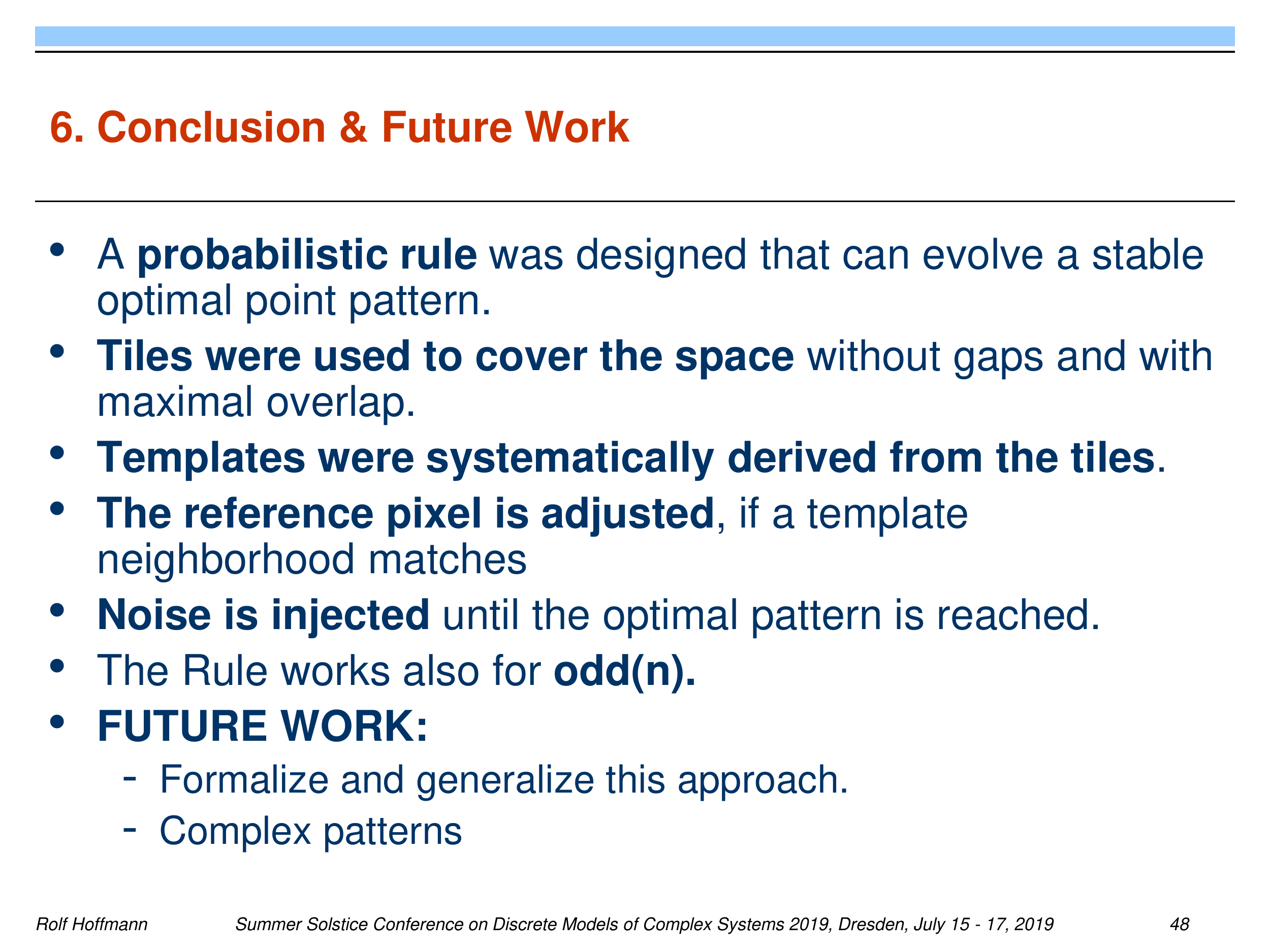}
\caption{
Slides presented at Summer Solstice Conference on Discrete Models of Complex Systems 2019, Dresden, July 15 - 17, 2019 
}
\label{presentation}
\end{figure}

\newpage

\end{document}